\documentstyle[jkas,subfigure]{article}

\def\deg{^{\circ}}

\def\fdg{\hbox{$.\!\!^\circ$}}

\beginpage{1}
\endpage{4}
\year{2013}\volume{00}\month{April}

\runningauthor {K.Y. KO ET AL.} 
\runningtitle{Local anomalies in the WMAP 7-year CMB angular power spectrum}

\month{April} \year{2013} \volume{00} \issueno{0}
\beginpage{1}\endpage{18}
\date{Received February 1, 2013; Accepted March 21, 2013}

\begin{document}
\title{Local anomalies around the third peak in the CMB angular power spectrum \\ of the WMAP 7-year data} 
\author{Kyeong Yeon Ko$^{1,3}$, Chan-Gyung Park$^{2}$, and Jai-chan Hwang$^{3}$   } 
\address{$^1$ Korea Astronomy and Space Science Institute, 
                     Daejeon, Korea \\}
\address{$^2$ Division of Science Education and Institute of Fusion Science,
                     Chonbuk National University, Jeonju, Korea \\}
\address{$^3$ Department of Astronomy and Atmospheric Sciences,
                     Kyungpook National University, Daegu, Korea \\            
{\it E-mail : kyeongyeon.ko@gmail.com, parkc@jbnu.ac.kr, jchan@knu.ac.kr}}
\address{\normalsize{\it (Received February 1, 2013; Accepted March 21, 2013)}}
\offprints{C.-G. Park}
\abstract{
We estimate the power spectra of CMB temperature anisotropy in localized
regions on the sky using the WMAP 7-year data. Here, we report that the
north hat and the south hat regions at the high Galactic latitude
($|b|\ge 30\deg$) show anomaly in the power spectrum amplitude around
the third peak, which is statistically significant up to $3\sigma$.
We try to figure out the cause of the observed anomaly by analyzing
the low Galactic latitude ($|b|<30\deg$) regions where the galaxy
contamination is expected to be stronger, and regions that are weakly
or strongly dominated by the WMAP instrument noise.
We also consider the possible effect of unresolved radio point sources.
We found another but less statistically significant anomaly in the low
Galactic latitude north and south regions whose behavior is opposite
to the one at the high latitude.
Our analysis shows that the observed north-south anomaly at high latitude
becomes weaker on the regions with high number of observations
(weak instrument noise), suggesting that the anomaly is significant
at sky regions that are dominated by the WMAP instrument noise.
We have checked that the observed north-south anomaly has
weak dependences on the bin-width used in the power spectrum estimation
and the Galactic latitude cut. We have also discussed the possibility that
the detected anomaly may hinge on the particular choice of the multipole
bin around the third peak.
We anticipate that the issue of whether the anomaly is intrinsic one
or due to the WMAP instrument noise will be resolved by the forthcoming
Planck data.}

\maketitle

\section{Introduction}
The cosmic microwave background radiation (hereafter CMB) provides us
with a wealth of information on the history of the universe. The thermal
black body nature of the CMB energy spectrum is now considered as the firm
evidence of the hot big bang scenario for the beginning of the universe
\citep{Alpher-etal-1948,Dicke-etal-1965,Penzias-etal-1965}.
The existence of large-scale structure in the universe also implies that
there were primordial density perturbations as the seeds for structure
formation. It was expected that these inhomogeneities would have the
imprint on the CMB as the minute temperature fluctuations (anisotropy)
\citep{Sachs-etal-1967,Peebles-etal-1970,Bond-etal-1987}.
The CMB anisotropy was discovered by the Cosmic Background Explorer (COBE)
Differential Microwave Radiometers experiment \citep{Smoot-etal-1992}
and has been confirmed by many ground-based and balloon-borne
experiments (see \citealt{Hu-etal-2002,Scott-etal-2006} for reviews
and references therein).

Recently, the Wilkinson Microwave Anisotropy Probe (WMAP) has opened
a new window to the precision cosmology by measuring the CMB temperature
anisotropy and polarization with high resolution and sensitivity
\citep{Bennett-etal-2003,Jarosik-etal-2011}.
For every data release, the WMAP team presented their estimation of
the angular power spectra for temperature and polarization anisotropy
\citep{Hinshaw-etal-2003,Hinshaw-etal-2007,Page-etal-2003,
Page-etal-2007,Nolta-etal-2009,Larson-etal-2011}.
By comparing the measured CMB power spectra with the theoretical prediction,
the WMAP team determined the cosmological parameters with a few \% precision
\citep{Spergel-etal-2003,Spergel-etal-2007,Komatsu-etal-2009,Komatsu-etal-2011},
and found that the observed CMB fluctuations are consistent with predictions
of the concordance $\Lambda\textrm{CDM}$ model with scale-invariant adiabatic
fluctuations generated during the inflationary epoch
\citep{Spergel-etal-2003,Peiris-etal-2003,Komatsu-etal-2011}.
Recent ground-based and balloon-borne experiments that have performed
the CMB power spectrum measurement and the cosmological parameter estimation
include the South Pole Telescope (SPT) \citep{Keisler-etal-2011},
the QUaD experiment \citep{Brown-etal-2009},
Arcminute Cosmology Bolometer Array Receiver (ACBAR) \citep{Reichardt-etal-2009},
the Cosmic Background Imager (CBI) \citep{Mason-etal-2003},
the Atacama Cosmology Telescope (ACT) \citep{Fowler-etal-2010},
the Degree Angular Scale Interferometer (DASI) \citep{Carlstrom-etal-2003},
BOOMERANG \citep{Jones-etal-2006}, Archeops \citep{Benoit-etal-2003},
and MAXIMA \citep{Lee-etal-2001}.

In the CMB data analysis the two-point statistics such as the correlation
function and the power spectrum has been widely used.
In particular, the relation between the CMB
angular power spectrum and the cosmological physics is well understood,
and the tight constraints on the cosmological parameters can be directly
obtained by comparing the measured power spectrum with the theoretical
prediction. Therefore, the accurate estimation of the angular power spectrum
from the observed CMB maps is the essential step.
The efficient techniques to measure the angular power spectrum from the
CMB temperature fluctuations with incomplete sky coverage have been
constantly developed (e.g., \citealt{Gorski-1994,Tegmark-1997,Bond-etal-1998,Oh-etal-1999,
Szapudi-etal-2001,Wandelt-etal-2001,Hansen-etal-2002,Hivon-etal-2002,Mortlock-etal-2002,Hinshaw-etal-2003,
Wandelt-etal-2003,Chon-etal-2004,Efstathiou-2004,Eriksen-etal-2004a,Wandelt-etal-2004,Brown-etal-2005,
Polenta-etal-2005,Fay-etal-2008,Dahlen-etal-2008,Das-etal-2009,Mitra-etal-2009,Ansari-etal-2010,Chiang-etal-2012}).

Until now, the analysis of the WMAP CMB data has been made using the
whole sky area except for strongly contaminated regions
\citep{Hinshaw-etal-2003,Hinshaw-etal-2007,Nolta-etal-2009,
Larson-etal-2011,Saha-etal-2006,Saha-etal-2008,Souradeep-etal-2006,
Eriksen-etal-2007a,Samal-etal-2010,Basak-etal-2012}.
On the other hand, there was only a small number of studies on the
power spectrum measurement on the partial regions of the sky
(e.g., \citealt{Eriksen-etal-2004b,Hansen-etal-2004a,Hansen-etal-2004b,
Ansari-etal-2010,Yoho-etal-2011,Chiang-etal-2012}).
Especially, \citet{Yoho-etal-2011} detected degree-scale
anomaly around the first acoustic peak of the CMB angular power spectrum
measured on a small patch of the north ecliptic sky.
In this work, we measure the angular power spectra from the WMAP
7-year temperature anisotropy data set on some specified regions of the sky.
We found that at high Galactic latitude regions there is the north-south
anomaly or asymmetry in the power amplitude around the third peak of
the angular power spectrum, which is the main result of this paper.
Our result differs from the well-known hemispherical asymmetry in
the angular power spectrum and the genus topology at large angular scales
\citep{Park-2004,Eriksen-etal-2004b,Eriksen-etal-2004c,Eriksen-etal-2007b,
Hansen-etal-2004b,Bernui-2008,Hansen-etal-2009,Hoftuft-etal-2009}.

This paper is organized as follows.
In Sec.\ \ref{sec:angular}, we describe the method of how to measure
the angular power spectrum. The application of the method to the WMAP 7-year
data and the comparison with the WMAP team's measurement are shown in
Sec.\ \ref{sec:application}.
In Sec.\ \ref{sec:power}, we measure the power spectra on various sky regions
defined with some criteria, and present our discovery of the north-south
anomaly around the third peak of the angular power spectrum at high Galactic
latitude regions. We also discuss the potential origin of such a north-south
anomaly. Section \ref{sec:conclusions} is the conclusion of this work.
Throughout this work we have used the HEALPix software
\citep{Gorski-etal-1999,Gorski-etal-2005} to measure the pseudo angular power
spectra and to generate the WMAP mock data sets.

\section{Angular power spectrum estimation method}
\label{sec:angular}

In this section we briefly review how the angular power spectrum is measured
from the observed CMB temperature anisotropy maps. Throughout this work
we do not consider the CMB polarization.

In the ideal situation where the temperature fluctuations on the whole
sky are observed, we can expand the temperature distribution $T(\mathbf{n})$
in terms of spherical harmonics as
\begin{equation}
   T({\mathbf{n}})
       = \sum_{l=2}^{\infty}\sum^{l}_{m=-l} a_{lm}Y_{lm}({\mathbf{n}}),
\label{eq:Tn}
\end{equation}
where $\mathbf{n}$ denotes the angular position on the sky,
$Y_{lm}(\mathbf{n})$ is the spherical harmonic basis function,
and the monopole ($l=0$) and dipole ($l=1$) components have been neglected.
From the orthogonality condition of the spherical harmonics, the coefficients
$a_{lm}$ can be obtained from
\begin{equation}
   a_{lm} = \int d\Omega T({\mathbf{n}})Y^{*}_{lm}({\mathbf{n}}),
\label{eq:alm}
\end{equation}
where $d\Omega$ denotes the differential solid angle on the sky.
If the temperature anisotropy is statistically isotropic,
then the variance of the harmonic coefficients $a_{lm}$ is independent
of $m$ and the angular power spectrum $C_l$ is given as the ensemble average
of two-point product of the harmonic coefficients,
\begin{equation}
\langle a_{lm}a^{*}_{l'm'} \rangle = C_l \delta_{ll'}\delta_{mm'},
\end{equation}
where the bracket represents the ensemble average and $\delta_{ll'}$
is the Kronecker delta symbol ($\delta_{ll'}=1$ for $l=l'$, and
$\delta_{ll'}=0$ for $l\ne l'$).
For a Gaussian temperature distribution, it is known that all the statistical
information is included in the two-point statistics like the angular power
spectrum $C_l$.

For the single realization of temperature fluctuations on the sky,
the ensemble average is replaced with the simple average of independent modes
belonging to the same multipole $l$ and the angular power spectrum becomes
\begin{equation}
   C_{l}^{\textrm{\scriptsize sky}} = \frac{1}{2l+1} \sum^{l}_{m=-l} |a_{lm}|^{2},
\label{eq:C_l}
\end{equation}
with uncertainty due to the cosmic variance
\begin{equation}
    \Delta C_l = \sqrt{\frac{2}{2l+1}} C_l.
\label{eq:Delta_Cl}
\end{equation}

In the practical situation, however, we usually exclude some portion of
the sky area where the contamination due to the Galactic emission
and the strong radio point sources is expected to affect the statistics of
the CMB anisotropy significantly.
The limited angular resolution, the instrument noise, and the finite pixels
of the processed map also prevent us from using the formulas,
Eqs. (\ref{eq:Tn})--(\ref{eq:Delta_Cl}), which are valid only
in the ideal situation.
The incomplete sky coverage with an arbitrary geometry and the
limited instrumental performance break the orthogonality relation between
the spherical harmonic functions. Thus, measuring angular power spectrum
from the realistic observational data is more complicated.

There are two popular ways of measuring the angular power spectrum
from the observed CMB data with incomplete sky coverage.
One is the maximum likelihood estimation method based on the Bayesian
theorem \citep{Bond-etal-1998,Tegmark-1997},
and the other is the pseudo power spectrum method that applies the
Fourier or harmonic transformation directly \citep{Hivon-etal-2002}.
For the latter, \citet{Hivon-etal-2002} developed
the so called MASTER (Monte Carlo Apodised Spherical Transform Estimator)
method which estimates the angular power spectrum from the high resolution
CMB map data. For mega-pixelized data sets like those from WMAP or Planck
\citep{Planck-etal-2011a}, the MASTER method is faster and more efficient than
the maximum likelihood method.

We refer to the pseudo angular power spectrum as $\tilde{C_{l}}$
to distinguish from the true angular power spectrum $C_l$ which we try
to estimate. Two quantities differ from each other due to the
incomplete sky coverage, the limited angular resolution and pixel size, and
the instrument noise, and they are related by \citep{Hansen-etal-2002,Hivon-etal-2002}
\begin{equation}
   \langle \tilde{C_{l}} \rangle
      = \sum_{l'}M_{ll'}F_{l'}B^{2}_{l'} \langle C_{l'} \rangle
      + \langle \tilde{N_{l}} \rangle,
\label{eq:pseudo cl}
\end{equation}
where
$B_{l}$ is the beam transfer function, $F_{l}$ is the pixel transfer
function due to finite pixel size of the CMB map,
$\langle \tilde{N}_{l}\rangle$ is the average pseudo noise power spectrum
due to the instrument noise, and $M_{ll'}$ is the mode-mode coupling
matrix that incorporates all the effect due to the incomplete sky coverage.
The mode-mode coupling matrix is written as
\begin{equation}
   M_{l_{1}l_{2}}
      = \frac{2l_{2} + 1}{4\pi} \sum_{l_{3}} (2l_{3} + 1)
        {\mathcal{W}}_{l_{3}} \left \lgroup \begin{array}{ccc}
        l_{1} & l_{2} & l_{3} \\
        0 & 0 & 0
        \end{array} \right\rgroup^{2},
\label{eq:mode_coupling}
\end{equation}
where ${\mathcal{W}}_l$ is the power spectrum of the window function
for incomplete sky coverage (see Sec.\ \ref{sec:weighting}),
and the last factor on the right-hand side enclosed with the parenthesis
represents the Wigner 3-$j$ symbol. We use the Fortran language and
the Mathematica software\footnote{Wolfram Research Inc., http://wolfram.com/mathematica}
to calculate the Wigner 3-$j$ symbols;
see also \citet{Hivon-etal-2002} and \citet{Brown-etal-2005} for the numerical
computation of the Wigner 3-$j$ symbols.

To reduce the statistical variance of the measured angular power spectrum
due to the cosmic variance and the instrument noise, we usually average
powers within the appropriate $l$ bins. According to \citet{Hivon-etal-2002},
the binning operator is defined as
\begin{displaymath}
   P_{bl}= \bigg\{ \begin{array}{ll}
            \frac{1}{2\pi}\frac{l(l+1)}{l^{(b+1)}_\textrm{\tiny low}
               -l^{(b)}_\textrm{\tiny low}},  &
            \textrm{if $2 \leq l^{(b)}_\textrm{\scriptsize low} \leq
                 l < l^{(b+1)}_\textrm{\scriptsize low}$}  \\
            0, & \textrm{otherwise}.
\label{binning operator}
\end{array}
\end{displaymath}
The reciprocal operator is defined as
\begin{displaymath}
   Q_{lb} = \bigg\{ \begin{array}{ll}
            \frac{2\pi}{l(l+1)}, & %
            \textrm{if $2 \leq l^{(b)}_\textrm{\scriptsize low}
            \leq l < l^{(b+1)}_\textrm{\scriptsize low}$} \\
            0, & \textrm{otherwise},
\label{reciprocal operator}
\end{array}
\end{displaymath}
where $b$ is the bin index. For ease of comparison, we use exactly
the same $l$-binning adopted by the WMAP teams ($b$ runs from 1 to 45)
\citep{Larson-etal-2011}. The true power spectrum can be estimated from
\begin{equation}
   \hat{C}_b = (K^{-1})_{bb'}P_{b'l}(\tilde{C}_{l}-\langle\tilde{N}_{l}\rangle),
\label{binned_power_spectrum}
\end{equation}
where
\begin{equation}
   K_{bb'}=P_{bl} M_{ll'}F_{l'}B_{l'}^2 Q_{l'b'}
\label{kernel}
\end{equation}
is a kernel matrix that describes the mode-mode couplings
due to the survey geometry, the finite resolution and pixel size,
and the binning process.

Ideally the average pseudo noise power spectrum $\langle \tilde{N}_{l}\rangle$
can be estimated by the Monte Carlo simulation that mimics the instrument
noise.
However, because of the statistical fluctuations in the noise power spectra
itself, it is rather difficult to obtain an accurate estimation of
the power contribution due to the instrument noise from the single
observed data.
In our analysis we evade this problem by measuring the cross-power
spectra between different channel data sets, in the same way as adopted
by the WMAP team (see the next section).

\section{Application to the WMAP 7-year data}
\label{sec:application}

\subsection{WMAP 7-year data}

The WMAP mission was designed to make the full sky CMB temperature and
polarization maps with high accuracy, precision, and reliability
\citep{Bennett-etal-2003}.
The WMAP instrument has 10 differencing assemblies (DAs) spanning five
frequency bands from 23 to 94 GHz:
one DA each at 23 GHz (K1) and 33 GHz (Ka1), two DAs each at 41 GHz (Q1, Q2)
and 61 GHz (V1, V2), and four DAs at 94 GHz (W1, W2, W3, W4),
with $0\fdg82$, $0\fdg62$, $0\fdg49$, $0\fdg33$, and $0\fdg21$ FWHM beam
widths, respectively.
In this work we use the foreground-reduced WMAP 7-year temperature fluctuation
maps that are prepared in the HEALPix format with
$N_\textrm{\scriptsize side}=512$ (resolution 9; r9).
The total number of pixels of each map is $12\times N_\textrm{\scriptsize side}^2
= 3,145,728$. Following the WMAP team's analysis, we use only V and W band
maps to reduce any possible contamination due to the Galactic foregrounds.
We use the WMAP beam transfer function ($B_l$) and the number of observations
($N_\textrm{\scriptsize obs}$) for each DA.
All the data sets used in this work are available
on the NASA's Legacy Archive for Microwave Background Data
Analysis (LAMBDA)\footnote{http://lambda.gsfc.nasa.gov}.

\subsection{Weighting schemes}
\label{sec:weighting}

\begin{figure}[!t]
\centering
\epsfxsize=8cm   \epsfbox{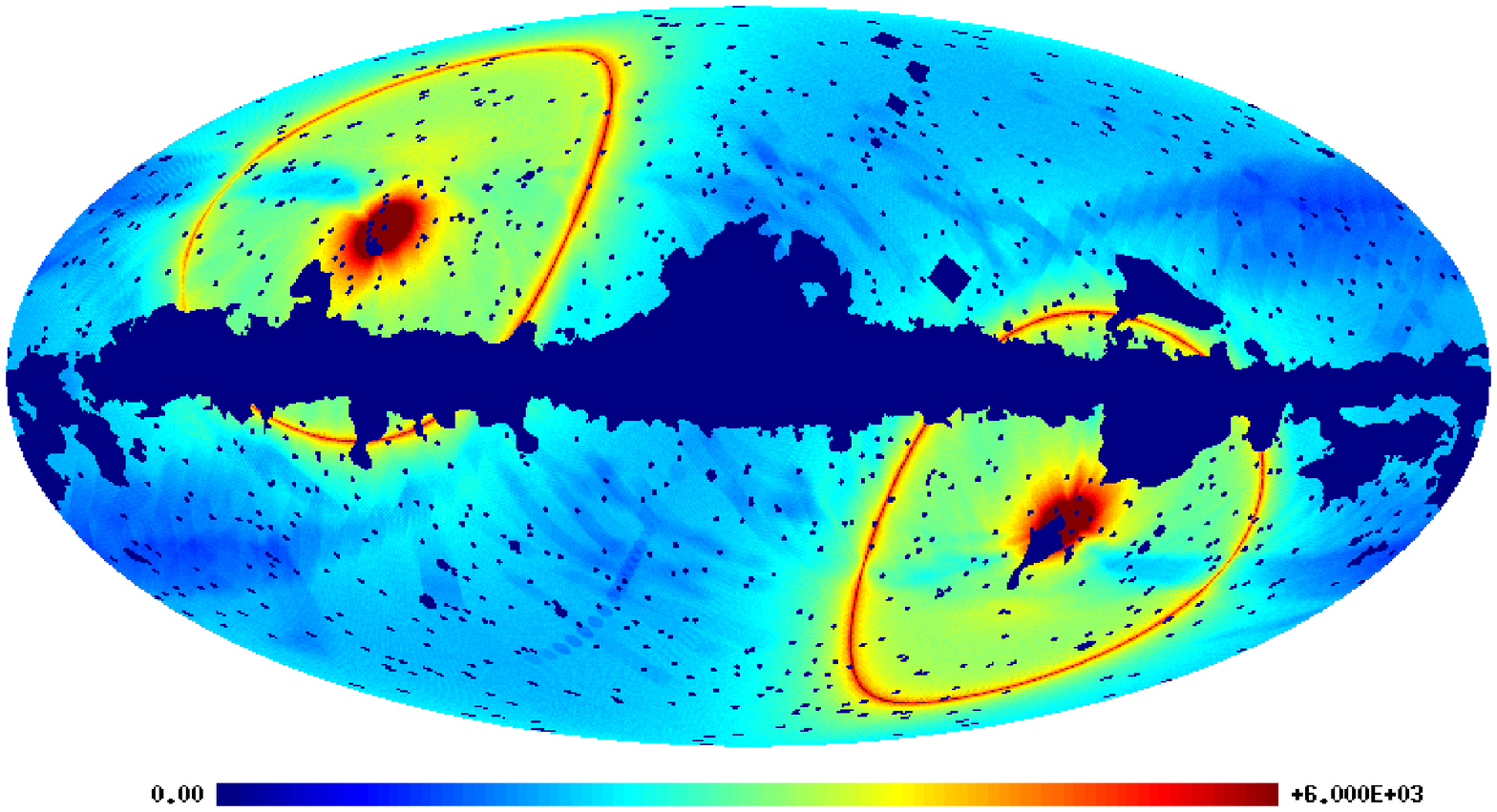}
\epsfxsize=8.5cm \epsfbox{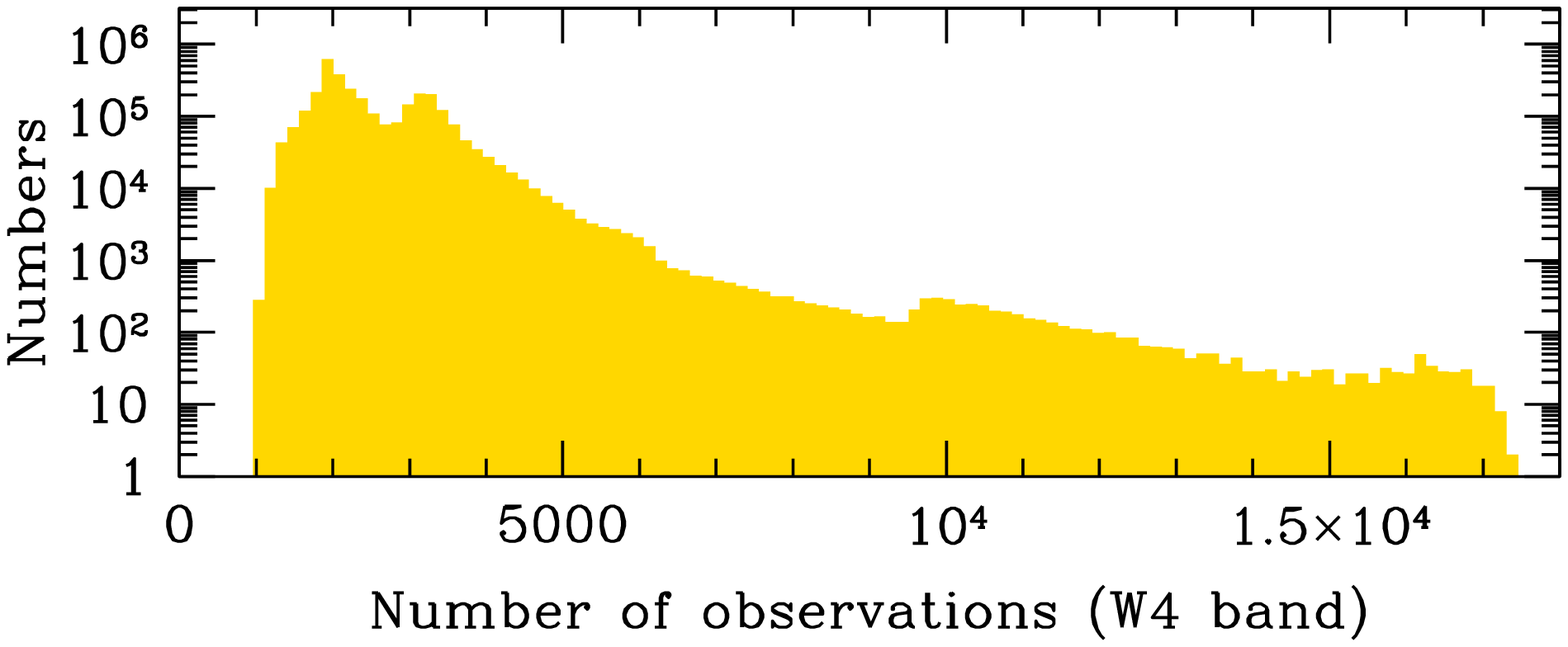}
\caption{(Top) A map of number of observations $N_\textrm{\scriptsize obs}(p)$
         at the WMAP 7-year W4 frequency channel (DA), multiplied with
         the KQ85 mask map $M(p)$ [see Eq.\ (\ref{window function})].
         Dark blue color corresponds to zero values
         and represents regions that are excluded by the KQ85 mask map,
         while dark red color denotes the value exceeding
         $N_\textrm{\scriptsize obs} \ge 6000$.
         The Mollweide projection in Galactic coordinates
         is used to display this map, where the Galactic center is located
         at the center, the Galactic longitude increases from center
         to left ($l=0\deg$--$180\deg$) and from right to center
         ($l=180\deg$--$360\deg$), and the Galactic latitude increases
         from bottom to top ($b=-90\deg$--$+90\deg$).
         The regions with large number of observations correspond
         to the ecliptic pole regions.
         (Bottom) A histogram of number of observations at the same
         frequency channel.
         }
\label{N_obs}
\end{figure}

The window function assigns a weight to the temperature fluctuation
on each pixel before the harmonic transformation is performed. In our analysis
the window function maps have been produced based on two weighting schemes,
the uniform weighting scheme and the inverse-noise weighting scheme
\citep{Hinshaw-etal-2003}.
The WMAP team provides mask maps which assigns unity to pixels that are used
in the analysis and zero to pixels that are excluded to avoid the foreground
contamination. We use the KQ85 mask map with resolution 9 that includes
about 78.3\% of the whole sky area.
The mask map excludes the Galactic plane region with the strong Galactic
emission and the circular areas with $0\fdg6$ radius centered on the strong
radio point sources.
The uniform weighting scheme uses the mask map as the window function
directly,
\begin{equation}
W(p)=M(p),
\label{eq:uniform_weighting}
\end{equation}
where $M(p)$ denotes a mask map on a pixel $p$ on the sky.
The power spectrum at high $l$ region is usually dominated by
the instrument noise. In the WMAP data the noise level on each pixel
is modeled by
\begin{equation}
\sigma (p) = \frac{\sigma_{0}}{\sqrt{N_\textrm{\scriptsize obs}(p)}},
\label{noise level}
\end{equation}
where $N_\textrm{\scriptsize obs}(p)$ is the number of observations on the pixel $p$
and $\sigma_0$ is the global noise level ($\sigma_0=3.319$, $2.955$, $5.906$,
$6.572$, $6.941$, $6.778$ mK for V1, V2, W1, W2, W3, W4 DAs, respectively).
To reduce the effect of the instrument noise, the inverse-noise weighting
scheme is defined as the product of the mask map and the map of number of
observations,
\begin{equation}
   W(p)=M(p) N_\textrm{\scriptsize obs}(p).
\label{window function}
\end{equation}
The window function at 94 GHz W4 frequency channel (DA) is shown
in Fig.\ \ref{N_obs}, together with a histogram of the number of observations.
Although the WMAP team applies the uniform weighting scheme for $l<600$
and the inverse-noise weighting scheme for $l>600$ \citep{Larson-etal-2011},
in this work we simply apply the single weighting scheme over the whole
range of $l$ considered ($2 \le l \le 1200$) during the power spectrum
estimation and compare the results based on the two weighting schemes.

\subsection{Angular power spectrum measured from the WMAP data}

In the WMAP team's data analysis, the power spectrum at low multipoles
($l \le 32$) was measured by a Blackwell-Rao estimator that is applied
to a chain of Gibbs samples obtained from the foreground-cleaned CMB map
while the power spectrum at high multipoles ($32 < l \le 1200$)
by the MASTER pseudo power spectrum estimation method
\citep{Larson-etal-2011}.
Throughout this work, the angular power spectrum over the whole $l$ range
($2 \le l \le 1200$) is measured based on the pseudo power spectrum estimation
method because our primary attention is paid on the high $l$ region where
the peaks are located.
Thus, as shown below our results at low multipoles are somewhat different
from the WMAP team's results.

The WMAP V and W bands have two (V1 and V2) and four (W1, W2, W3, and W4)
separate frequency channels (DAs).
Using the Anafast program in the HEALPix package, we measured 15 pseudo
cross power spectra ($\tilde{C}_l$) for different channel combinations
(V1W1, V1W2, and so on) based on the uniform and the inverse-noise weighting
schemes.
During the production of each pseudo cross power spectrum,
Anafast program estimates two separate sets of harmonic coefficients
for each DA using the formula Eq.\ (\ref{eq:alm}) where $T({\mathbf{n}})$
is now replaced with $W({\mathbf{n}})T({\mathbf{n}})$ and $W({\mathbf{n}})$
is the window function defined in Eqs.\ (\ref{eq:uniform_weighting})
and (\ref{window function}) for the corresponding DA.
We set the maximum multipole as $l=1200$.

For each channel combination, we calculate the mode-mode coupling matrix
$M_{ll'}$ and the kernel matrix $K_{bb'}$ using the beam transfer,
the pixel transfer functions, and the same $l$-binning as defined by the WMAP
team \citep{Larson-etal-2011}.
For the power spectrum of window function in Eq.\ (\ref{eq:mode_coupling})
we use the cross power spectrum of window functions for the corresponding DA
combination, which is obtained from the Anafast program by inserting
the window functions at two different frequency channels as the input data.
The squared beam transfer function $B_l^2$ appearing in
Eq.\ (\ref{eq:pseudo cl}) is also modified into the product of beam transfer
functions from the two different frequency channels.

The final binned cross power spectrum $\hat{C}_b$ for each channel combination
is obtained from Eq.\ (\ref{binned_power_spectrum}) neglecting the term
for the average noise power spectrum $\langle \tilde{N}_l \rangle$.
Since the properties of the instrument noise for each channel are independent,
the process of cross-correlation statistically suppresses the contribution
of the instrument noise to the power spectrum estimation.
This technique has an advantage that our power spectrum estimator is not biased
by noise if the noise in the two independent channels is uncorrelated;
the cross-correlation between the uncorrelated noise vanishes statistically.
Thus, the measured cross-power spectra are independent of instrument noise
from the individual channels in every practical sense (see \citealt{Hinshaw-etal-2003}).
Note that the self-combination of each frequency channel data (like V1V1, V2V2, W1W1)
is not considered during the analysis because it is essentially needed to
subtract the contribution of noise power from the measured auto-power spectrum and
it is a difficult task to precisely estimate the noise power from the single observed
map (see the discussion at the end of Sec.\ \ref{sec:angular}).
For each binned cross power spectrum we subtract the contribution due to the
unresolved radio point sources expected in the
WMAP temperature anisotropy maps by using the information presented in
\citet{Nolta-etal-2009} and \citet{Larson-etal-2011}.
The amplitude of the binned power spectrum due to the unresolved point sources
can be written as
\begin{equation}
   C_{\textrm{\scriptsize ps},b}^i = A_\textrm{\scriptsize ps} P_{bl}S_l^i.
\end{equation}
Here $S_l^i$ is the point-source spectral function given by
\begin{equation}
   S_l^{i} \equiv
       r(\nu_k) r(\nu_{k'})
       \left(\frac{\nu_{k}\nu_{k'}}{\nu_{\textrm{\scriptsize Q}}^2}\right)^{\beta}
\label{point-source spectral function}
\end{equation}
where $\nu_{k}$ and $\nu_{k'}$ denote the individual frequency channel
belonging to $i$-th channel combination, $r(\nu_k)$ is a conversion factor
from antenna to thermodynamic temperature,
$\nu_\textrm{\scriptsize Q}=40.7~\textrm{GHz}$ is the Q-band central frequency,
$10^3 A_\textrm{\scriptsize ps}=9.0 \pm 0.7~\mu\textrm{K}^2$, and $\beta=-2.09$
(see also \citealt{Huffenberger-etal-2006,Souradeep-etal-2006}).
The subindex $l$ is dummy because the spectral function does not depend on it.

To combine these cross power spectra into the single angular power spectrum,
in fact we need the Monte Carlo simulation data sets with the similar
properties of the WMAP observational data.
Using the Synfast program of HEALPix and assuming the concordance
flat $\Lambda\textrm{CDM}$ model with parameters,
$\Omega_{b}h^{2}=0.02260$, $\Omega_{c}h^{2}=0.1123$,
$\Omega_\Lambda=0.728$, $n_s=0.963$, $\tau=0.087$,
$\Delta_{\mathcal{R}}^2(k=0.002~\textrm{Mpc}^{-1})=2.441\times 10^{-9}$
(Table 14 of \citealt{Komatsu-etal-2011}),
we have made 1000 mock data sets that mimic the WMAP beam
resolution and instrument noise for each channel.
We use the CAMB software to obtain the theoretical model power spectrum
\citep{Lewis-etal-2000}.
We assumed that the temperature fluctuations and instrument noise follow
the Gaussian distribution. In the simulation data sets we do not consider
any effect due to the residual foreground contamination.
We have analyzed the one thousand WMAP simulation mock data sets
in the same way as the real data set is analyzed.

\begin{figure*}[t!]
\centering
\epsfxsize=8.0cm \epsfbox{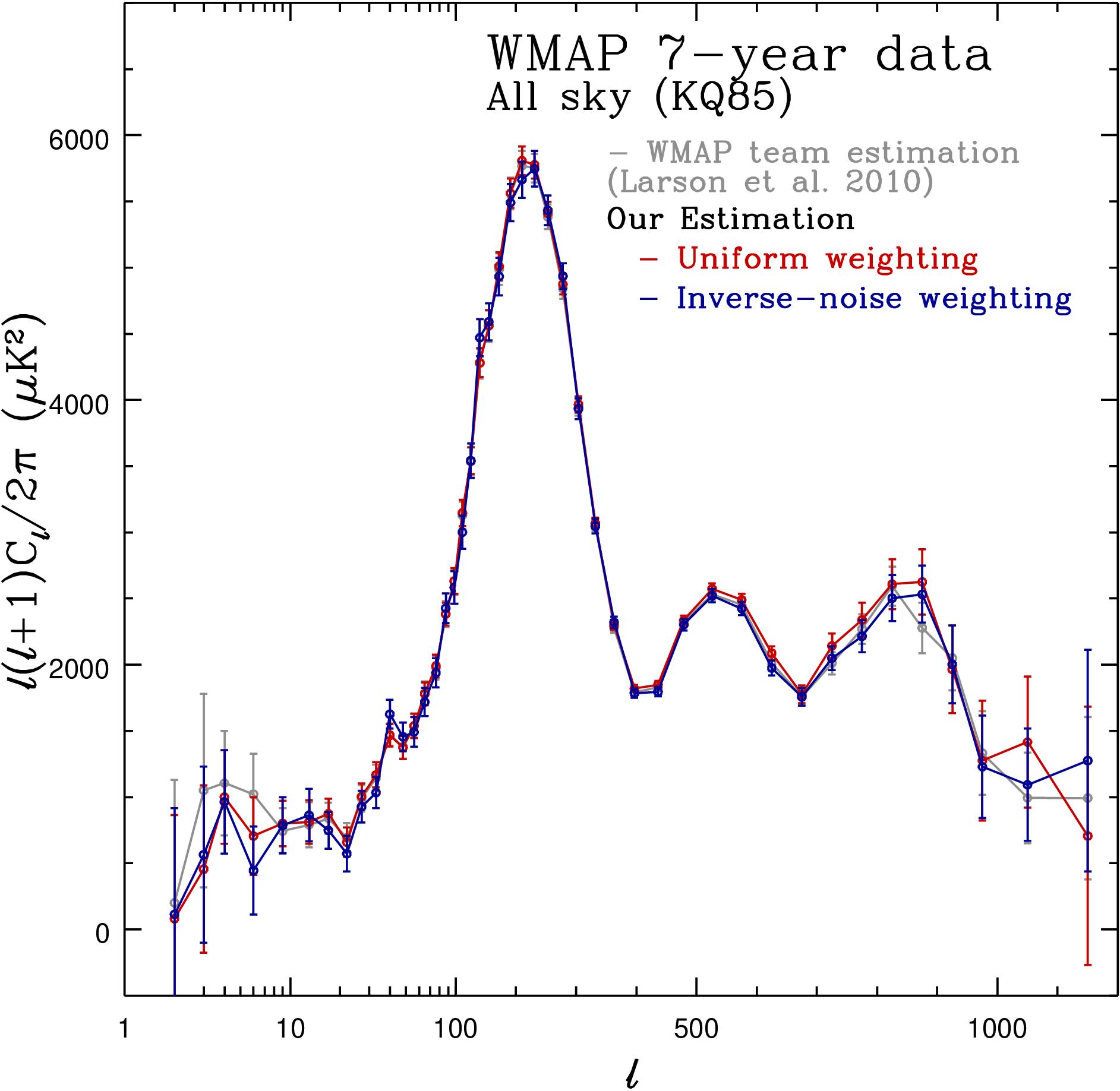}
\epsfxsize=8.0cm \epsfbox{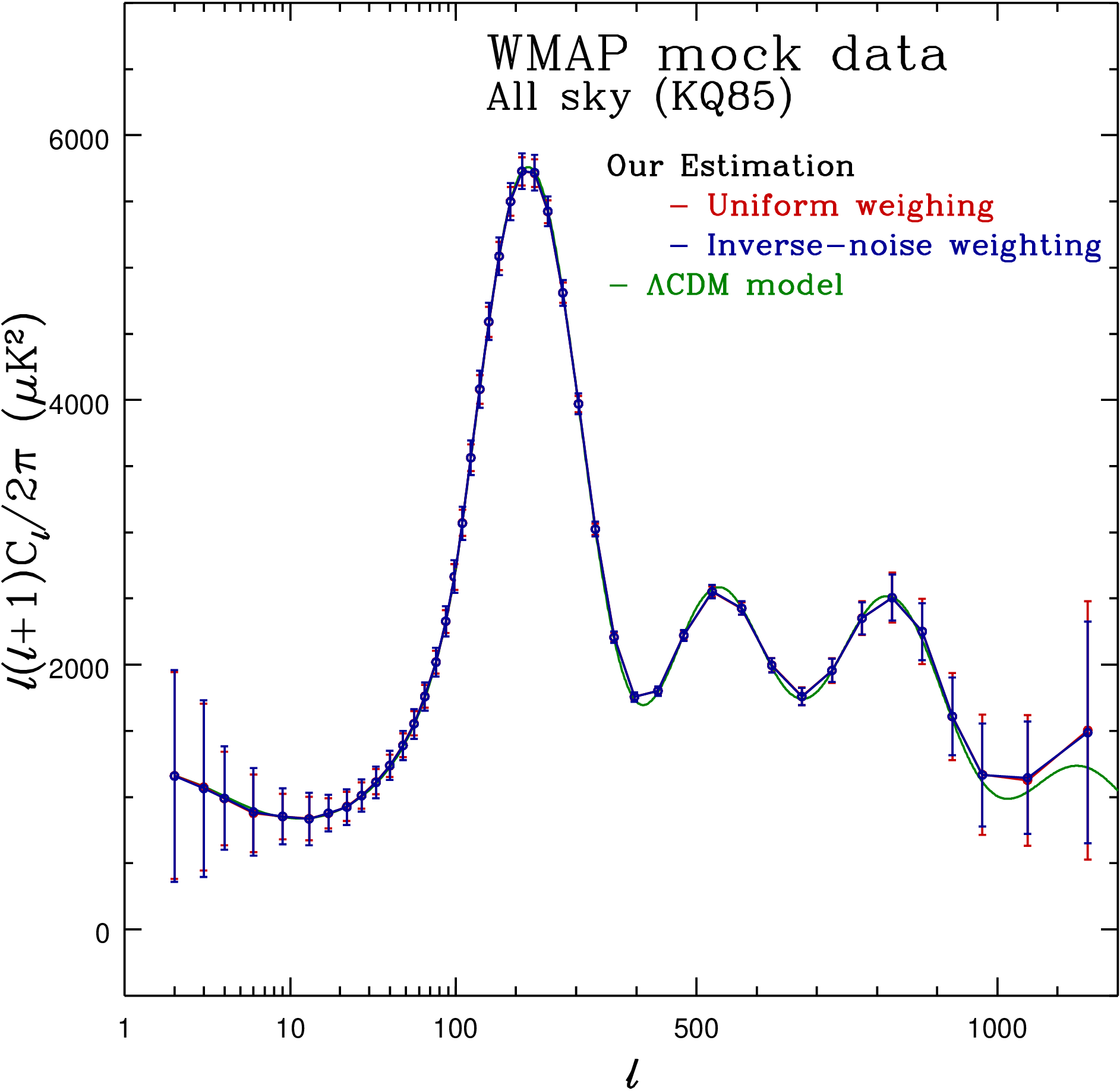}
\caption{
    (Left) Angular power spectra measured from the WMAP 7-year temperature
    anisotropy maps. Our results are shown as red and blue curves with dots
    and error bars for uniform and inverse-noise weighting schemes,
    respectively. The sky area (78.27\% of the whole sky) defined in the KQ85
    mask map has been used.
    For a comparison, the WMAP team's result is also shown as
    grey curve with dots and error bars \citep{Larson-etal-2011}.
    (Right) Averaged angular power spectra measured from the one thousand
    WMAP mock data sets that are consistent with the concordance
    $\Lambda\textrm{CDM}$ model \citep{Komatsu-etal-2011}.
    Red color denotes the result based on the uniform weighting scheme
    while blue based on the inverse-noise weighting scheme.
    The power spectrum of the assumed concordance $\Lambda\textrm{CDM}$ model
    is shown as green curve. }
\label{all sky power spectrum}
\end{figure*}

Before combining the cross power spectra into the single power spectrum,
each cross power spectrum has been averaged within $l$-bins specified
to reduce the statistical fluctuations due to the cosmic variance and the
instrument noise. We use the same $l$-bins defined by the WMAP 7-year data
analysis \citep{Larson-etal-2011}.
Thus, in a case when we measure the power spectrum from the WMAP data with
a sky fraction much smaller than the area enclosed by KQ85 mask,
the measured powers at low and high $l$ regions become uncertain and
fluctuate significantly with strong cross-correlations between
adjacent bins (see below).
We average the 15 binned cross power spectra into the single angular
power spectrum by applying the combining algorithm which is similar to
that used in the WMAP first year data analysis \citep{Hinshaw-etal-2003}.
For each $l$-bin, we construct a covariance matrix defined as
\begin{equation}
   (\Sigma_\textrm{\scriptsize full})^{ij}_{b}
      = \langle [C_{b}^i-\bar{C}^{i}_{b}][C_{b}^j-\bar{C}^{j}_{b}]\rangle
        + (\Sigma_\textrm{\scriptsize src})^{ij}_{b},
\label{full_covariance_matrix}
\end{equation}
where $i$ and $j$ denote the cross combination of the WMAP frequency channels
(V1V2, V1W1, W1W3, and so on), $b$ is the binning index ($b=1$,\ldots,$45$),
$C_b^i$ is the measured cross power spectrum for channel combination $i$,
$\bar{C}_b^i$ is the ensemble averaged cross power spectrum for the same
channel combination expected in the concordance $\Lambda\textrm{CDM}$
model (we have used one thousand WMAP mock data sets to estimate
the ensemble averaged quantities), and
\begin{equation}
   (\Sigma_\textrm{\scriptsize src})^{ij}_{b}
       = (P_{bl} S_l^{i}) (P_{bl'} S_{l'}^{j}) \sigma^{2}_\textrm{\scriptsize src}
\end{equation}
is the covariance matrix between errors expected to be caused during
the subtraction of powers due to unresolved point sources.
We use ${\sigma^{2}_\textrm{\scriptsize src}}=(\delta A_\textrm{\scriptsize ps})^{2}
=(0.0007~\mu\textrm{K}^2\textrm{sr})^2$.
The final combined angular power spectrum is obtained by
\begin{equation}
   \hat{C}_{b}
      = \frac{\sum_{i=1}^{15}\sum_{j=i}^{15} \hat{C}^{i}_{b}
         (\Sigma^{-1}_\textrm{\scriptsize full})^{ij}_{b}}
         {\sum_{i=1}^{15}\sum_{j=i}^{15}
           (\Sigma^{-1}_\textrm{\scriptsize full})^{ij}_{b}} .
\label{final_power_spectrum}
\end{equation}
To estimate the uncertainty for the combined power spectrum $\hat{C}_b$,
we obtain one thousand combined power spectra ($\hat{C}_b^{\textrm{\scriptsize sim}}$)
together with their average and standard deviation at each bin.
The standard deviation measured at each bin is used as the uncertainty
(or error bar) for the measured combined power spectrum.
By comparing the theoretical model power spectrum with the average
power spectrum obtained from the simulation data sets, we can check
whether our power spectrum measurement algorithm works correctly or not.

\begin{figure*}[t!]
\centering%
\subfigure[High latitude, north hat with $b\ge 30\deg$ (24.11\%)]
{\epsfxsize=4.07cm \epsfbox{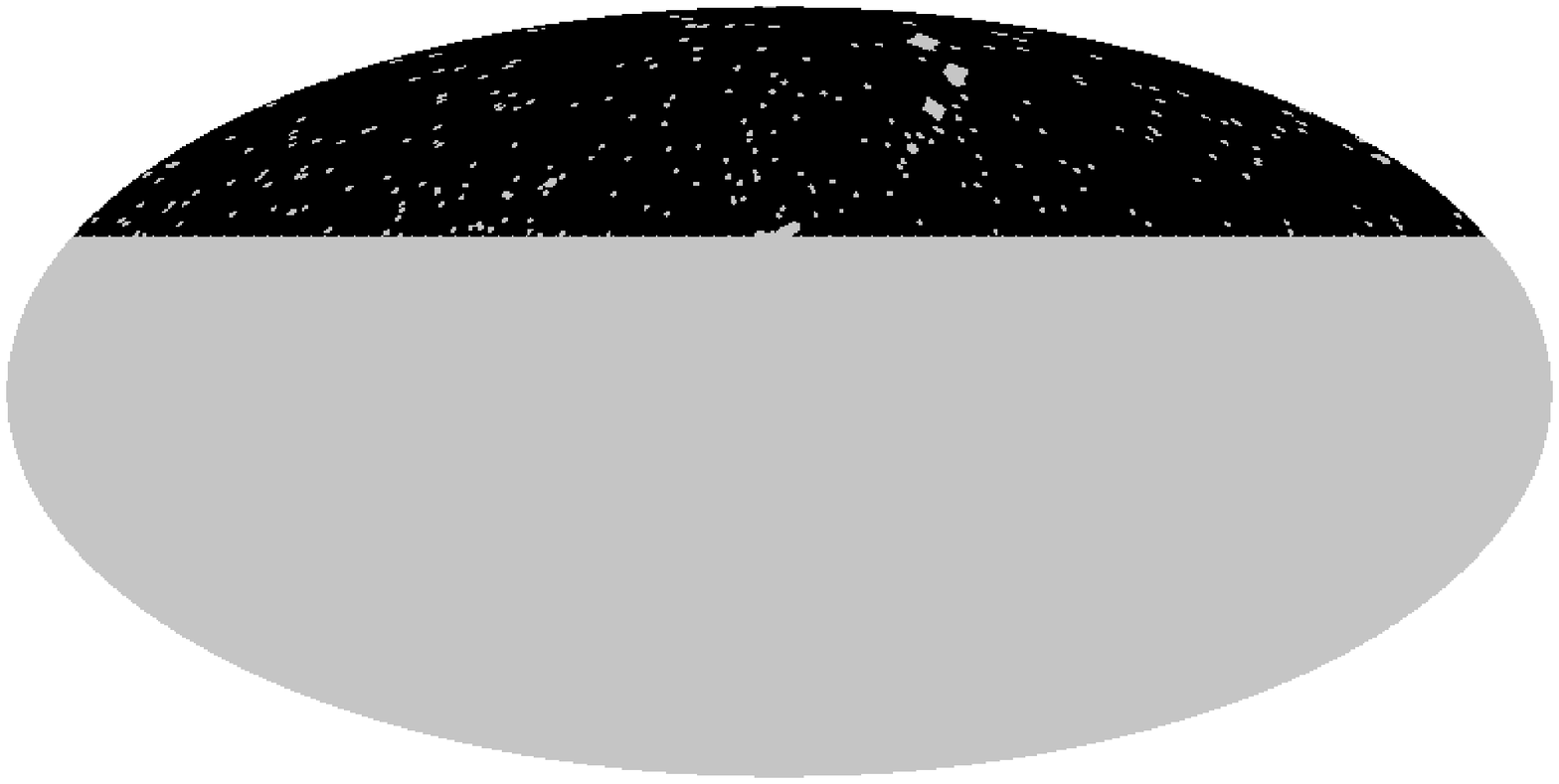}}
\subfigure[High latitude, south hat with $b \le -30\deg$ (23.98\%)]
{\epsfxsize=4.07cm \epsfbox{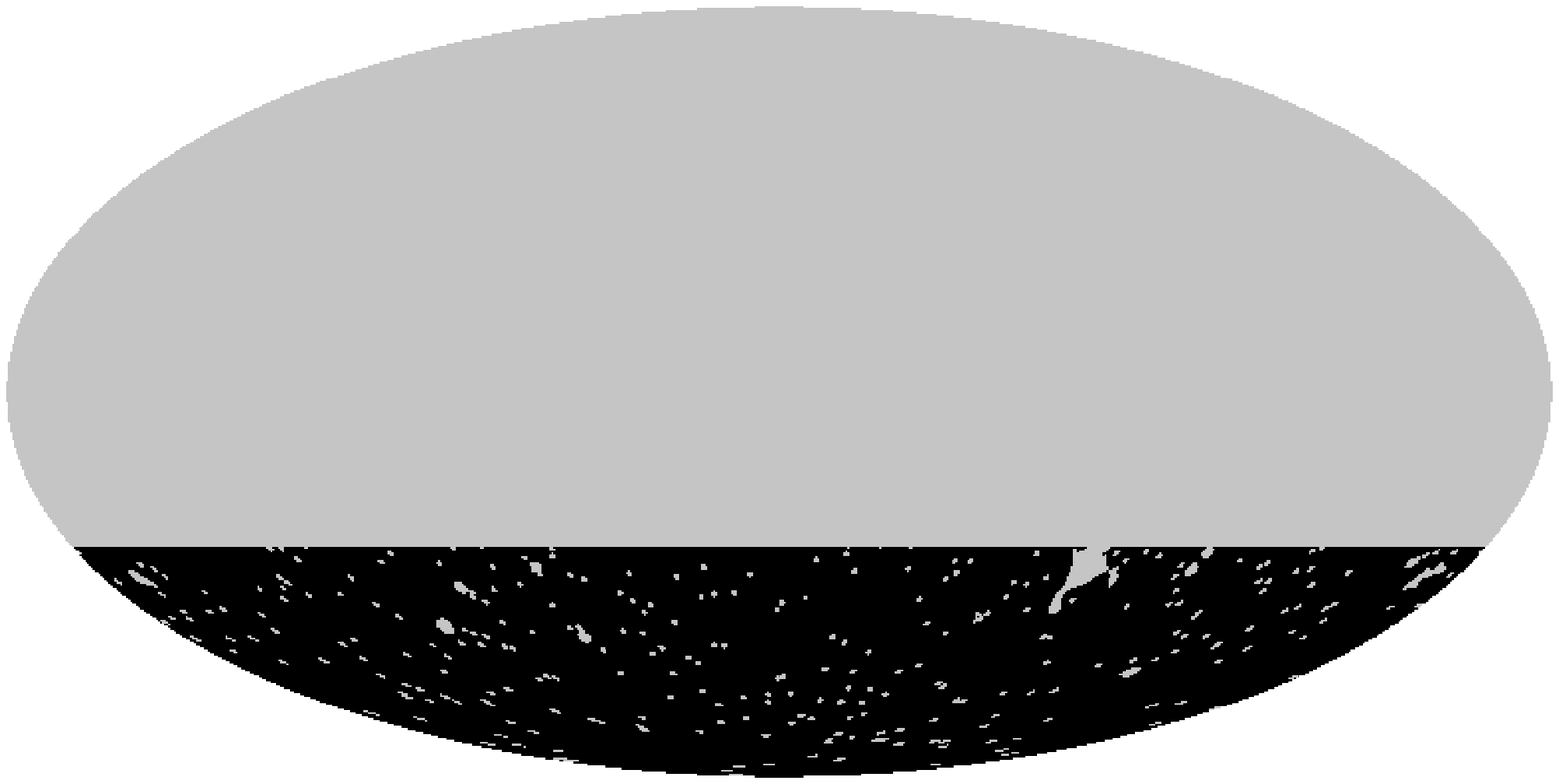}}
\subfigure[Low latitude, north region with $0\deg < b < 30\deg$ (15.15\%)]
{\epsfxsize=4.07cm \epsfbox{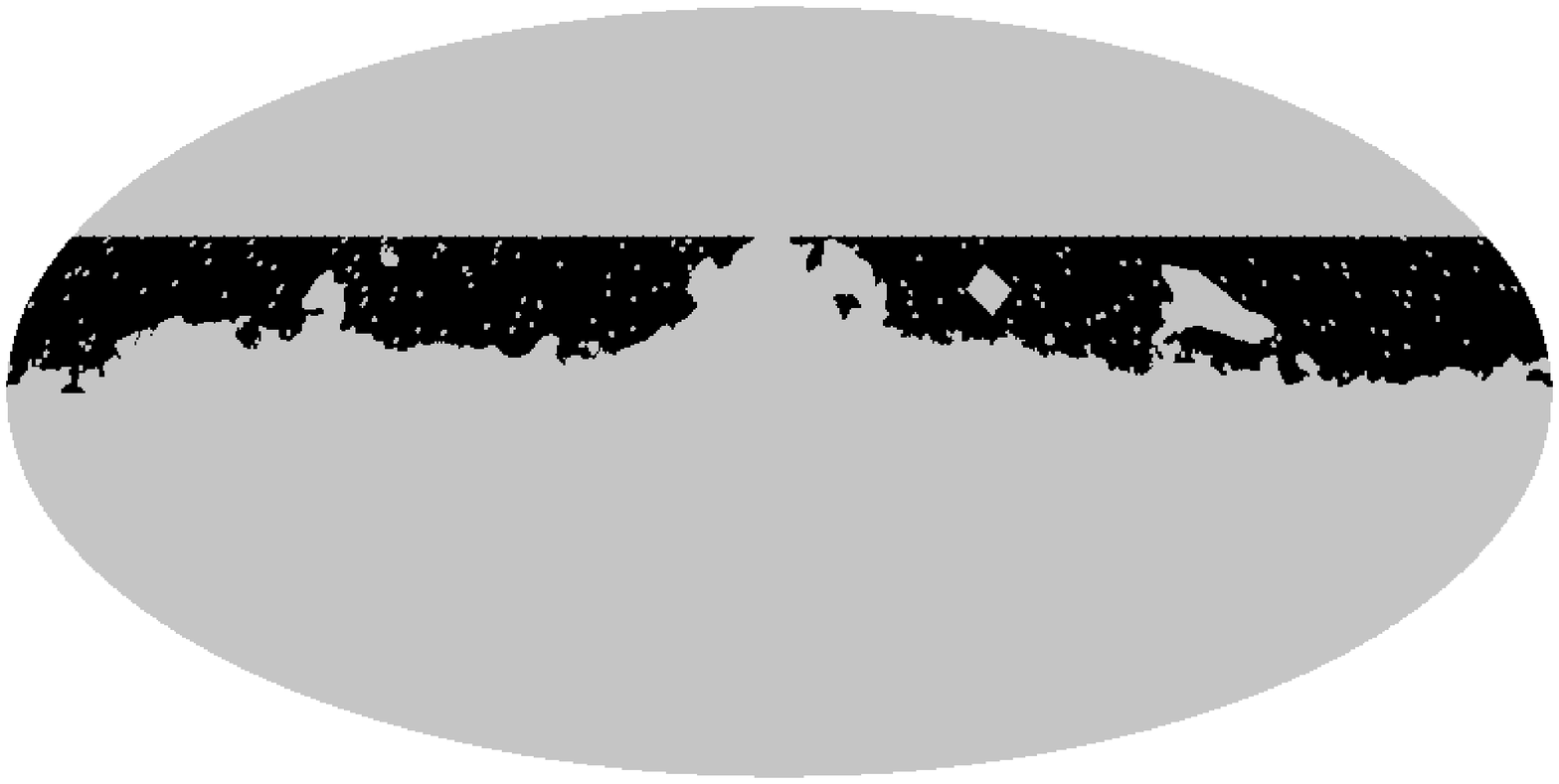}}
\subfigure[Low latitude, south region with $-30\deg < b < 0\deg$ (15.03\%)]
{\epsfxsize=4.07cm \epsfbox{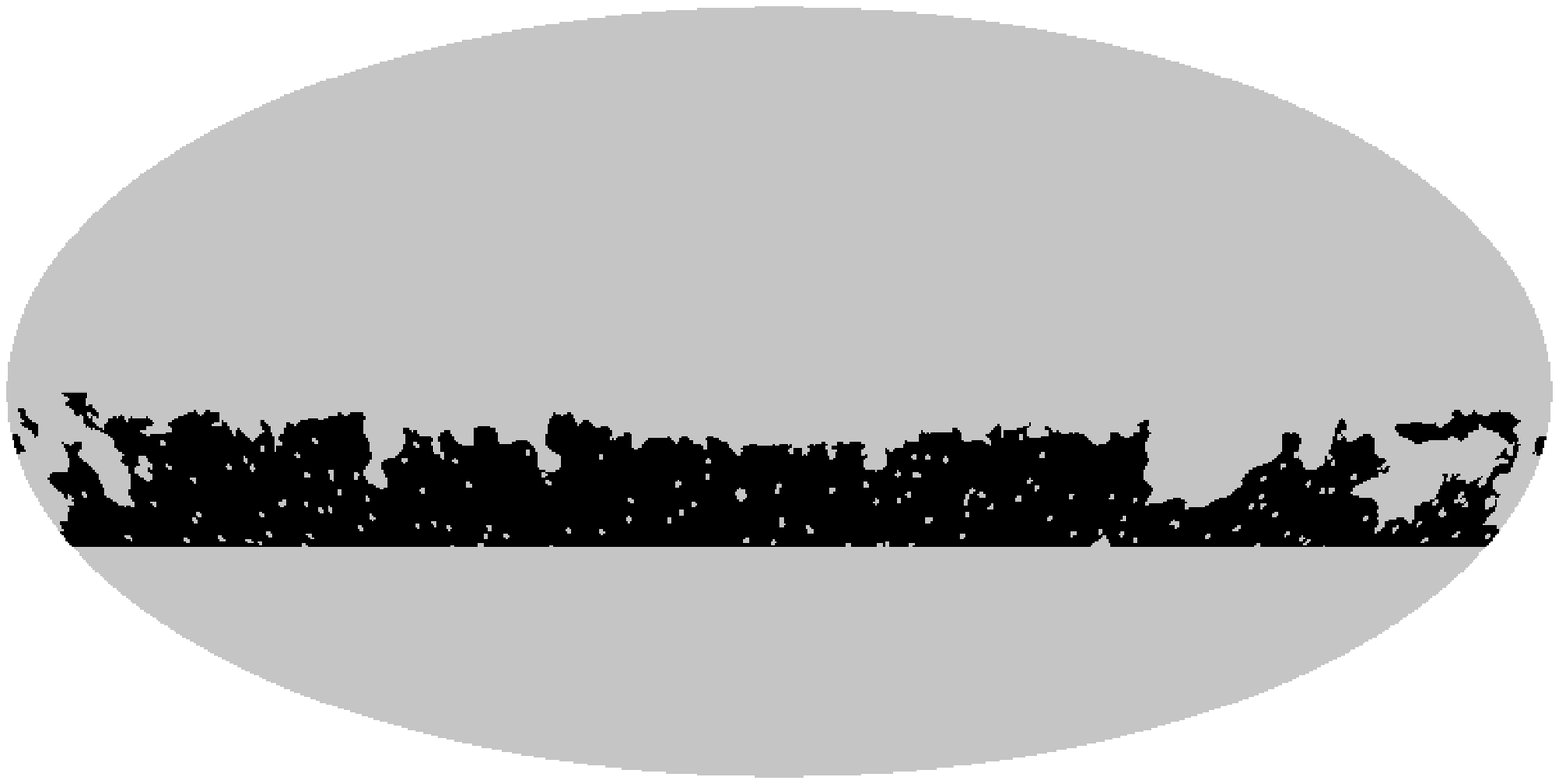}}
\subfigure[North hat with high instrument noise (15.20\%)]
{\epsfxsize=4.07cm \epsfbox{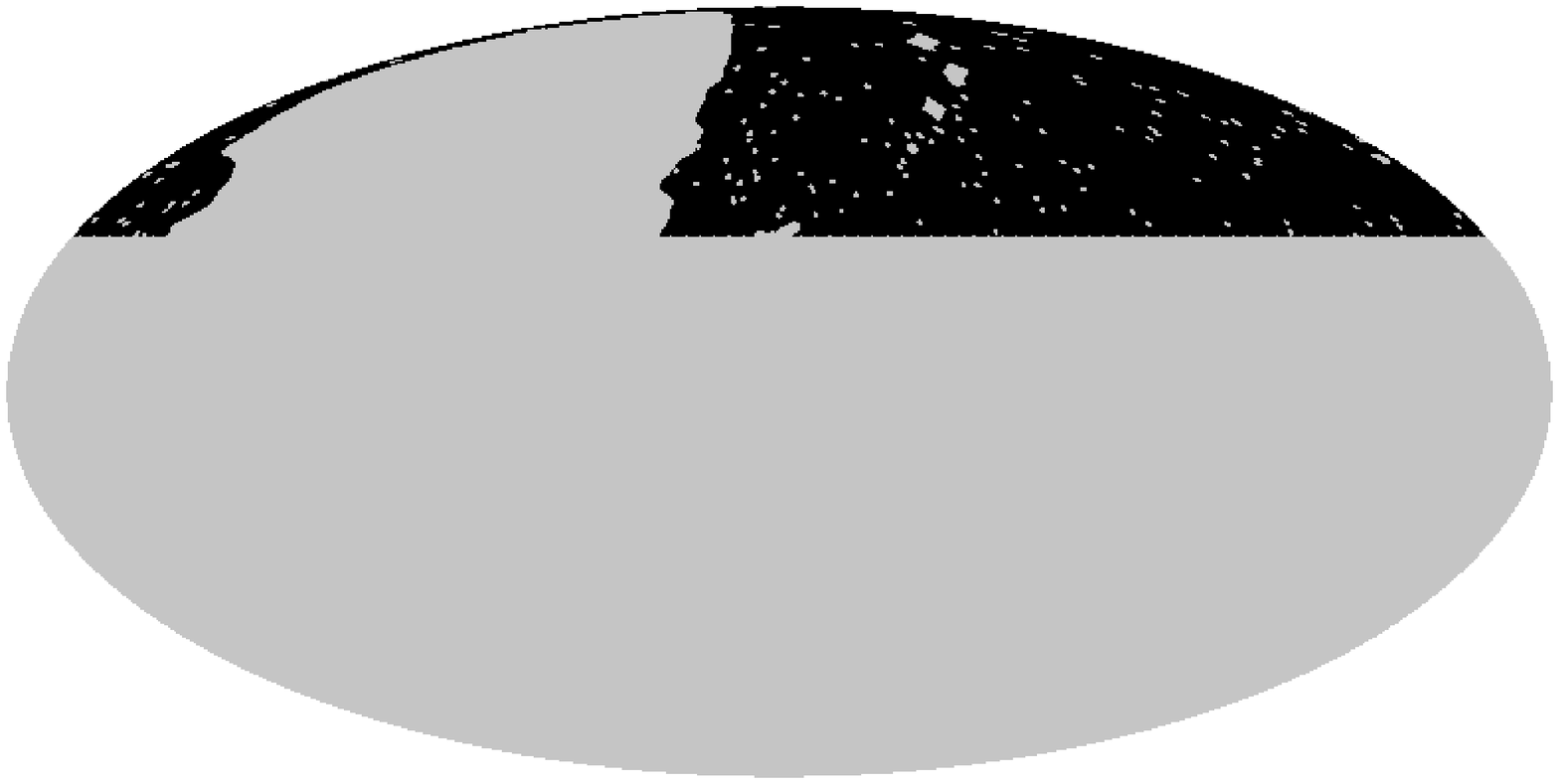}}
\subfigure[South hat with high instrument noise (15.14\%)]
{\epsfxsize=4.07cm \epsfbox{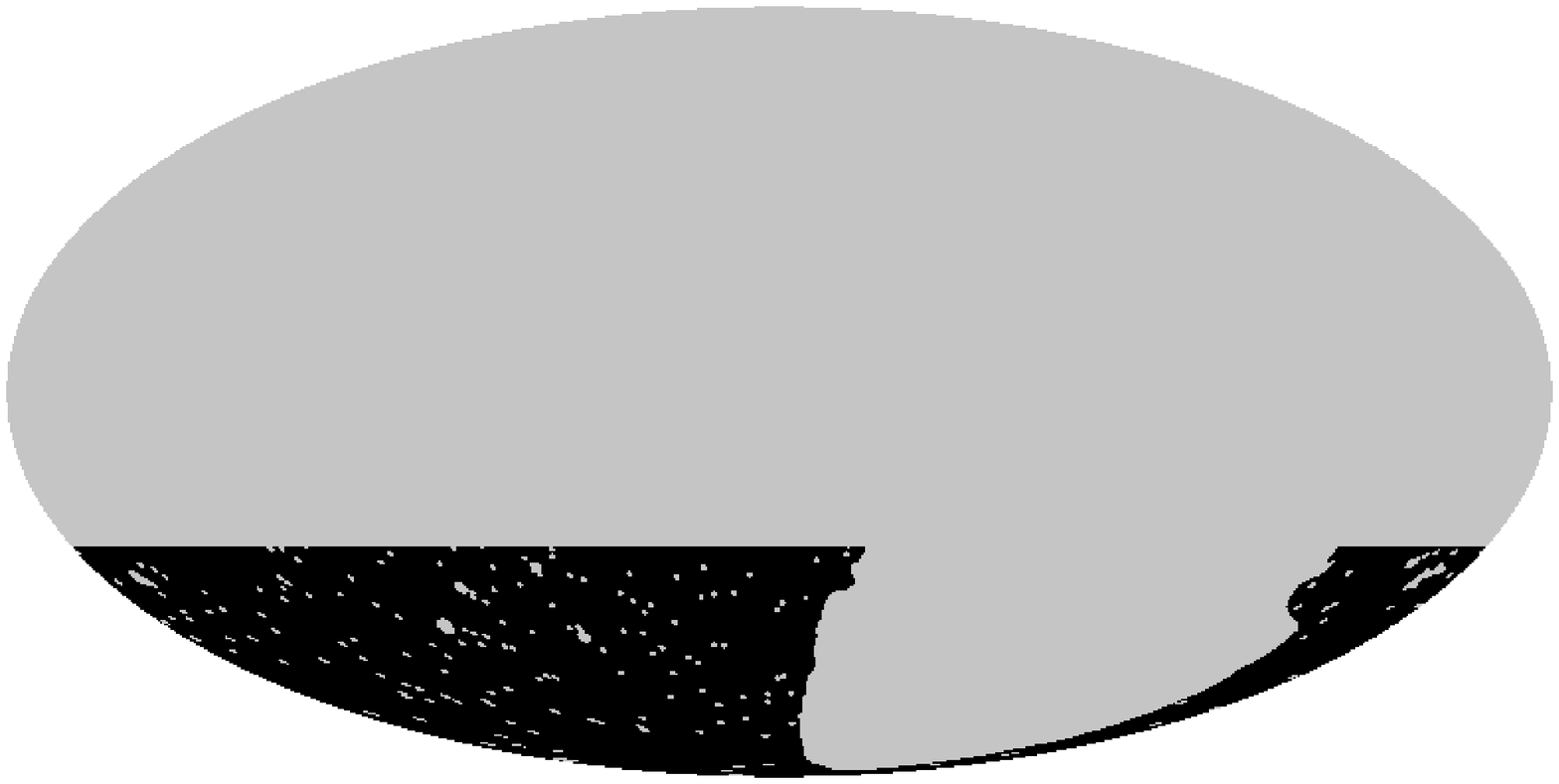}}
\subfigure[North hat with low instrument noise (13.05\%)]
{\epsfxsize=4.07cm \epsfbox{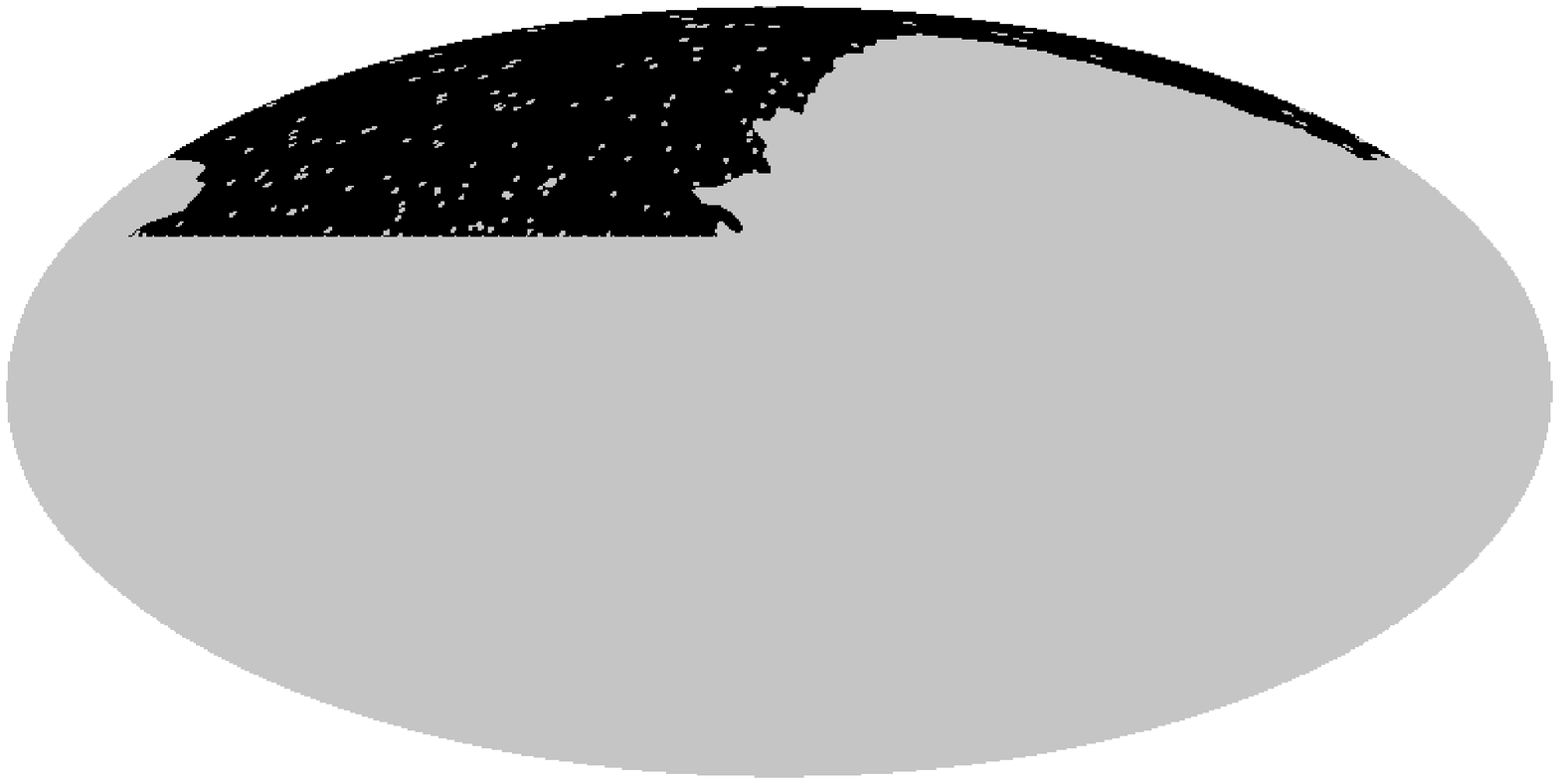}}
\subfigure[South hat with low instrument noise (13.01\%)]
{\epsfxsize=4.07cm \epsfbox{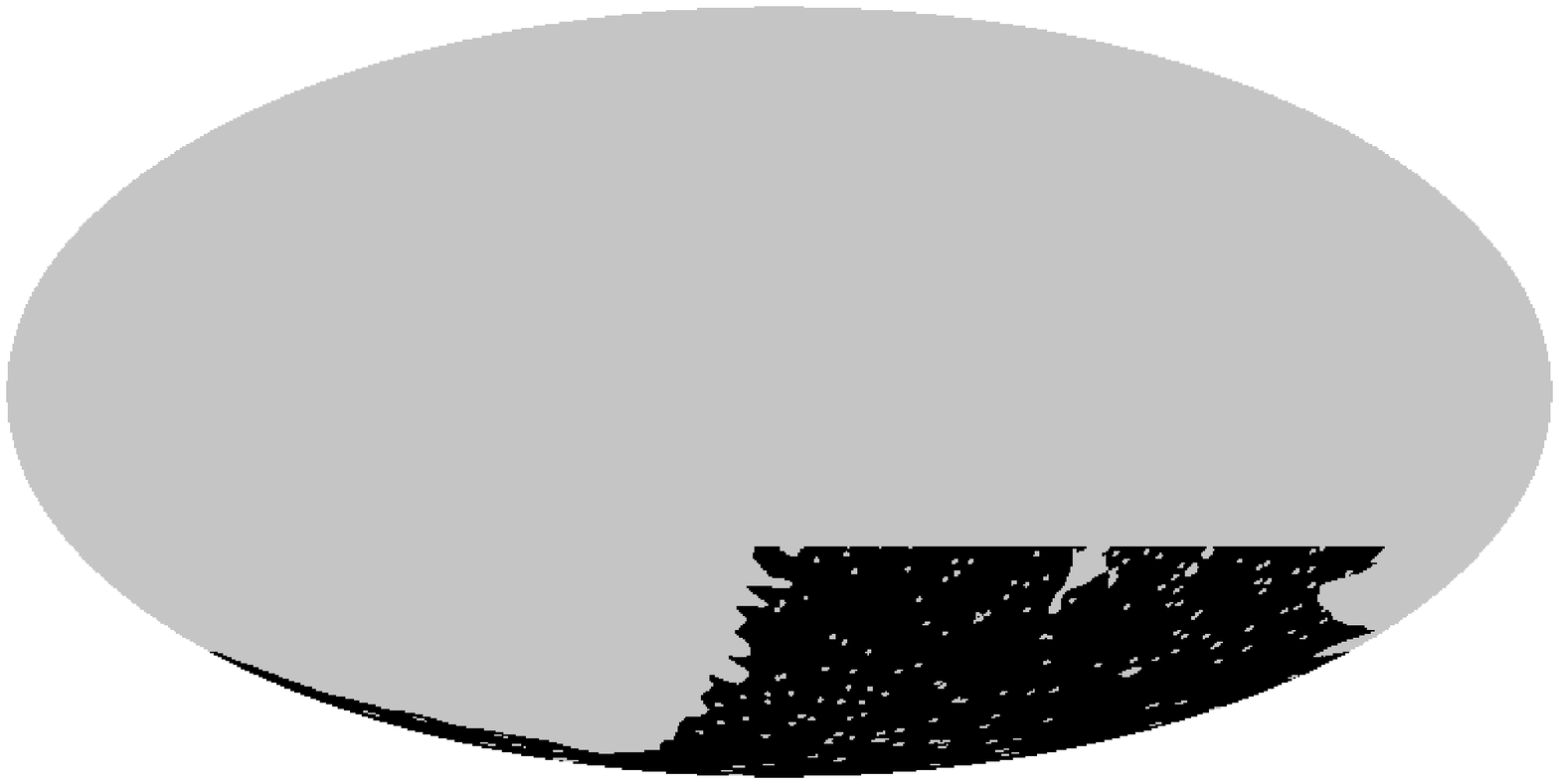}}
\caption{Mask maps of various local areas defined by applying the Galactic
         latitude cuts [($a$)--($d$)] and the thresholds on the (smoothed)
         map of number-of-observations [($e$)--($h$)] on the KQ85 mask map.
         The black area corresponds to $M(p)=1$ while the grey area to
         $M(p)=0$.
         Each number in the parenthesis indicates the fraction of sky area
         with $M(p)=1$.
        }
\label{mask maps of local areas}
\end{figure*}

The left panel of Fig.\ \ref{all sky power spectrum} shows the angular
power spectra of the WMAP 7-year temperature maps measured with the uniform
(red) and the inverse-noise (blue dots with error bars) weighting schemes.
We have used the sky area defined by the KQ85 mask that was adopted
in the WMAP team's temperature analysis.
For a comparison, the WMAP team's measurement has been shown together
as grey dots with error bars \citep{Larson-etal-2011}.
Our estimations of the angular power spectrum in both the uniform and
the inverse-noise weighting schemes are consistent with the WMAP team's
estimation. We note that the amplitude of the power spectrum at the 41th bin
($851 \le l \le 900$) is slightly larger than the WMAP team's estimation,
but they are statistically consistent with each other.
Although statistically consistent, our result is a bit different from
the WMAP team result at low $l$ region because the WMAP team applied
the different estimation method (Blackwell-Rao estimator based on Gibbs
sampling) at this low $l$ region while we simply used the pseudo power
spectrum estimation method over the whole $l$ range.

The right panel of Fig.\ \ref{all sky power spectrum} shows the averaged
angular power spectra measured from the 1000 mock data sets that mimic
the WMAP instrument performance based on the concordance $\Lambda\textrm{CDM}$
model. The averaged power spectra based on both the uniform and the
inverse-noise weighting schemes coincide with each other and
restore the assumed model power spectrum (green curve) within measurement
uncertainties, which demonstrates that our measuring algorithm works correctly.
However, our algorithm gives a slightly positive bias with respect to
the true value at the last two bins at the highest multipoles, where
the uncertainty due to the finite size of the pixels is significantly large.
Comparing the results based on the two weighting schemes, we can see that
in a case of the inverse-noise weight scheme the size of error bars is
bigger (smaller) at lower (higher) multipoles.

\section{Power spectra of various local areas on the sky}
\label{sec:power}

Here we present the angular power spectra measured on
various local regions on the sky defined based on the simple criteria
such as the Galactic latitude or the number-of-observations cuts.
To define a partial sky region, we basically use the KQ85 mask map and
the number of observations at the W band (W4 DA).
By putting a limit on the Galactic latitude or the number of observations,
we have obtained several local regions with different characters.

Figure \ref{mask maps of local areas} summarizes the mask maps (or
the window functions in the uniform weighting scheme) that are used
in our power spectrum measurements. We estimate the angular power spectra
on each local area using the corresponding mask map as a window function.
Especially, the north hat ($b \ge 30\deg$; $b$ is the Galactic latitude)
and the south hat ($b \le -30\deg$) regions defined by the simple Galactic
latitude cut show a difference in the power amplitude around
the third peak of the angular power spectrum, which can be interpreted as
the anomaly effect (see Fig.\ \ref{hats difference histogram} below).
To assess the statistical significance of the observed anomaly,
we compare the difference of power
spectrum amplitudes with the prediction of the fiducial flat
$\Lambda\textrm{CDM}$ model. Then, we search for the origin of the observed
anomaly by considering possibilities due to the residual Galaxy contamination,
the WMAP instrument noise, and unresolved point sources.

\subsection{North hat and south hat}

\begin{figure*}[t!]
\centering
\epsfxsize=8cm \epsfbox{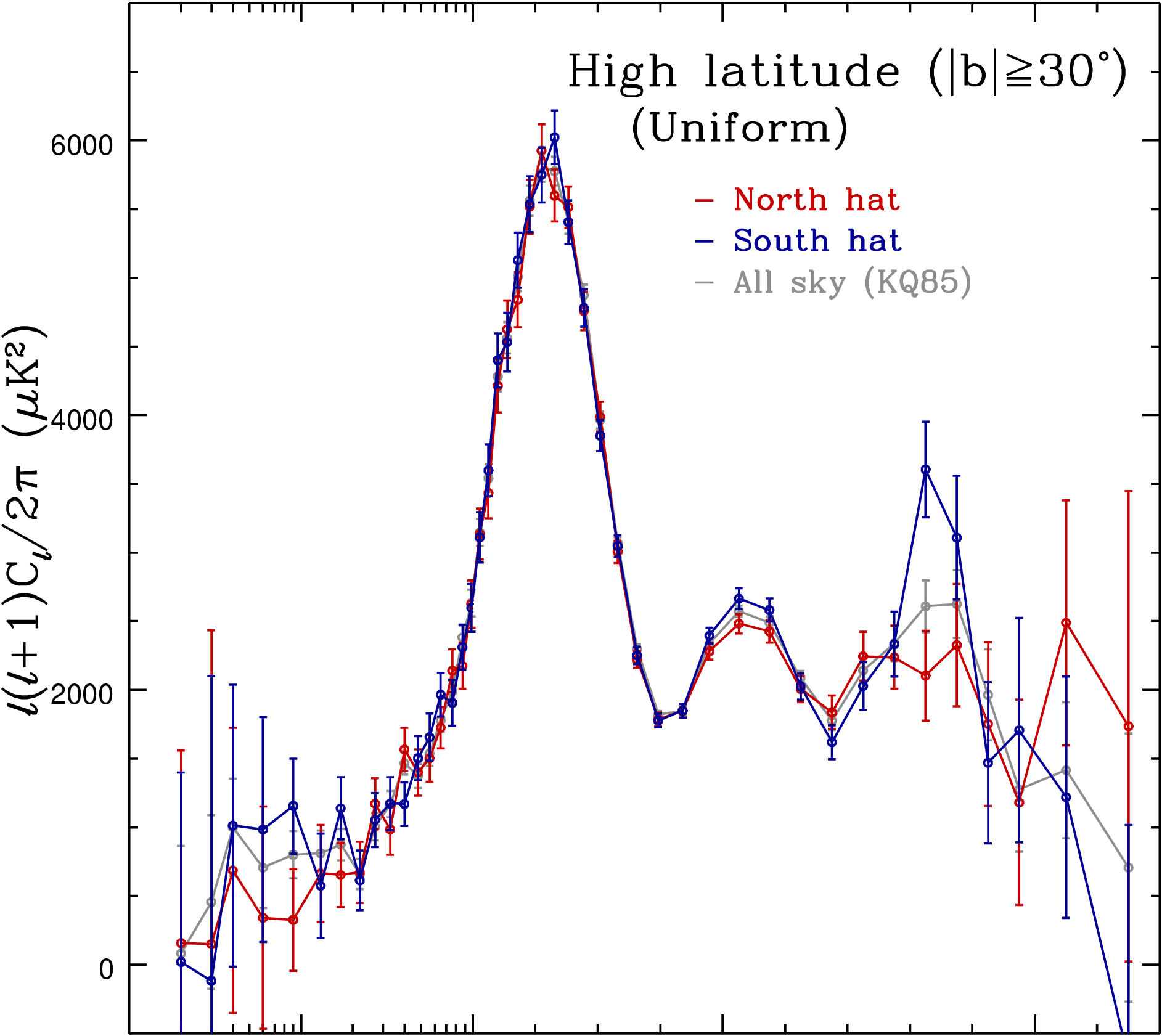}
\hspace{3mm}
\epsfxsize=8cm \epsfbox{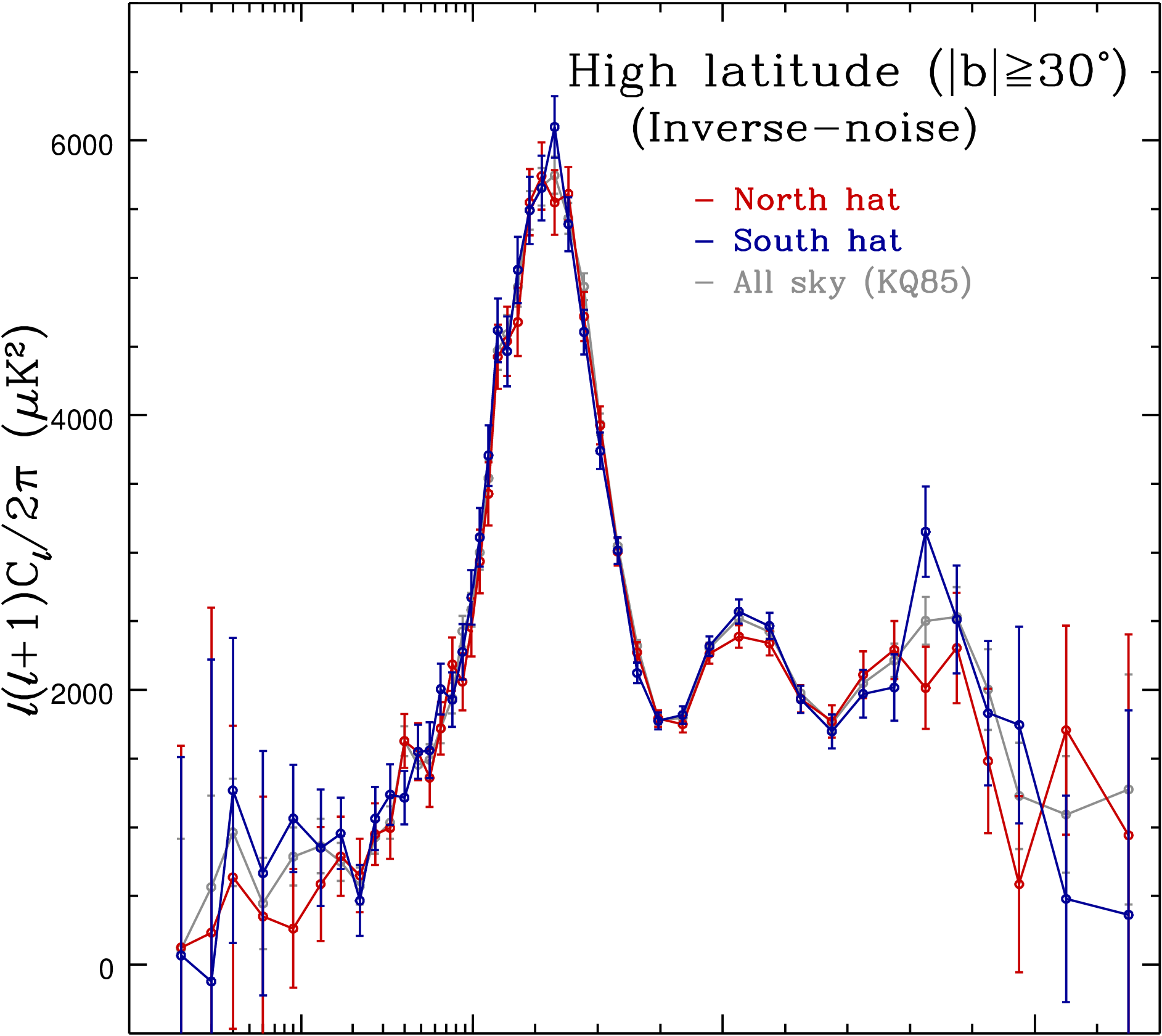} \\
\epsfxsize=8cm \epsfbox{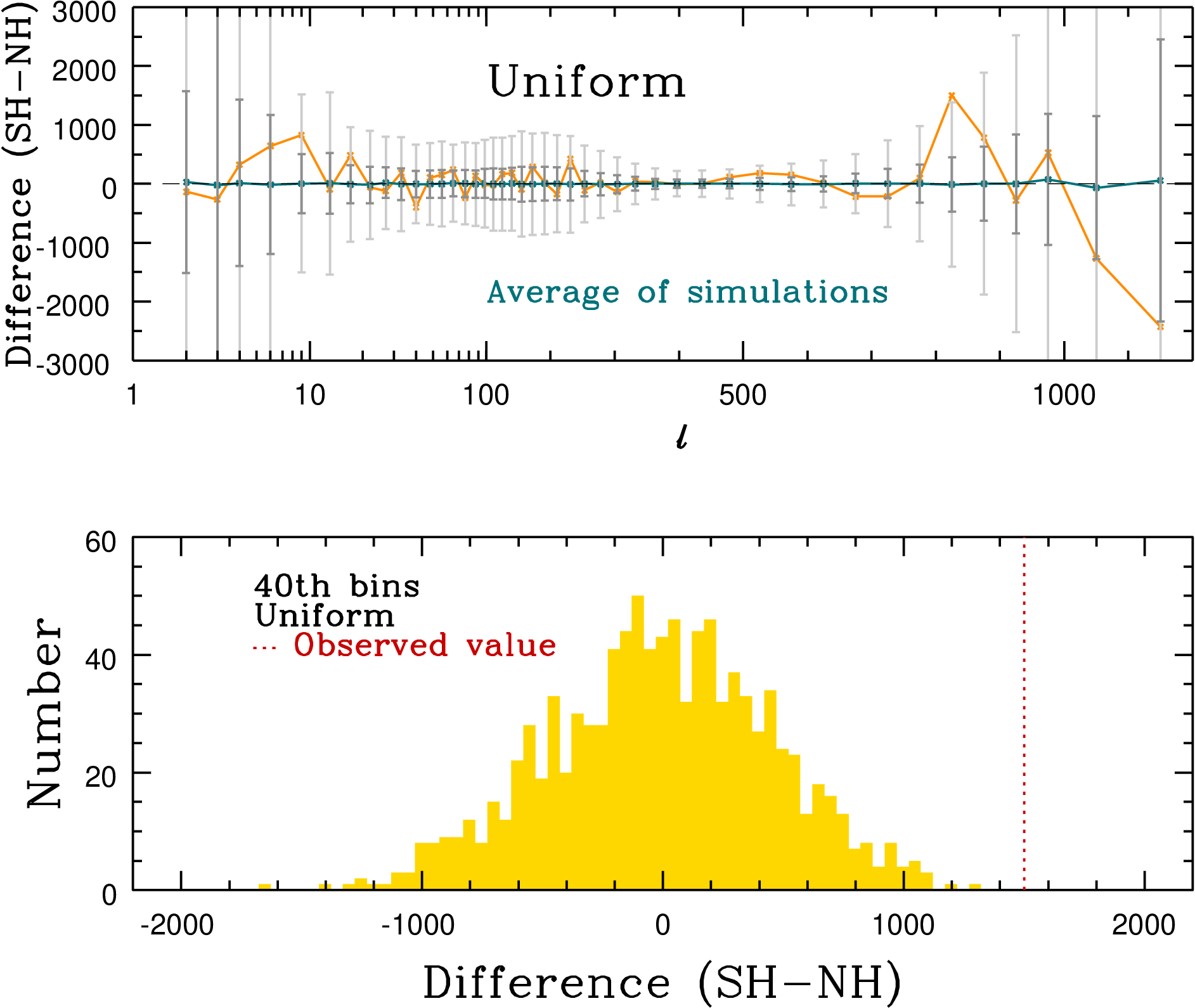}
\hspace{3mm}
\epsfxsize=8cm \epsfbox{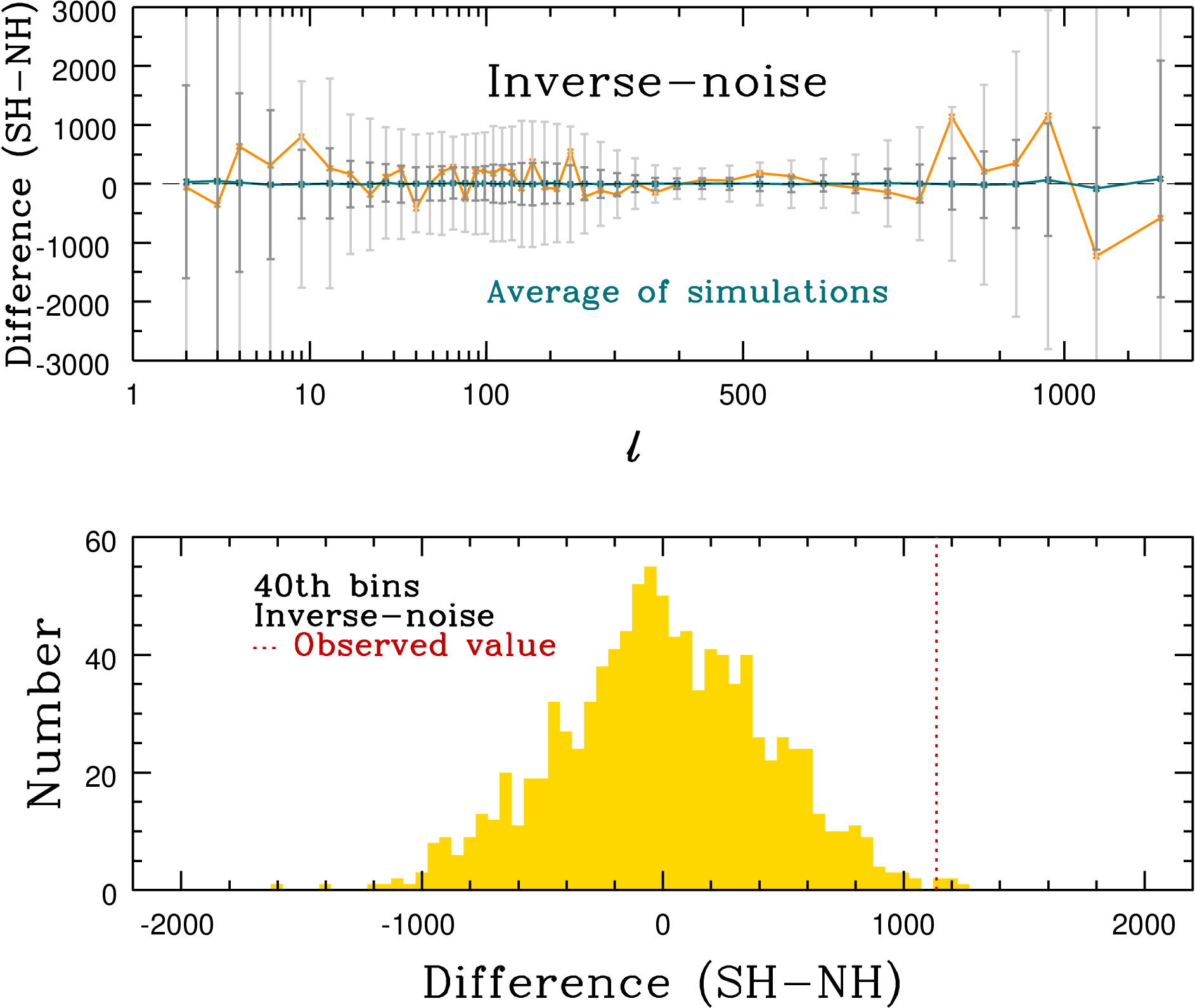}
\caption{
     (Top) Angular power spectra measured on the north hat and the
     south hat regions based on the uniform (left) and the inverse-noise
     (right panel) weighting schemes.
     The power spectra measured on the north hat ($b \ge 30\deg$) are shown
     as red color while those on the south hat ($b \le -30\deg$) as
     blue color. The grey color denotes the result measured on the whole sky
     enclosed by KQ85 mask map. (Middle) Difference of powers between
     the south hat (SH) and north hat (NH) are shown (orange color).
     The averaged difference estimated from the one thousand
     WMAP mock data sets are shown as dark green dots with $1\sigma$ (dark
     grey) and $3\sigma$ (light grey) error bars.
     (Bottom) Histograms of differences between SH and NH powers
     in the 40th bin that corresponds to $801 \le l \le 850$,
     measured from the $\Lambda\textrm{CDM}$ based WMAP mock data sets,
     for uniform (left) and inverse-noise (right panel) weighting
     schemes. The vertical red dashed lines indicate the difference values
     measured from the WMAP 7-year data.
     }
\label{hats difference histogram}
\end{figure*}

Figure \ref{hats difference histogram} displays the angular power spectra
measured on the north hat and the south hat regions.
Each power spectrum has been estimated in both the uniform and the
inverse-noise weighting schemes separately. The overall features in
the measured angular power spectra are similar to the power spectrum
measured on the whole sky region with KQ85 mask (grey curves).
However, the amplitudes of the power spectrum around
the third peak show opposite deviations in the south hat and the north hat.

We can see such a difference more dramatically in the middle panels
of Fig.\ \ref{hats difference histogram}, where the difference of powers
between the south hat (SH) and north hat (NH) are displayed for an ease
of comparison.
We have also estimated the average and the variance of differences between
SH and NH powers from the one thousand WMAP mock data sets,
which are shown as dark green dots with $1\sigma$ (dark grey) and
$3\sigma$ (light grey) error bars.
Since we have assumed the homogeneity and isotropy of the universe
during the production of the WMAP mock data sets,
the average of differences is expected to be zero over the whole $l$ range.
We note that in the third peak (that corresponds to the 40th bin;
$801 \le l \le 850$) the difference between SH and NH regions is statistically
significant because it is located around the $3\sigma$ deviating from the
zero value.
The histogram of differences between SH and NH powers in the 40th bin
measured from the $\Lambda\textrm{CDM}$ based WMAP mock data sets
and the value measured from the WMAP data (vertical red dashed line)
also confirms this result
(bottom panels of Fig.\ \ref{hats difference histogram}).

\subsection{Low latitude area}

\begin{figure*}[t!]
\centering
\epsfxsize=8cm \epsfbox{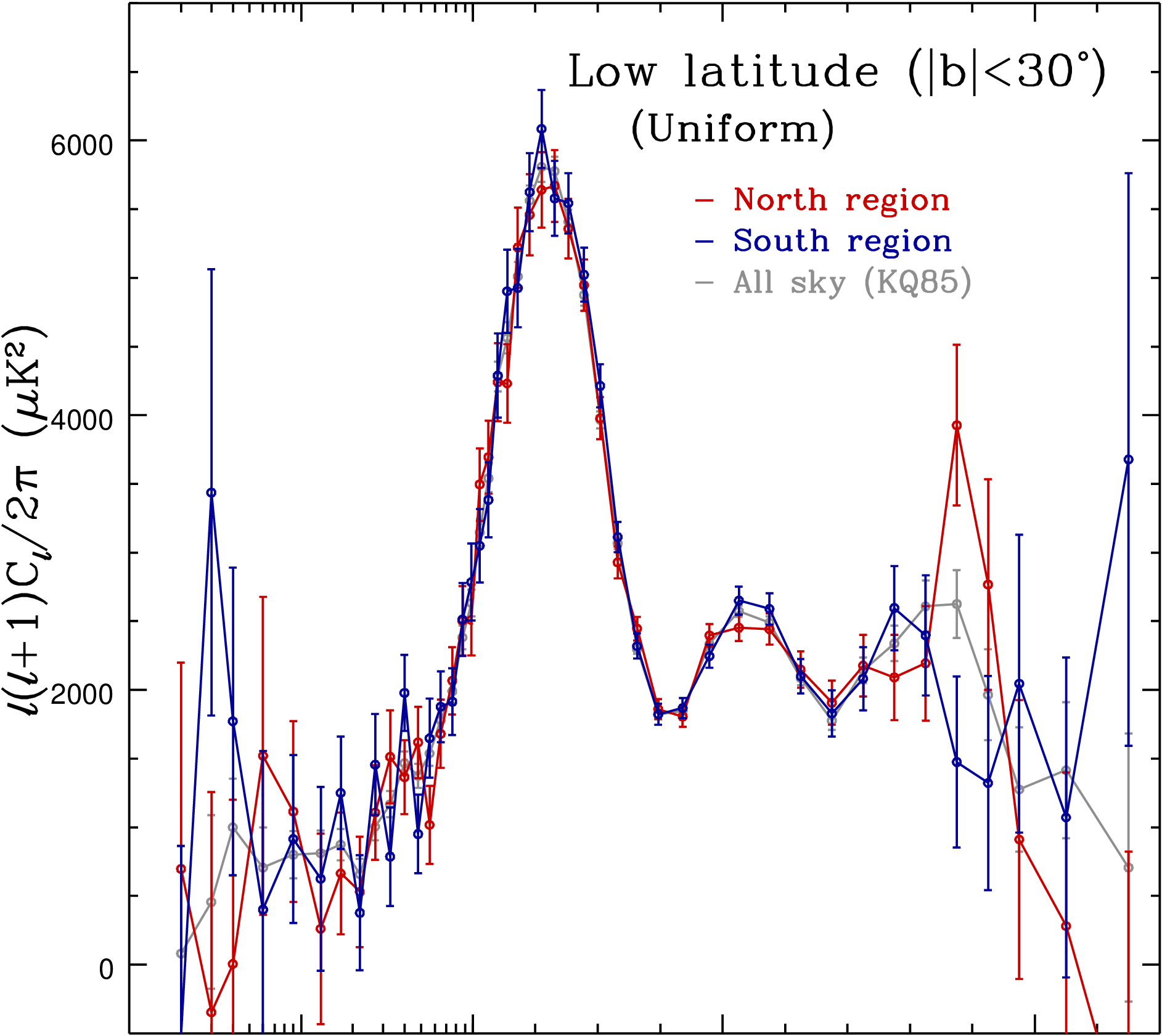}
\hspace{3mm}
\epsfxsize=8cm \epsfbox{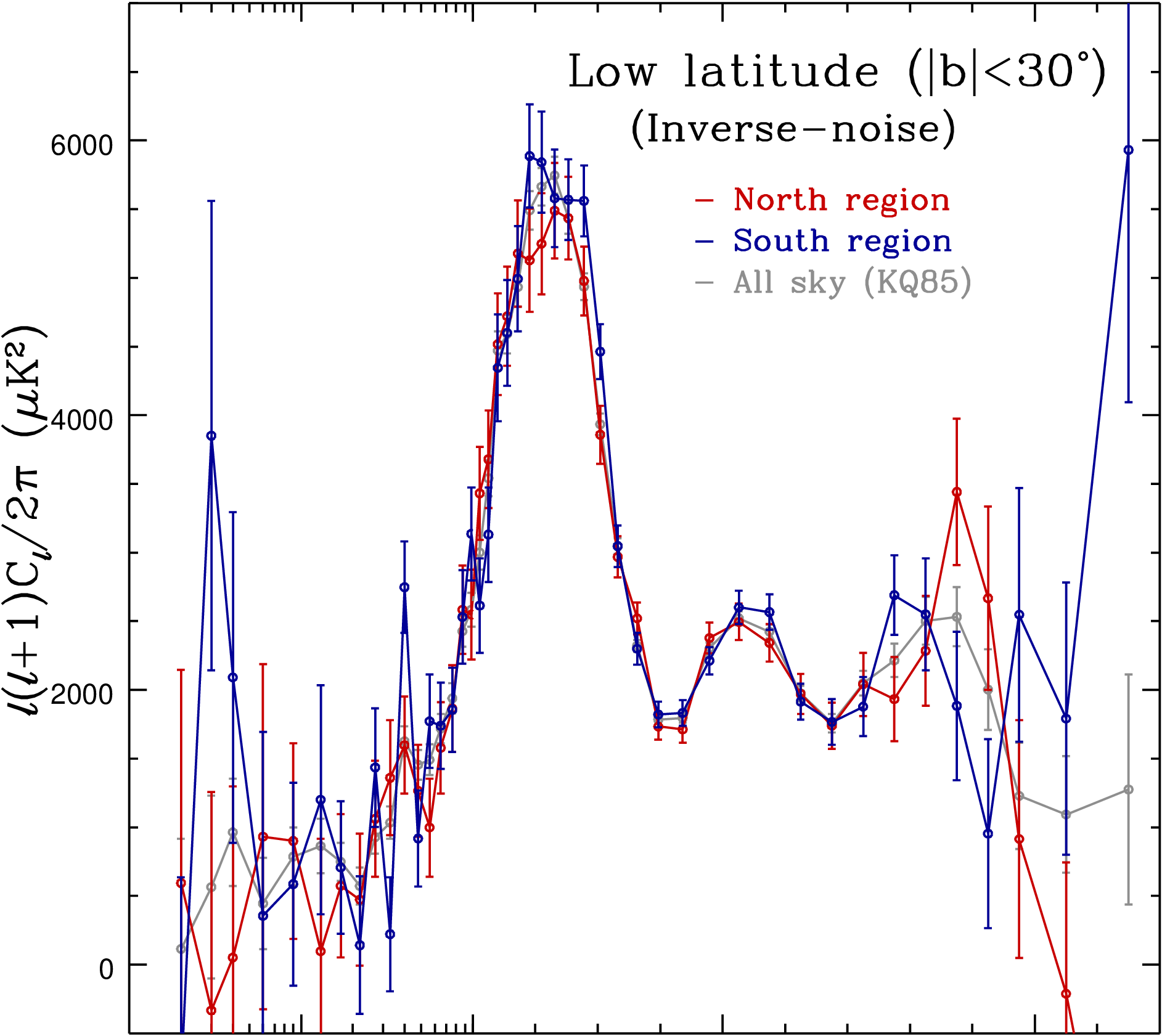} \\
\epsfxsize=8cm \epsfbox{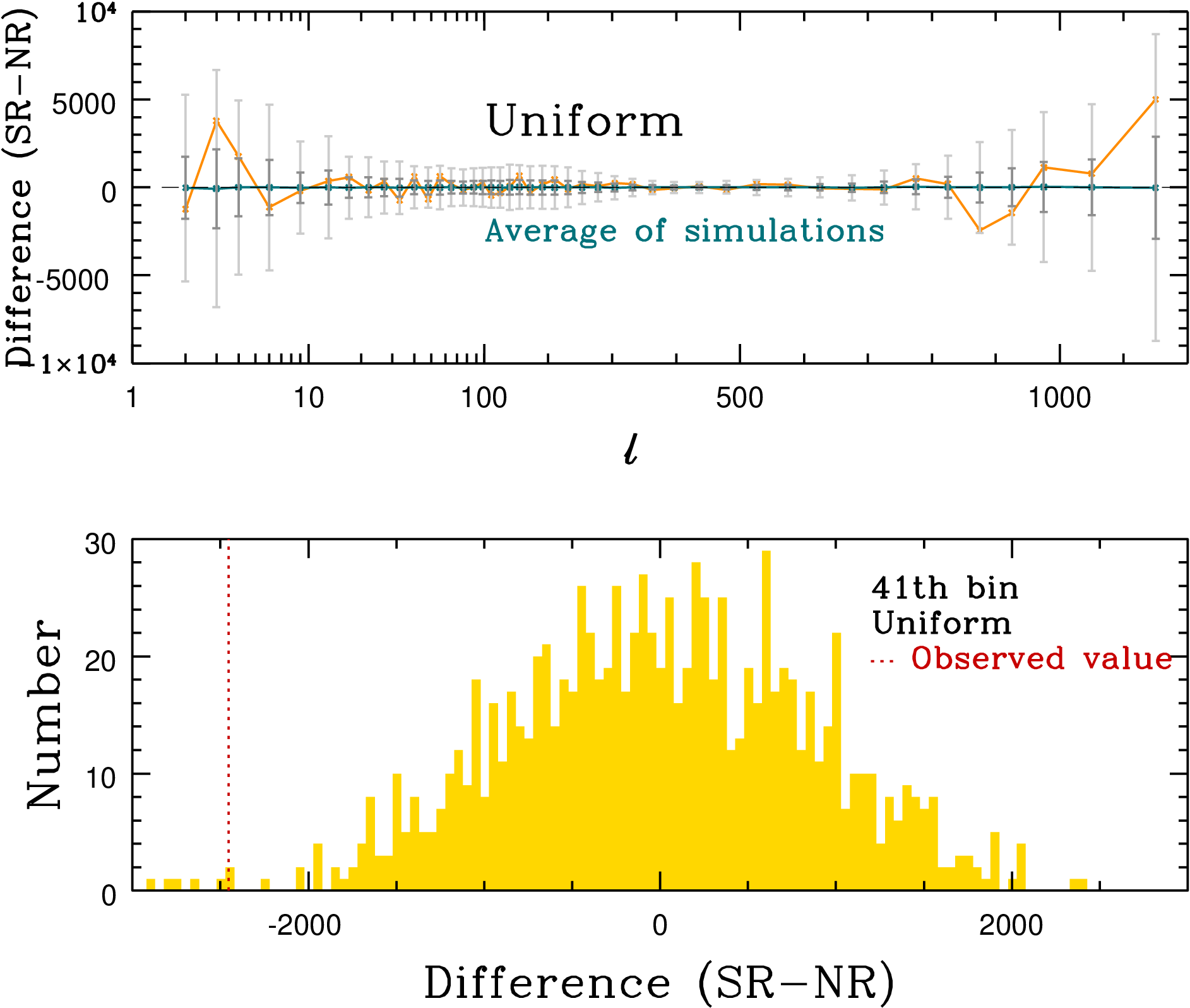}
\hspace{3mm}
\epsfxsize=8cm \epsfbox{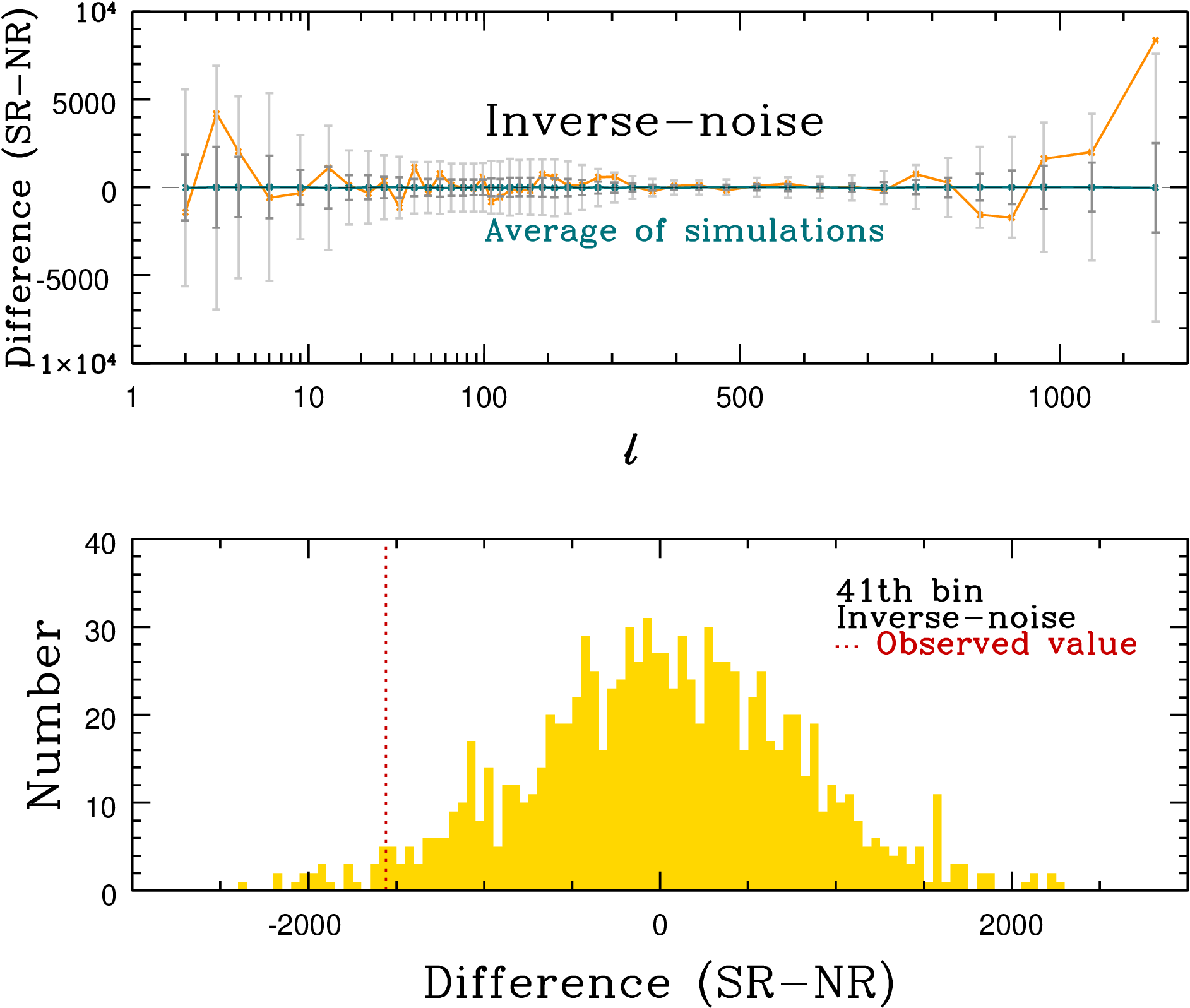}
\caption{
     Similar to Fig.\ \ref{hats difference histogram},
     but for regions at the low Galactic latitude areas ($|b|<30\deg$).
     The power spectra are measured on the north region
     (NR; $0\deg < b < 30\deg$; red color) and on the south region
     (SR; $-30\deg < b < 0\deg$; blue color).
     In the bottom panels, shown are histograms of power differences
     at the 41th $l$-bin ($851 \le l \le 900$) between south and north regions,
     expected in the concordance $\Lambda\textrm{CDM}$ model,
     together with observed values from the WMAP 7-year data
     (vertical red dashed lines).
}
\label{center difference histogram}
\end{figure*}

The foreground contamination at high Galactic latitude regions like the north
hat and the south hat is generally expected to be small.
Although the foreground model has been subtracted from the observed
temperature fluctuations and the sky region that are significantly
contaminated by the Galactic emission has been excluded by the KQ85 mask,
we can expect that the residual foreground emission at low latitude regions
may be stronger than high latitude regions such as the north hat and
the south hat.
We estimated angular power spectra on two separate regions at the low
Galactic latitude defined as low latitude north ($0\deg < b < 30\deg$)
and south ($-30\deg < b < 0\deg$) regions, where the Galactic plane
region has been excluded by the KQ85 mask.
The results are presented in Fig.\ \ref{center difference histogram}
with a similar format as in Fig.\ \ref{hats difference histogram}.

The angular power spectra measured on the sky regions that are expected
to be contaminated by the residual Galactic foreground emission
show {\it different} features from those seen in the cases of the north
and south hats.
The behavior of power spectrum amplitudes around the third peak
is {\it opposite} to the case of north versus south hat regions.
The power on the low latitude north region is larger than that on the
south region, and the difference between the north and the south regions
is more significant at the 41th bin ($851 \le l \le 900$).
The difference is again statistically significant with a $3\sigma$ deviation
from the zero value (for uniform weighting scheme),
which is another anomaly found in this work.
Although not shown here, if we consider the whole northern and southern
hemispheres as two localized regions, then the anomalies at high and low
Galactic latitudes are compensated and not noticeable.
Furthermore, there are two more noticeable differences between the north
and the south regions at the second bin ($l=3$) and at the last bin
($1101 \le l \le 1200$). For the latter, the measured difference exceeds
$3\sigma$ from the zero value for the inverse-noise weighting scheme
(see also Table \ref{tab:counts} below).
Therefore, based on this comparison, it seems that the north-south anomaly
around the third peak of the angular power spectrum observed on
the high Galactic latitude north and south hat regions is not due to
any possible residual Galactic foreground emission.

\subsection{Instrument noise}

\begin{figure*}[t!]
\centering
\epsfxsize=8cm \epsfbox{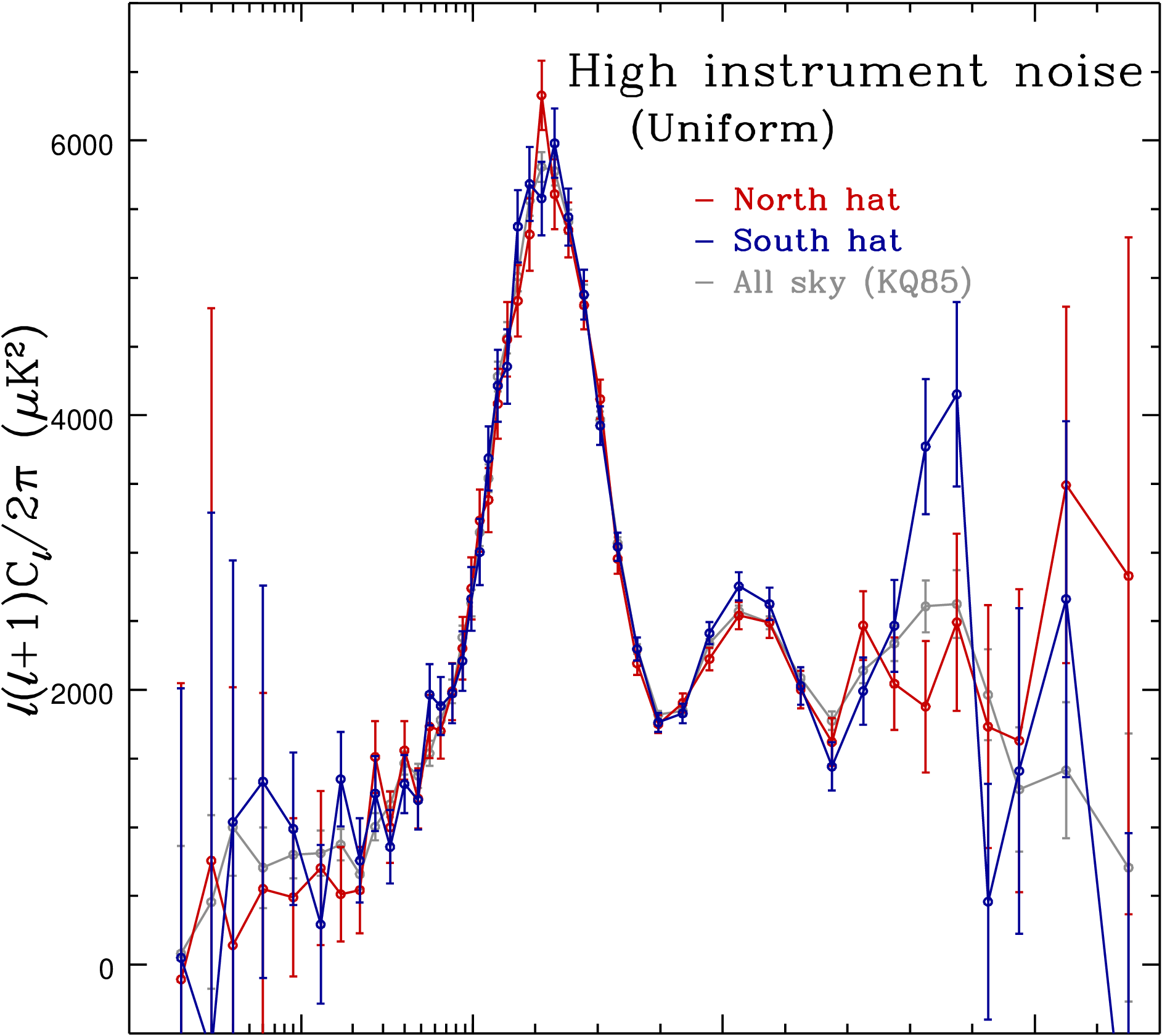}
\hspace{3mm}
\epsfxsize=8cm \epsfbox{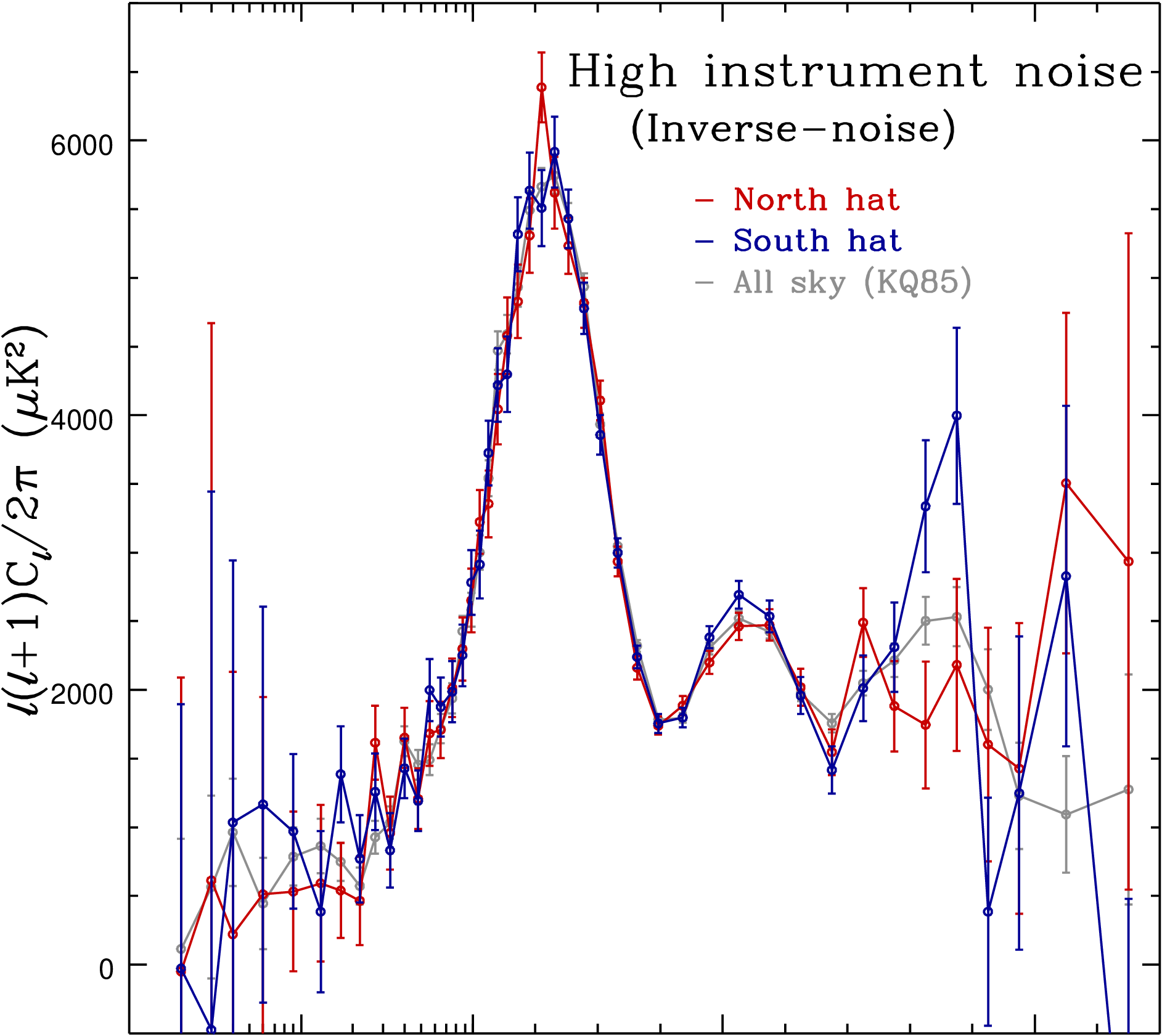} \\
\epsfxsize=8cm \epsfbox{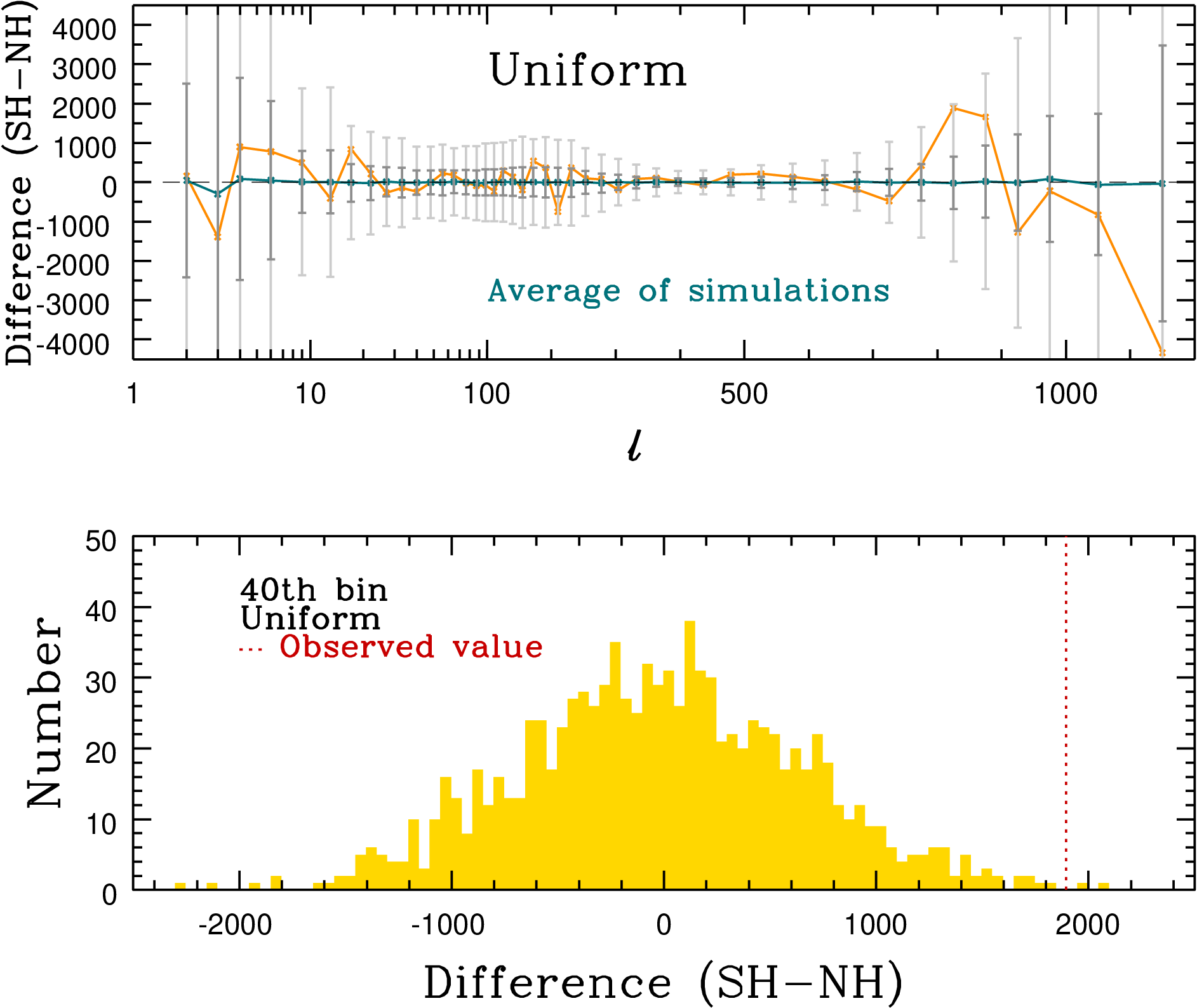}
\hspace{3mm}
\epsfxsize=8cm \epsfbox{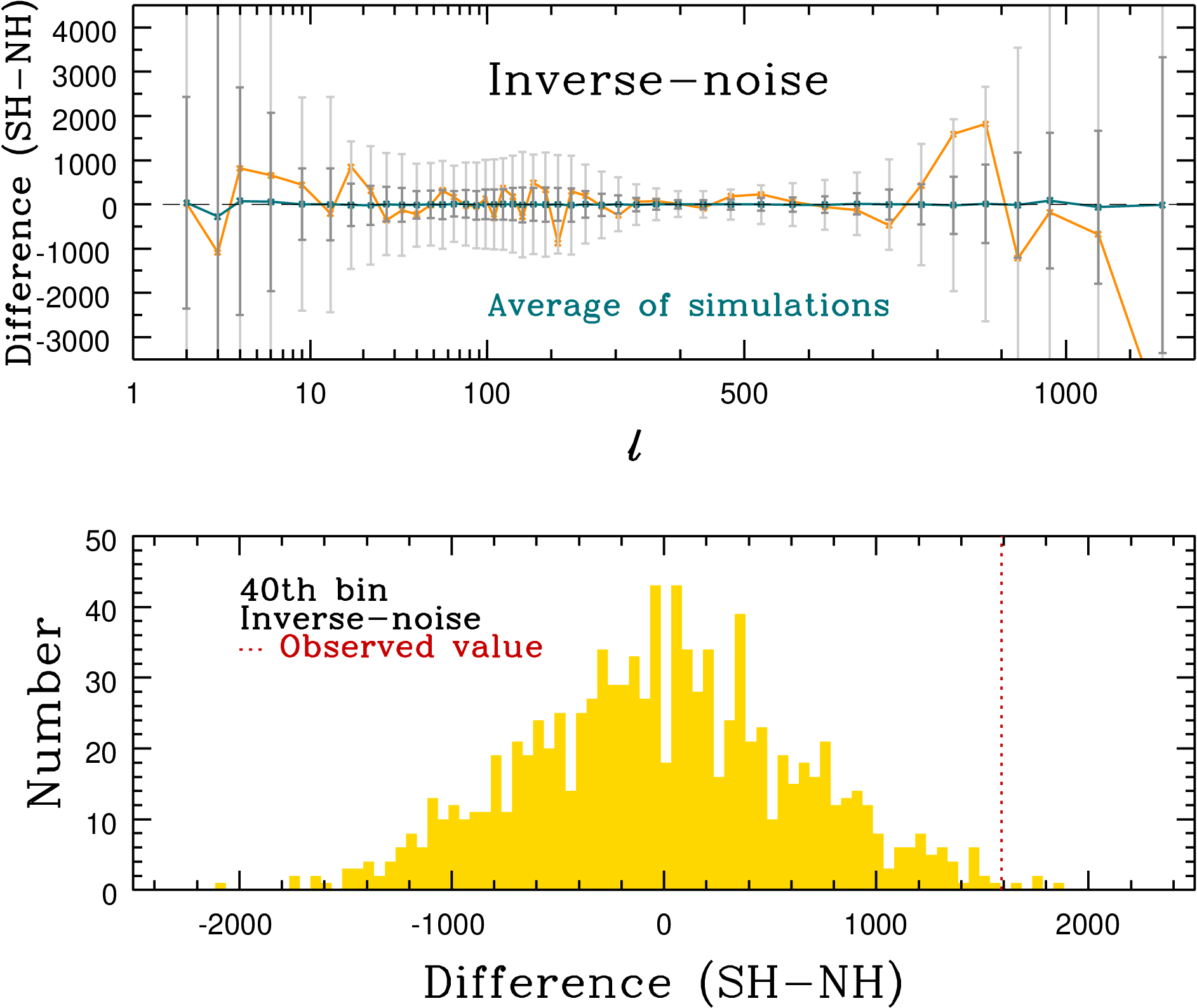}
\caption{
        Similar to Fig.\ \ref{hats difference histogram},
        but for regions with high instrument
        noise ($N_\textrm{\scriptsize obs}<2500$ on the number-of-observations
        map at W4 DA smoothed with $\textrm{FWHM}=1\deg$) that belong
        to the north hat (left) and the south hat (right).
        }
\label{low nobs power spectrum}
\end{figure*}

\begin{figure*}[t!]
\centering
\epsfxsize=8cm \epsfbox{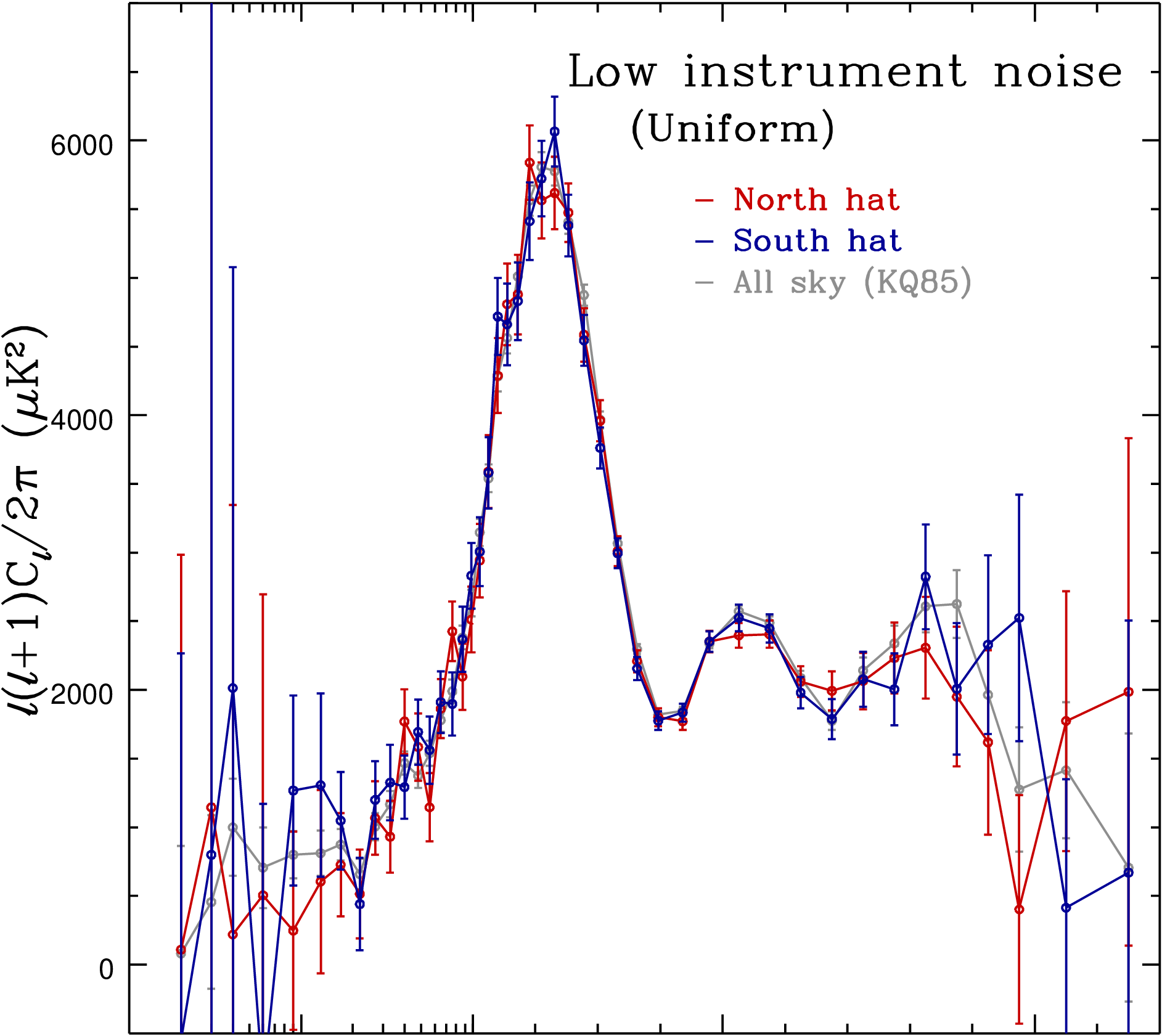}
\hspace{3mm}
\epsfxsize=8cm \epsfbox{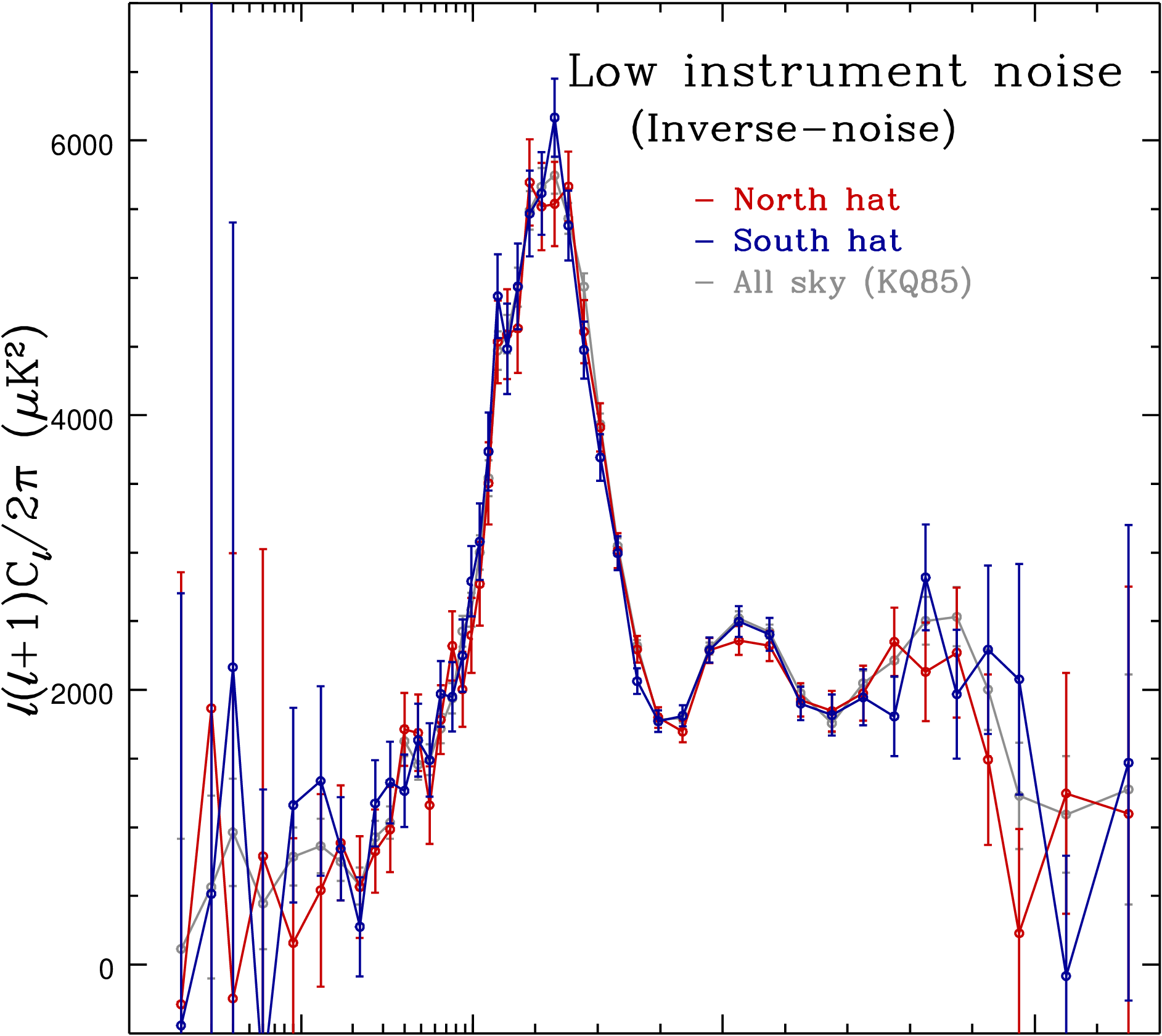} \\
\epsfxsize=8cm \epsfbox{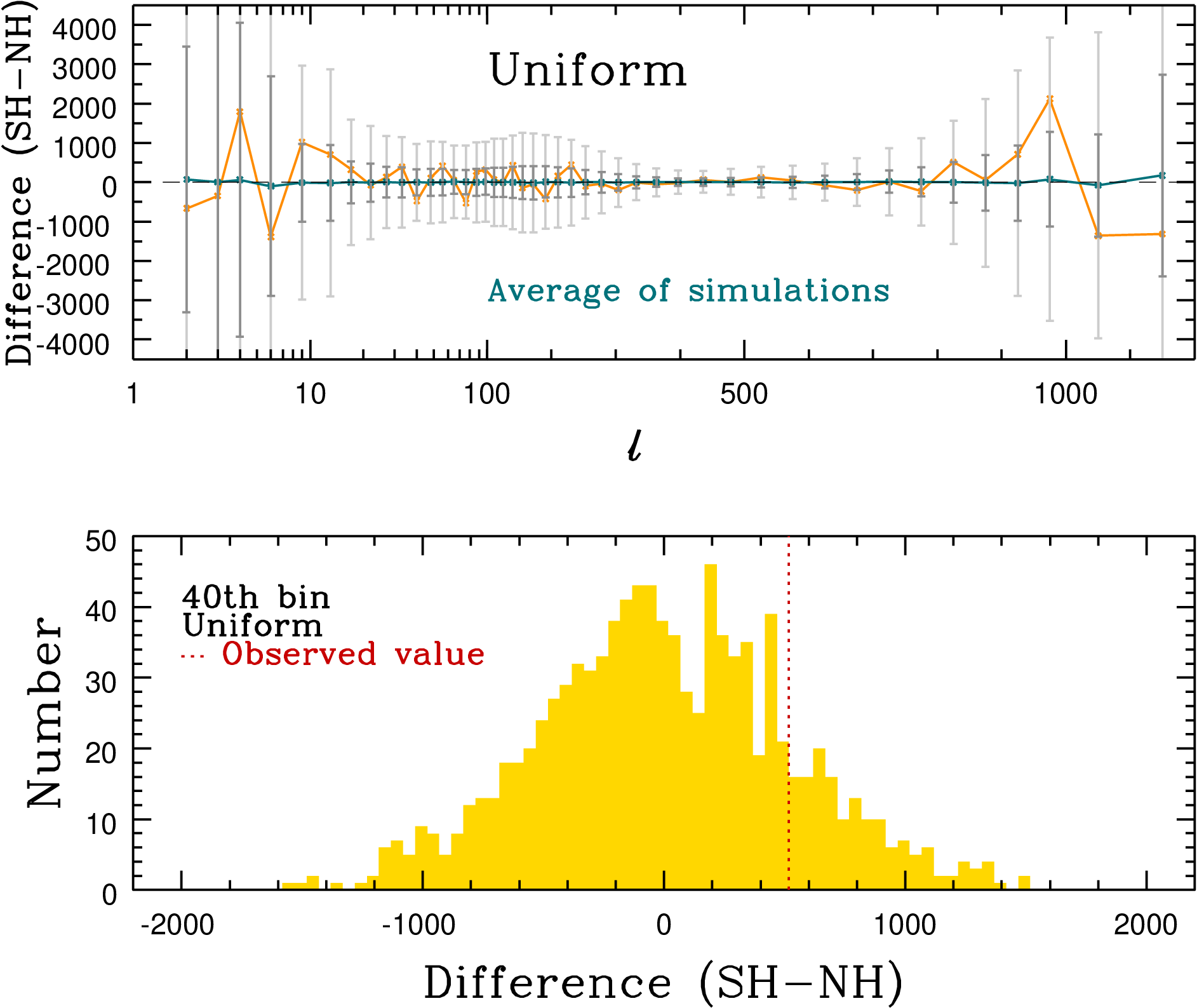}
\hspace{3mm}
\epsfxsize=8cm \epsfbox{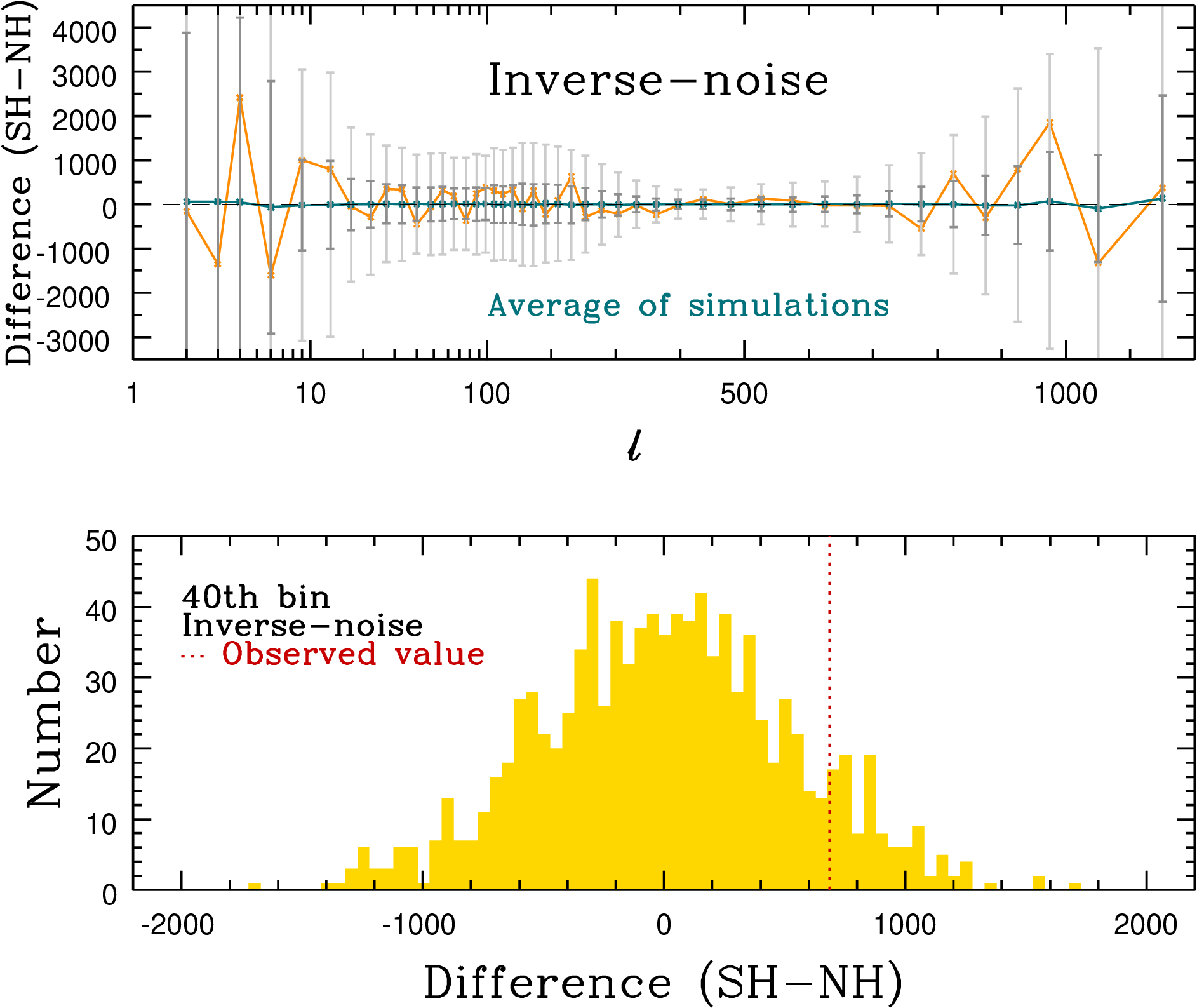}
\caption{
        Similar to Fig.\ \ref{hats difference histogram},
        but for regions with the low instrument noise ($N_\textrm{\scriptsize obs}>2035$)
        belonging to the north hat (left) and the south hat (right).
    }
\label{high nobs power spectrum}
\end{figure*}

Although the WMAP satellite probed the CMB temperature fluctuations on
the whole sky, its scanning strategy is somewhat inhomogeneous such that
the regions near the ecliptic poles were probed many times as compared to the
regions near the ecliptic plane.
Thus, it is expected that the CMB anisotropy is strongly affected
by the instrument noise in the ecliptic plane region.
To quantify such an effect due to the WMAP instrument noise,
we use the number of observations, $N_\textrm{\scriptsize obs}(p)$, at the WMAP W4
frequency channel, which is shown in Fig.\ \ref{N_obs}.

Defining regions with high and low instrument noise by simply putting
threshold limits on the number of observations results in the geometrical
sky regions whose boundary is not smooth and the pixels near the boundary
are not contiguous with many small islands outside the primary region,
which prevents us to obtain the unbiased estimation of the power spectrum.
To avoid this problem, we have used the Smoothing program of the HEALPix
software to smooth the map of number-of-observations at the W4 frequency
channel with a Gaussian filter of $1\deg$ FWHM.
Then, we define the north hat (or the south hat) regions with high instrument
noise by selecting pixels with $N_\textrm{\scriptsize obs}(p)<2500$ on the smoothed
number-of-observations map. We also exclude the island areas that are not
included in the primary big regions.
Similarly, to define the regions with low instrument noise, we set
$N_\textrm{\scriptsize obs}(p)>2035$ on the same map.
The threshold value for the number of observations at the region of
high (low) instrument noise has been set to obtain the sky area
of about 15\% (13\%) of the whole sky for $N_\textrm{\scriptsize obs}(p)<2500$
($N_\textrm{\scriptsize obs}(p)>2035$). For ecliptic plane regions with high instrument
noise (i.e., with low number of observations), we choose the threshold
value to include the larger sky area for the purpose of reducing the
statistical variance in the power spectrum measurement.
The sky regions defined in this way are shown in Fig.\
\ref{mask maps of local areas} $(e)$--$(h)$.

We have measured angular power spectra on the ecliptic plane regions
with small number of observations (defined as $N_\textrm{\scriptsize obs} < 2500$)
that belong to the north hat or the south hat regions.
The results are shown in Fig.\ \ref{low nobs power spectrum}.
The angular power spectra measured on the ecliptic plane regions
(that belong to the north hat or the south hat regions) show the behavior
that is similar to the case of the whole north hat and south hat regions
(see Fig.\ \ref{hats difference histogram}).
The difference of power spectrum amplitudes between the south and the north
hats still statistically significant, deviating from the zero value up to
$3\sigma$.

\begin{center}
\begin{table*}[t!]
\small
\caption{Probabilities of finding the north-south (absolute) difference
         larger than the observed value, estimated based on one thousand
         $\Lambda\textrm{CDM}$ based WMAP mock data sets.
         The $l_b$ and $\Delta l$ denote the multipole $l$ value
         at the center of the bin and the bin-width, respectively, used
         in the power spectrum estimation.}
\centering
\begin{tabular}{lccccc}
\hline\hline\\[-2mm]
     &     & & & \multicolumn{2}{c}{$p$ value (\%)} \\[-1mm]
\quad\quad\quad\quad area & bin number & $l_b$ & $\Delta l$ & & \\[-2mm]
\cline{5-6}\\[-2mm]
     &     & & & uniform weighting~ & ~inverse-noise weighting \\[+1mm]
\hline\\[-2mm]
High Galactic latitude ($|b|\ge 30\deg$) & 40 & 825 & 50 & $0.1$ & $0.7$ \\
~~ ---with high instrument noise         & 40 & 825 & 50 & $0.5$ & $1.0$ \\
~~ ---with low instrument noise          & 40 & 825 & 50 & $31$  & $19$  \\
~~ ---with 20\% increased bin-width      & 39 & 825 & 60 & $0.0$ & $0.8$ \\
~~ ---with 50\% increased bin-width      & 39 & 825 & 75 & $0.0$ & $2.5$ \\[+1mm]
\hline\\[-2mm]
High Galactic latitude ($|b|\ge 25\deg$) & 40 & 825 & 50 & $0.6$ & $5.9$  \\
High Galactic latitude ($|b|\ge 35\deg$) & 40 & 825 & 50 & $0.0$ & $0.0$  \\
High Galactic latitude ($|b|\ge 40\deg$) & 40 & 825 & 50 & $1.5$ & $1.1$  \\
High Galactic latitude ($|b|\ge 45\deg$) & 40 & 825 & 50 & $20$  & $12$  \\[+1mm]
\hline\\[-2mm]
Low Galactic latitude  ($|b| < 30\deg$)  & 41 & 875 & 50 & $0.5$ & $5.0$  \\
                                         & 45 & 1150& 100 & $8.5$ & $0.1$  \\[+1mm]
\hline\hline
\end{tabular}
\label{tab:counts}
\end{table*}
\end{center}

We have also measured angular power spectra on the ecliptic pole regions
with large number of observations (defined as $N_\textrm{\scriptsize obs}(p) > 2035$),
which are shown in Fig.\ \ref{high nobs power spectrum}.
Here, the power spectra measured on the ecliptic pole regions
in the north hat and the south hat are {\it consistent} with each other
in the $l$ range around the third peak.
They are also very consistent with the power spectrum measured on the whole
sky with KQ85 mask (grey dots with error bars).
Around the third peak, the power spectrum measured on the ecliptic pole
regions in the north hat is very similar to the case of the whole north hat
region, while the power spectrum measured on the ecliptic pole regions
in the south hat has decreased in amplitude compared with the case of the whole
south hat region (see Fig.\ \ref{hats difference histogram}).
As a result, the observed north-south difference in the 40th bin of the power
spectrum is around $1\sigma$ confidence limit and is not statistically
significant any more (bottom panels of Fig.\ \ref{high nobs power spectrum}).
These results strongly suggest that the north-south anomaly
around the third peak likely originates from the unknown systematic effects
contained on the sky regions that are affected by strong instrument noise.

Based on the one thousand WMAP mock data sets, we count the number of
cases where the magnitude of the north-south power difference is larger
than the observed value for several chosen local regions.
The results are summarized in Table \ref{tab:counts} for particular $l$-bins
where the high statistical significance is seen. The listed $p$-values are
probabilities of finding the north-south (absolute) difference larger than
the observed value.
Although the statistical significance for the north-south anomaly
at the high latitude regions becomes weaker for the inverse-noise weighting
scheme, the observed anomaly is still rare in the $\Lambda\textrm{CDM}$
universe with $p=0.7$\%. Comparing the cases for high and low instrument
noise at the high latitude regions strongly suggests that the north-south
anomaly around the third peak come from the region dominated by the WMAP
instrument noise.

\subsection{Point sources}

Recently, the Planck point source catalogue has been publicly available
\citep{Planck-etal-2011b}. We take the Planck Early Release Compact Source Catalogue
(ERCSC) at 100 GHz frequency band and exclude the sources at high Galactic
latitude ($|b|\ge 30\deg$) whose angular position on the sky is located
at the excluded region with $M(p)=0$ in the KQ85 mask map.
The remaining 163 radio point sources at $|b|\ge 30\deg$ are considered
as the sources that were newly discovered by the Planck satellite at 100 GHz.
By excluding (i.e., by assigning $M(p)=0$ at) the circular areas centered
on these point sources with radius $0\fdg6$ in the WMAP team's KQ85 mask map,
we have produced a new mask map (or window function) and
measured the angular power spectra based on this new window function.
Because the number of newly discovered sources are very small, the
additional exclusion of point sources almost does not affect the angular
power spectra on the north and the south hat regions.
Therefore, at the present stage we cannot find any evidence that
the north-south anomaly around the third peak of the angular power spectrum
originates from the unresolved point sources.

\subsection{Dependence of the north-south anomaly on the
            bin-width and the Galactic latitude cut}

\begin{figure*}[t!]
\epsfxsize=4.25cm \epsfbox{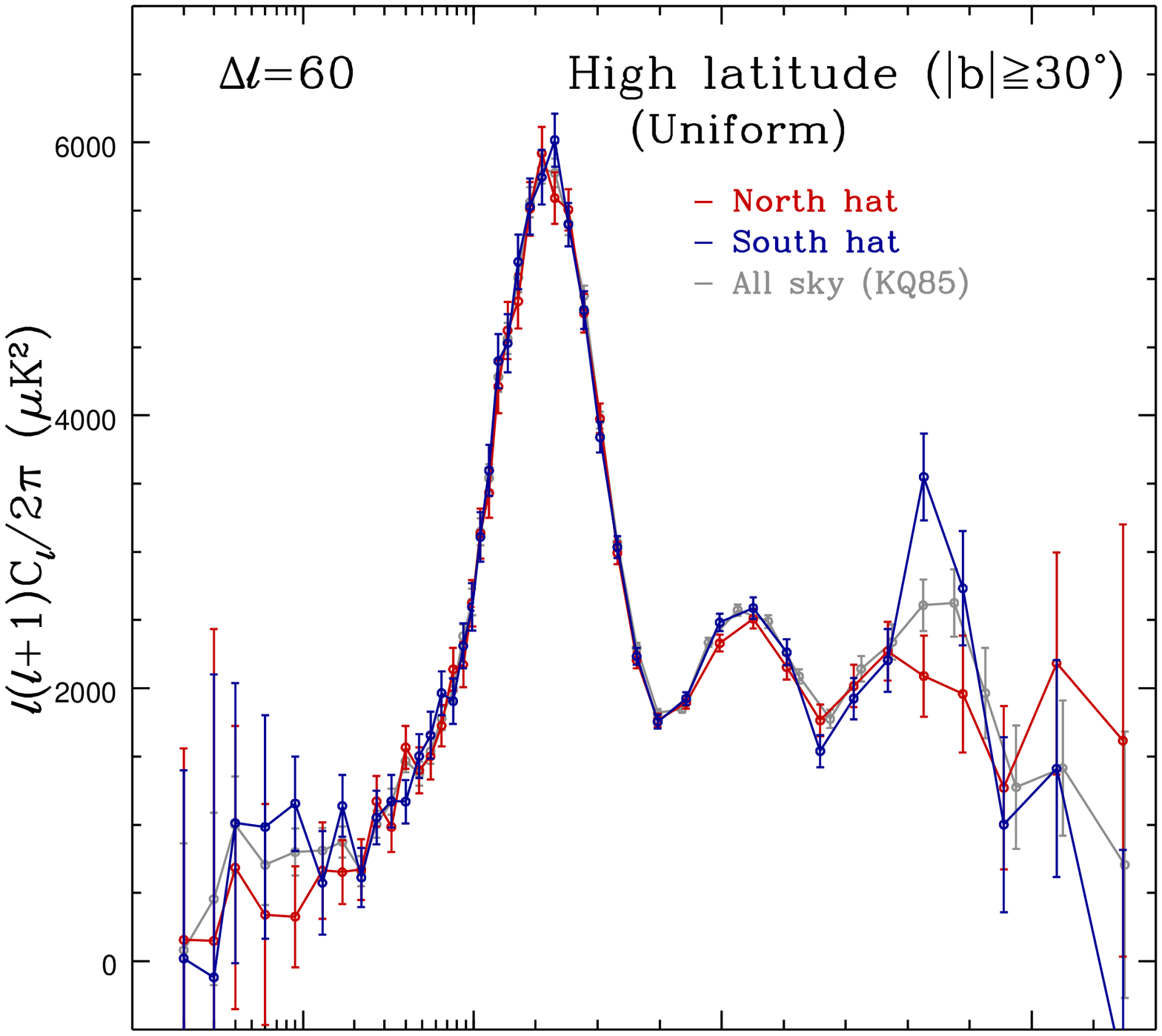}  %
\epsfxsize=4.25cm \epsfbox{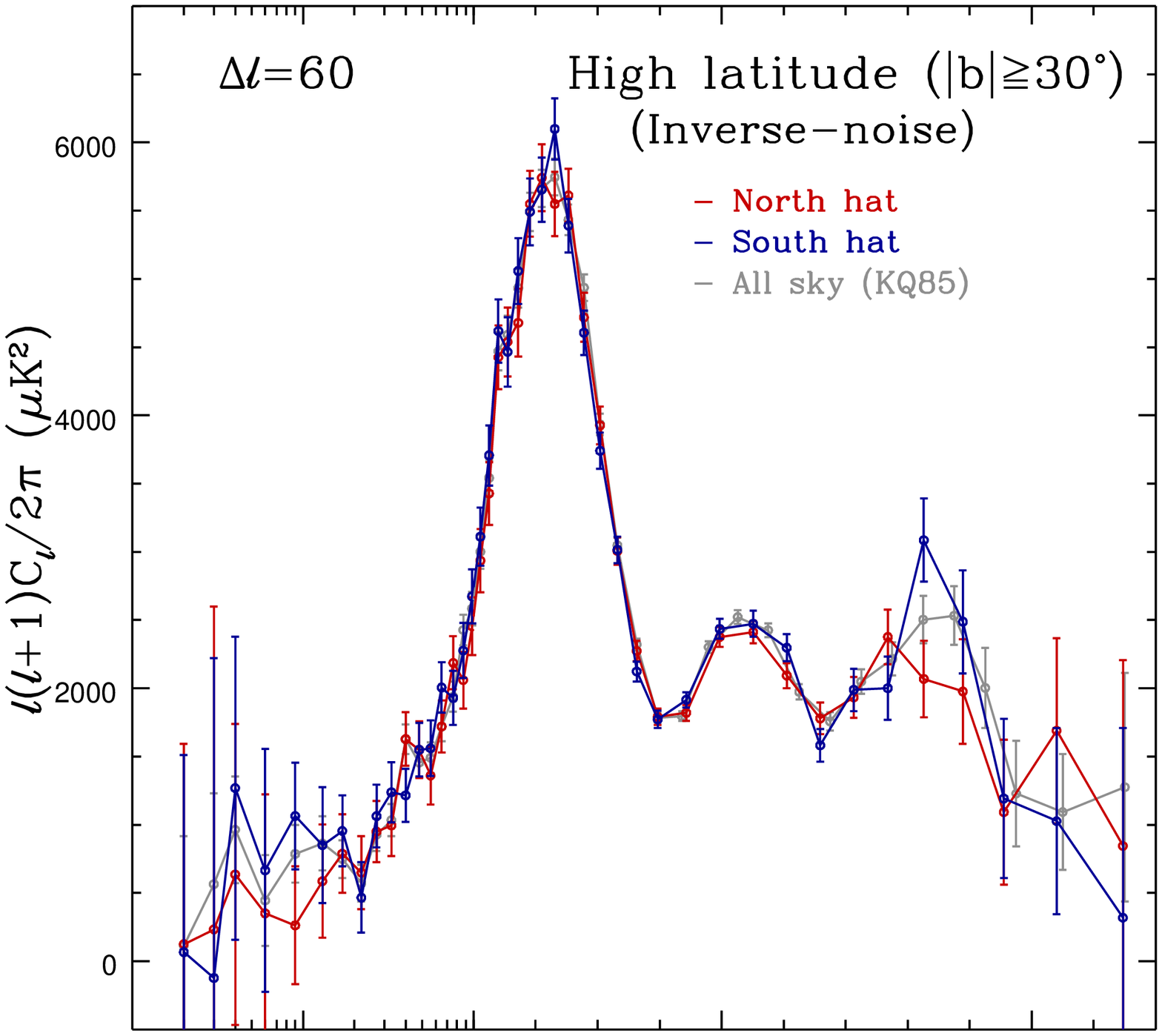}
\epsfxsize=4.25cm \epsfbox{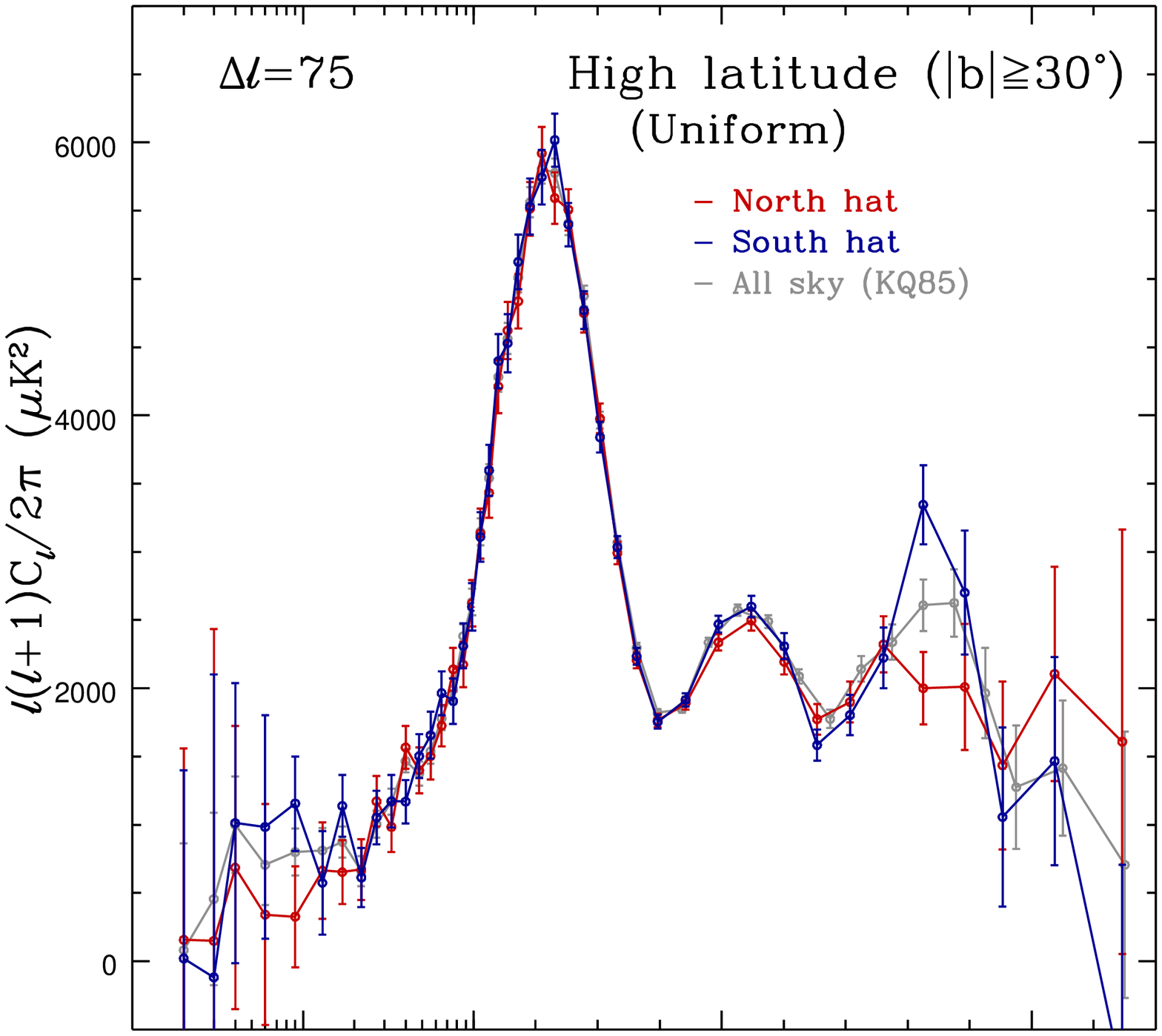} %
\epsfxsize=4.25cm \epsfbox{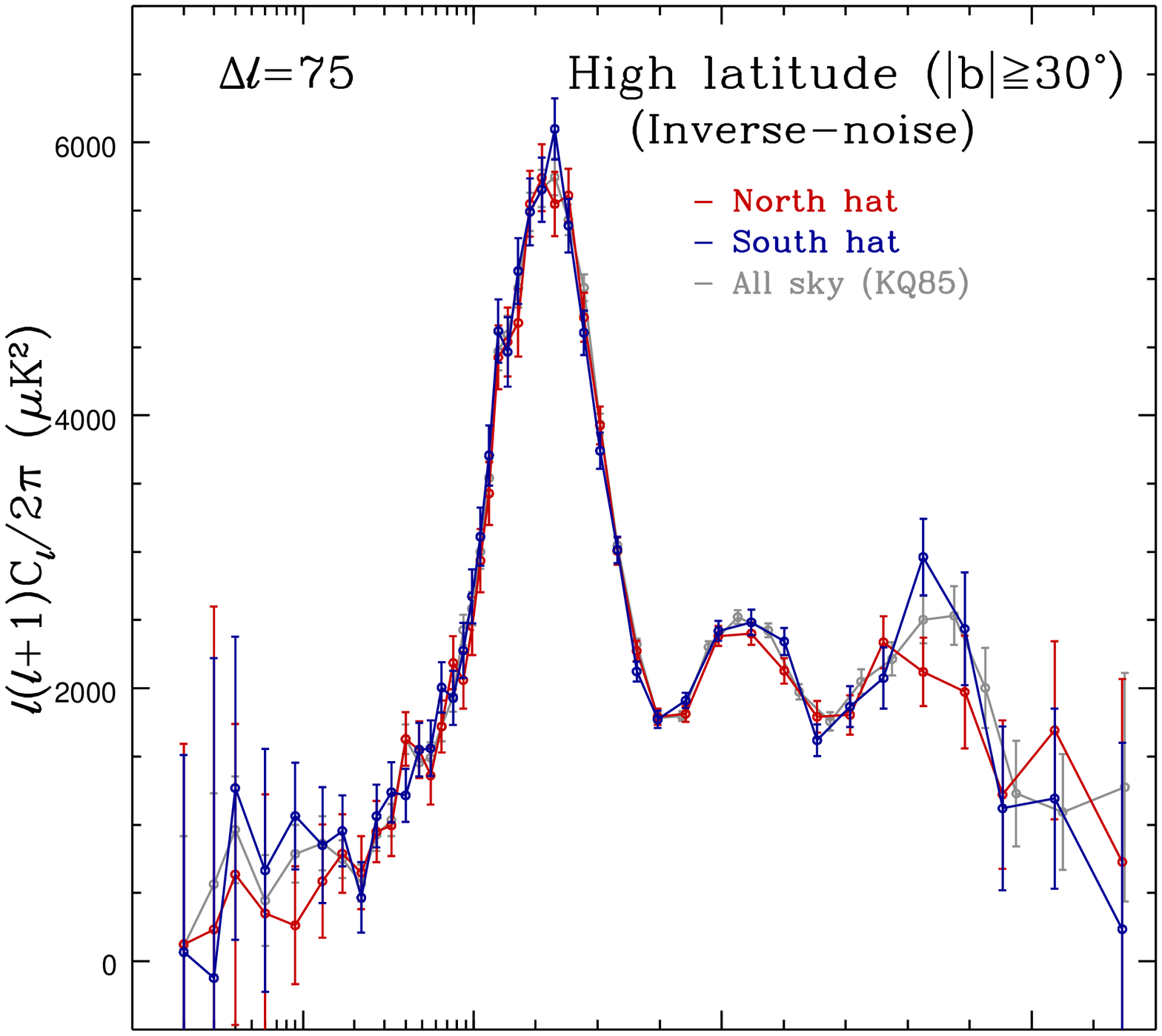} \\
\epsfxsize=4.25cm \epsfbox{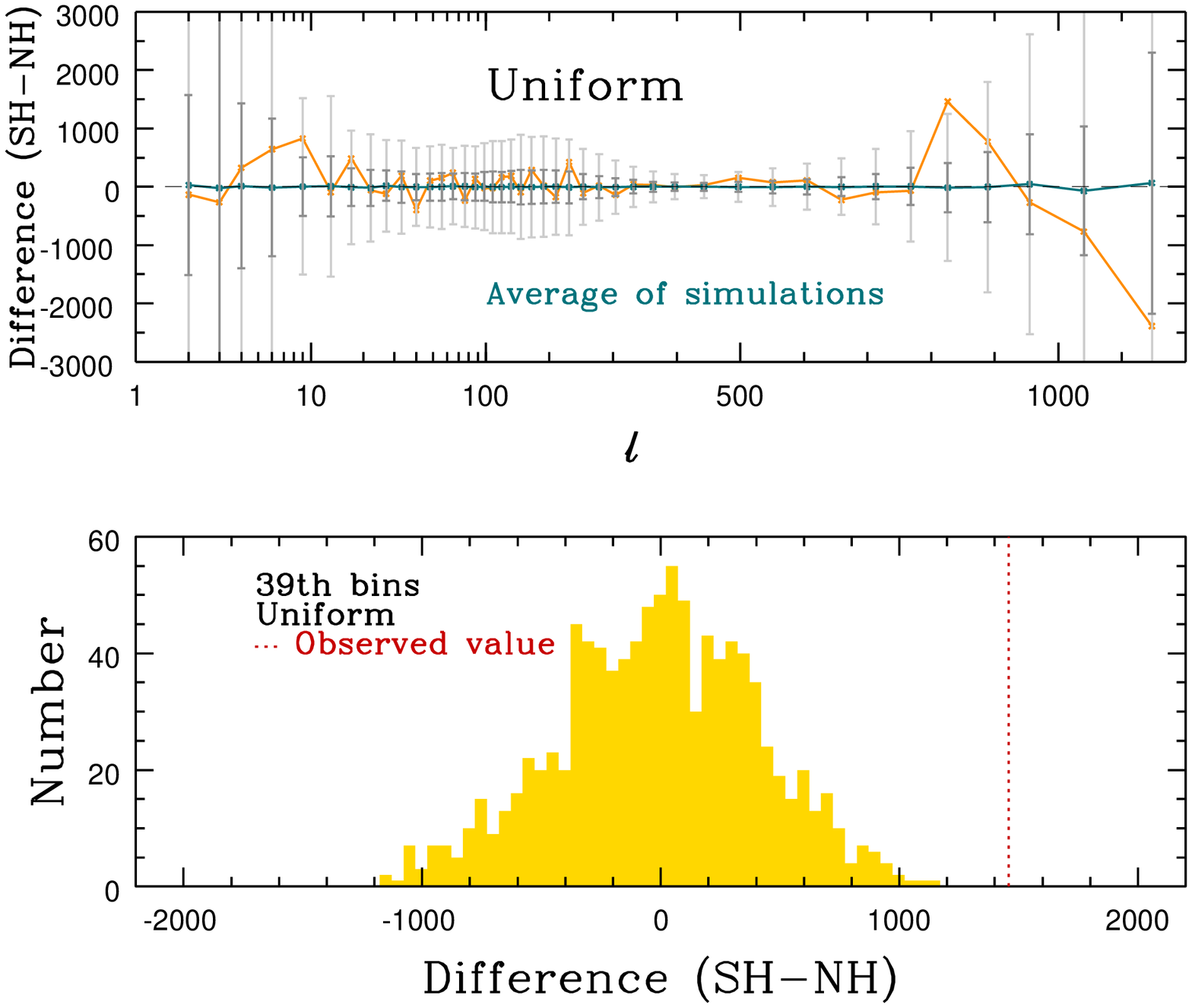} %
\epsfxsize=4.25cm \epsfbox{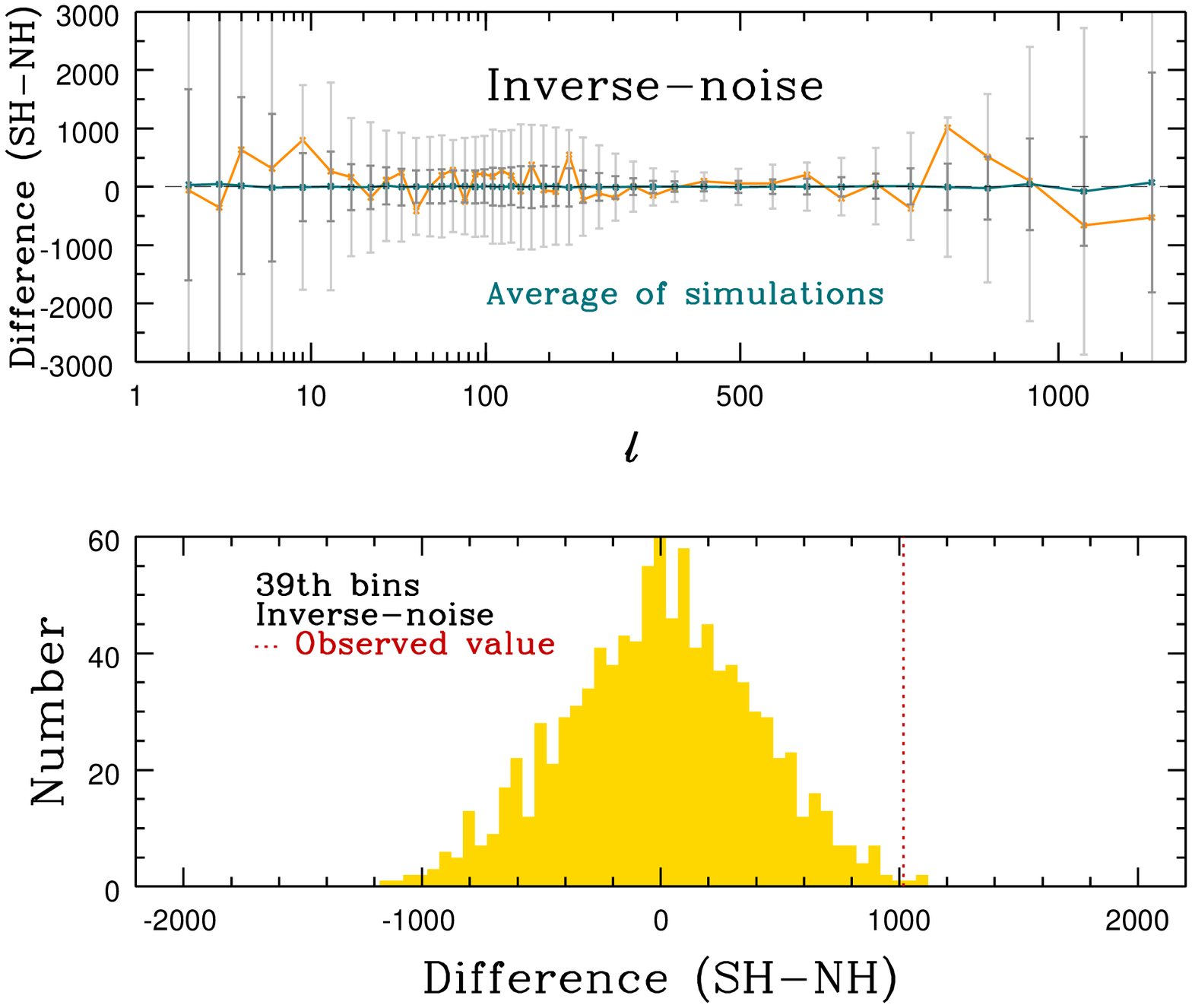}
\epsfxsize=4.25cm \epsfbox{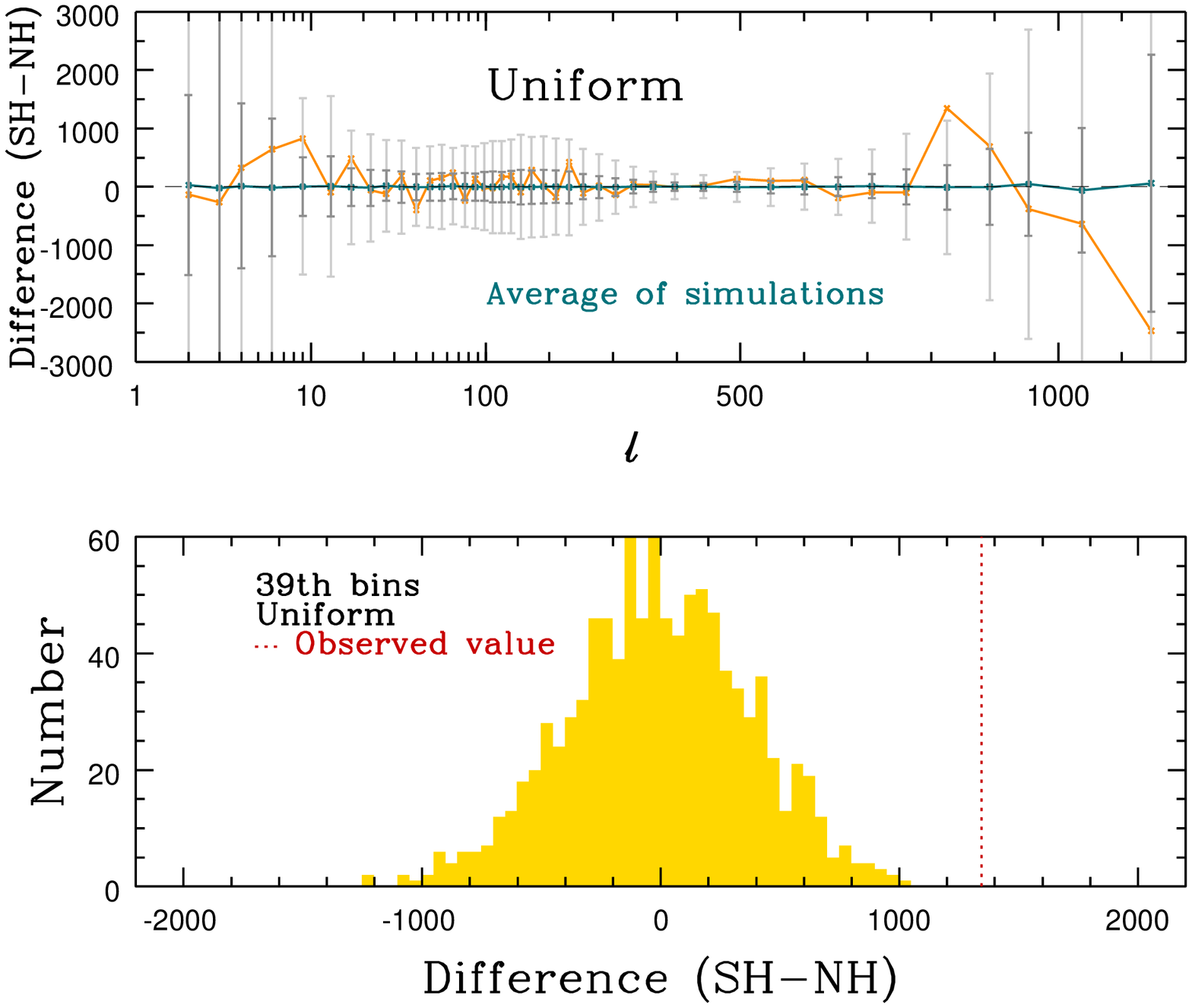} %
\epsfxsize=4.25cm \epsfbox{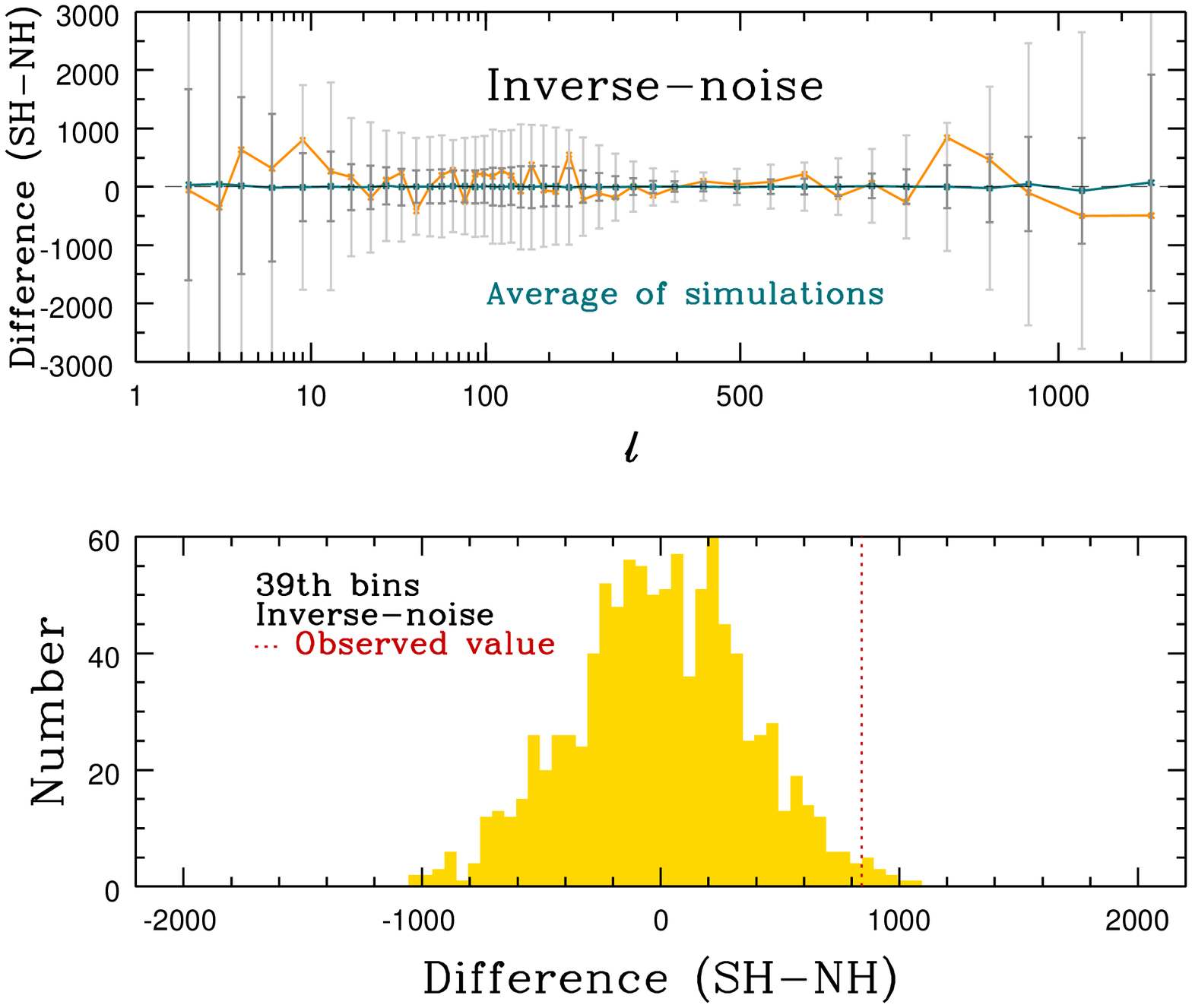}
\caption{Similar to Fig.\ \ref{hats difference histogram}
         but now the bin-width has been increased by a factor of $1.2$
         (first two panels) and $1.5$ (last two panels).
         The 39th bin centered at $l=825$ corresponds to our focused bin
         located around at the third peak with the bin ranging over 
         $796 \le l \le 855$ ($\Delta l=60$) for 20\% increased bin-width
         and $788 \le \ell \le 862$ ($\Delta l=75$) for 50\% increased one.
         The results obtained in the uniform weighting scheme are shown
         on the first and the third panels while those in the inverse-noise
         weighting scheme on the second and the fourth ones.
         In both cases, the high Galactic latitude regions with
         $|b| \ge 30\deg$ have been used.
         For the power spectrum measured on the whole sky enclosed by KQ85
         mask (denoted as grey color), the original $l$-binning has been
         used as in Fig.\ \ref{hats difference histogram}. 
         }
\label{fig:b30_1.2_1.5bin}
\end{figure*}

Here we investigate the effect of the bin-width and the 
Galactic latitude cut on the detected north-south anomaly around the
third peak in the angular power spectrum measured from the WMAP 7-year data.

Our detection of the north-south anomaly is based on the 40th bin around
the third peak which ranges over $851 \le l \le 900$ ($\Delta l=50$);
see Fig.\ \ref{hats difference histogram}.
By fixing the bin center at $l=825$, we have increased the bin-width
by 20\% ($796 \le l \le 855$; $\Delta l=60$) and
50\% ($788 \le l \le 862$; $\Delta l=75$).
During the re-binning process, we reduce the total number of bins into 43
and widen the width of adjacent bins appropriately; now our focused bin
centered at $l=825$ is located at the 39th bin. Based on the new $l$-binning,
we have measured angular power spectra from the WMAP 7-year data and one
thousand sets of WMAP mock observations.
The results are shown in Fig.\ \ref{fig:b30_1.2_1.5bin}. 
In the case of the uniform weighting scheme, the increase of the
bin-width increases the statistical significance of the north-south anomaly;
there is no occurrence of such an anomaly larger than the observed one
among the one thousand WMAP simulations.
However, in the case of the inverse-noise weighting scheme,
the statistical significance slightly decreases for the 50\% increased
bin-width (see Table \ref{tab:counts} for the summary of statistical
significances).

\begin{figure*}[t!]
\centering
\epsfxsize=4.25cm \epsfbox{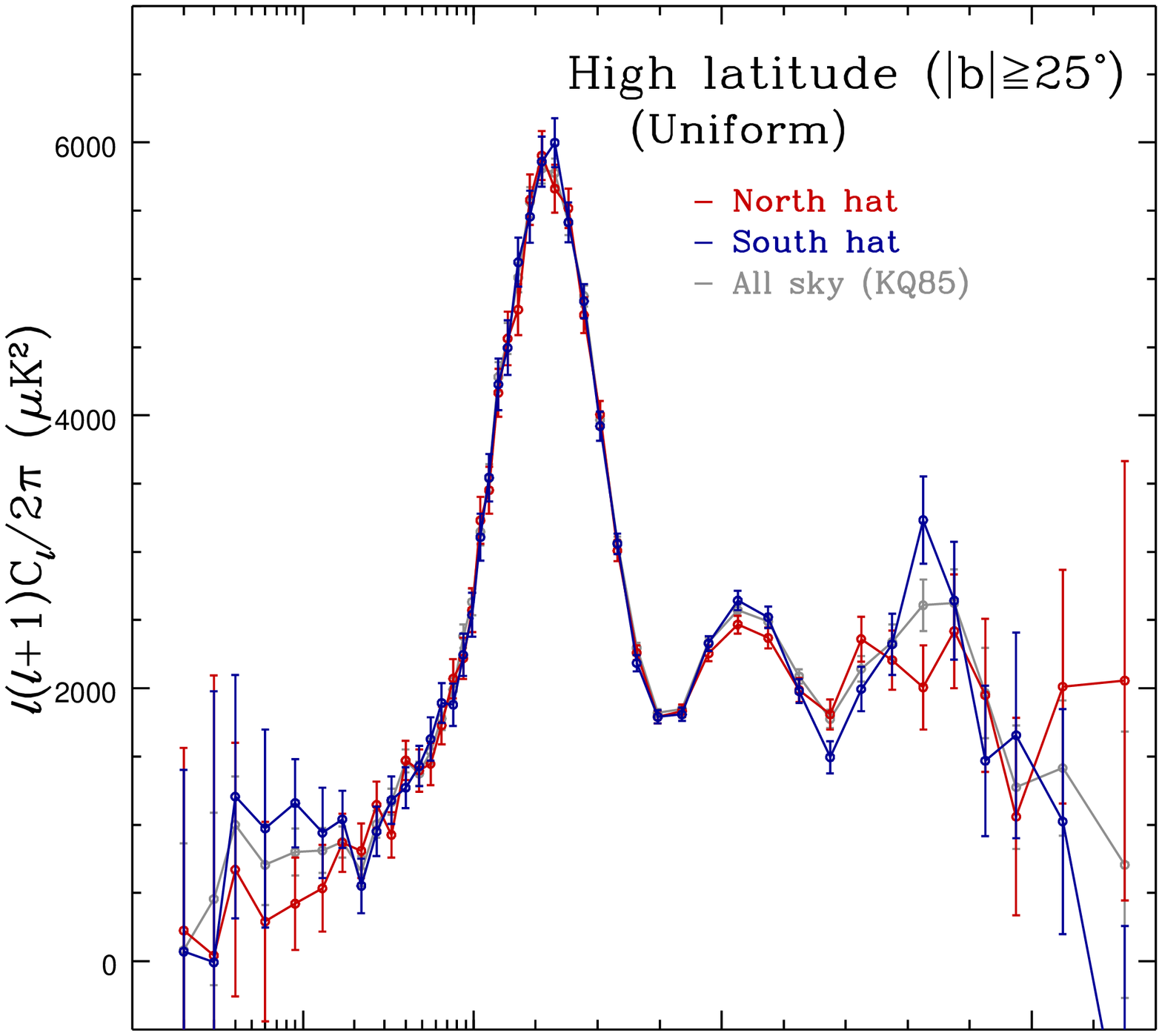} %
\epsfxsize=4.25cm \epsfbox{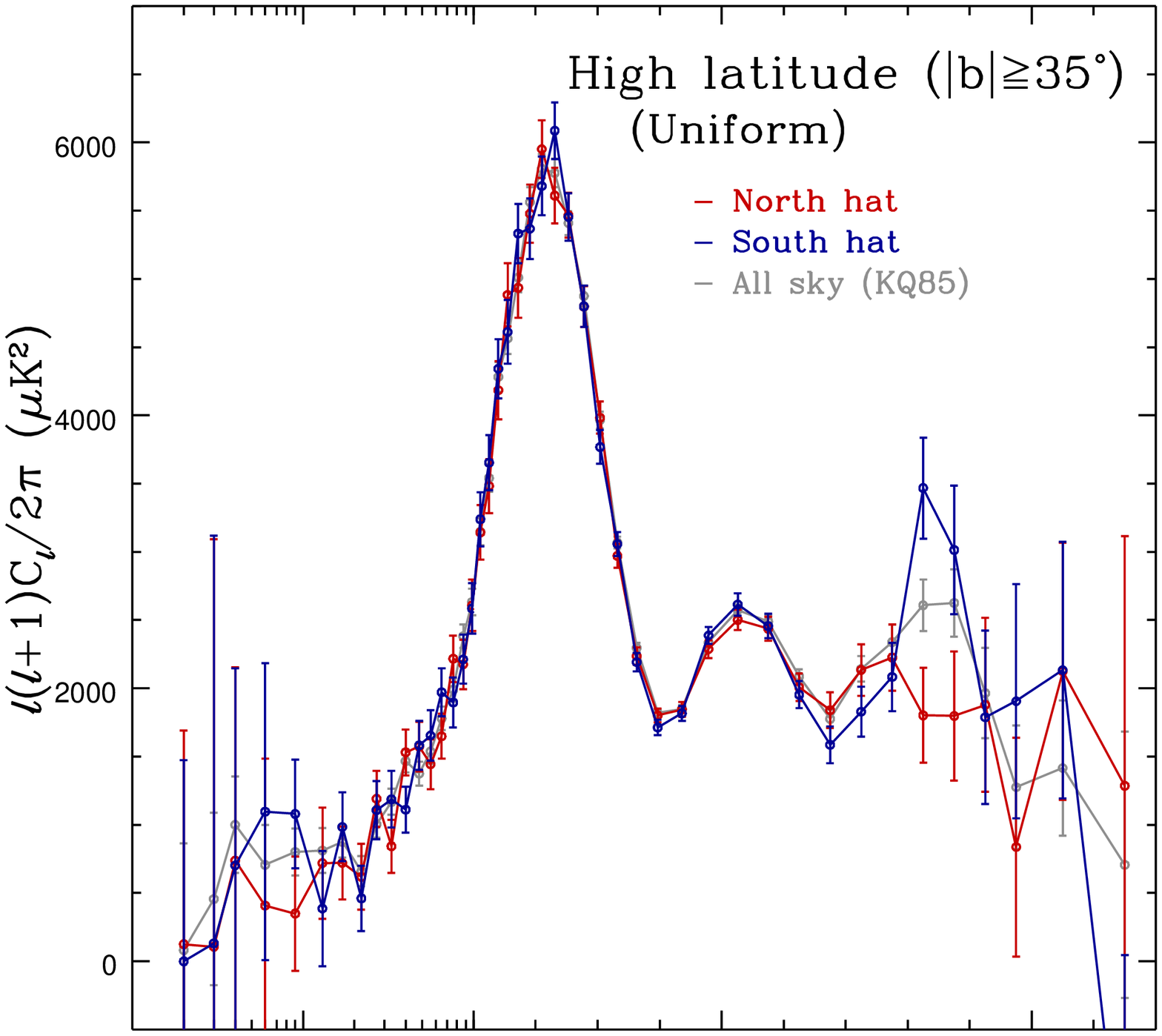}
\epsfxsize=4.25cm \epsfbox{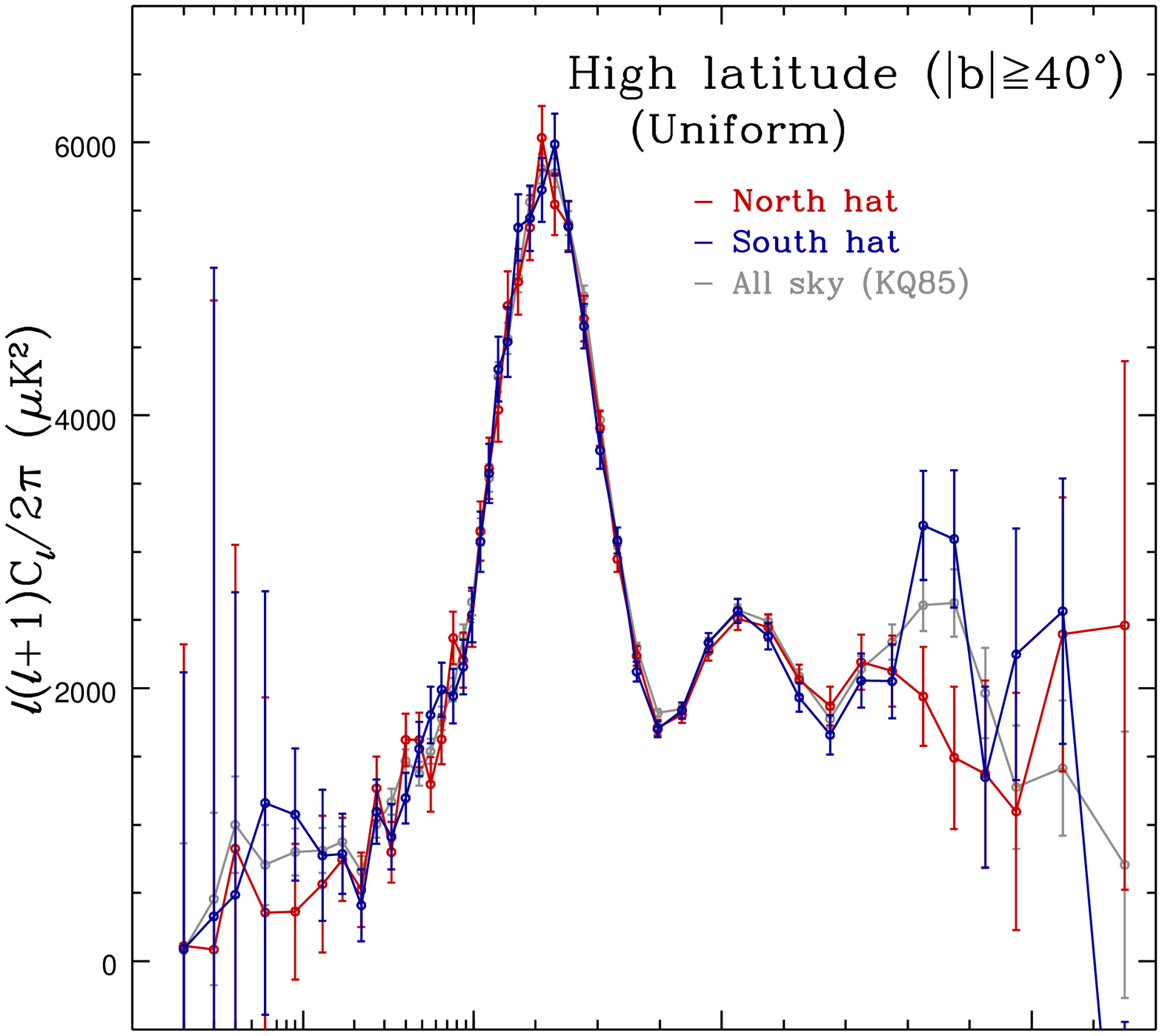}
\epsfxsize=4.25cm \epsfbox{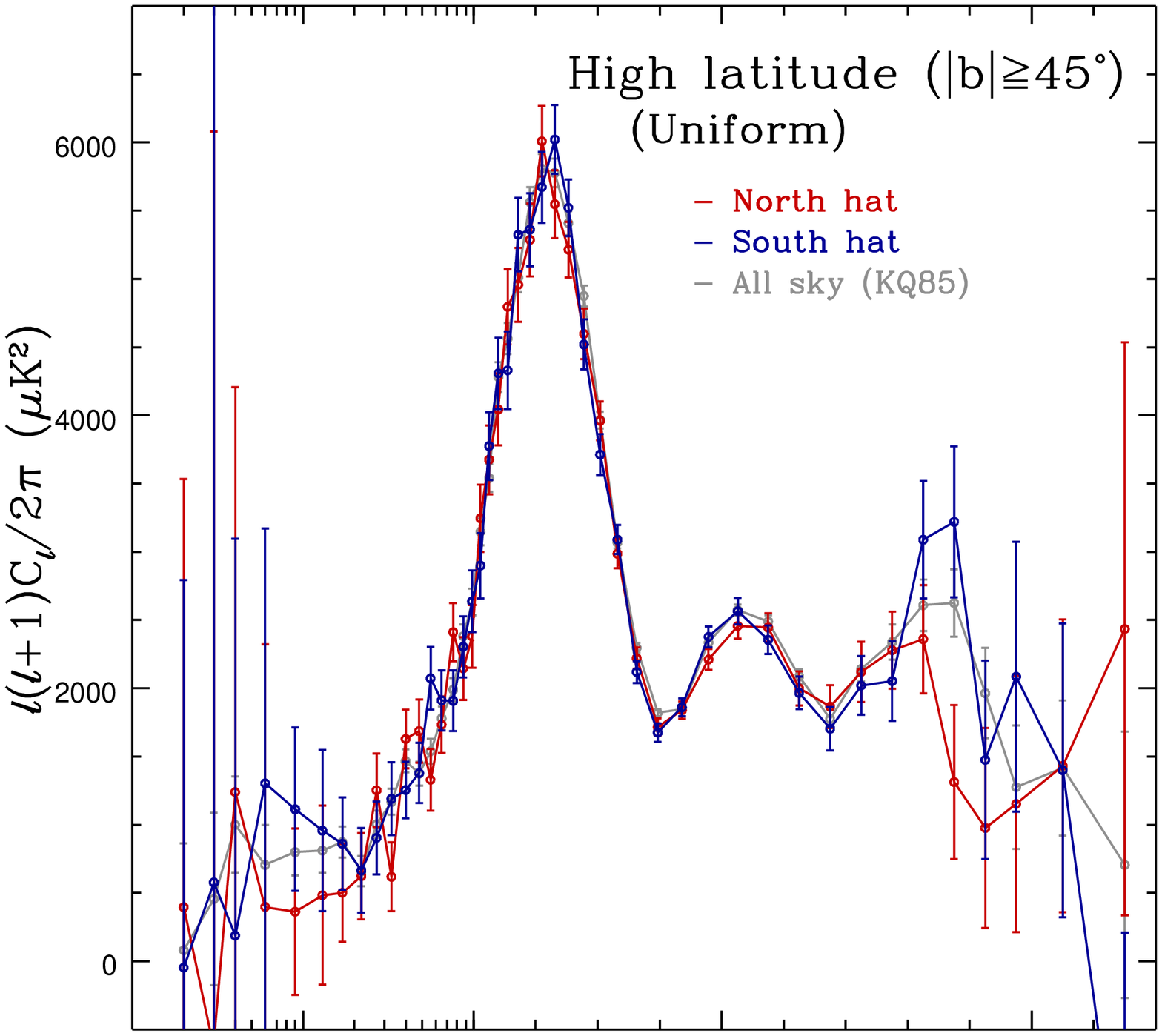} \\
\epsfxsize=4.25cm \epsfbox{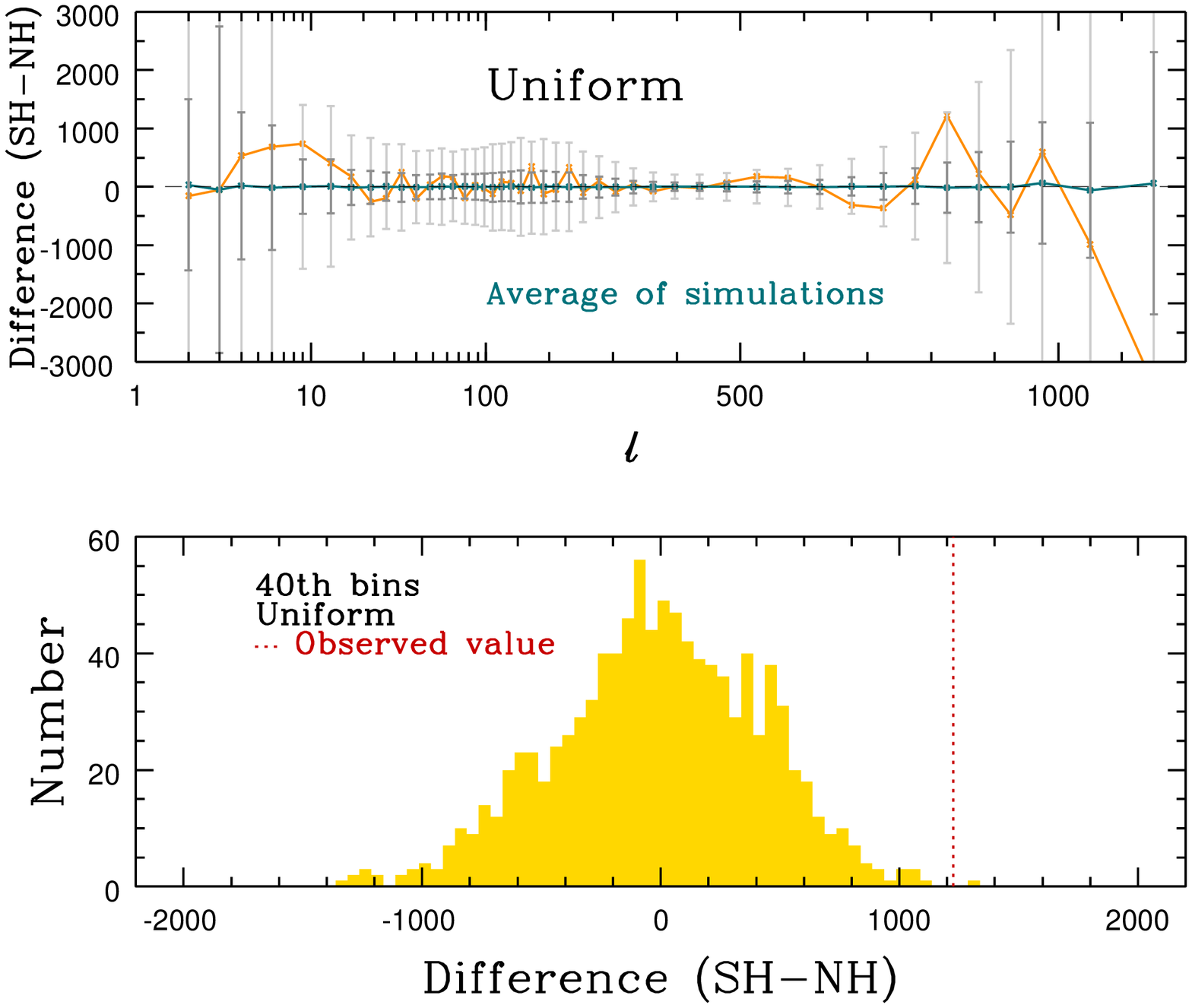} %
\epsfxsize=4.25cm \epsfbox{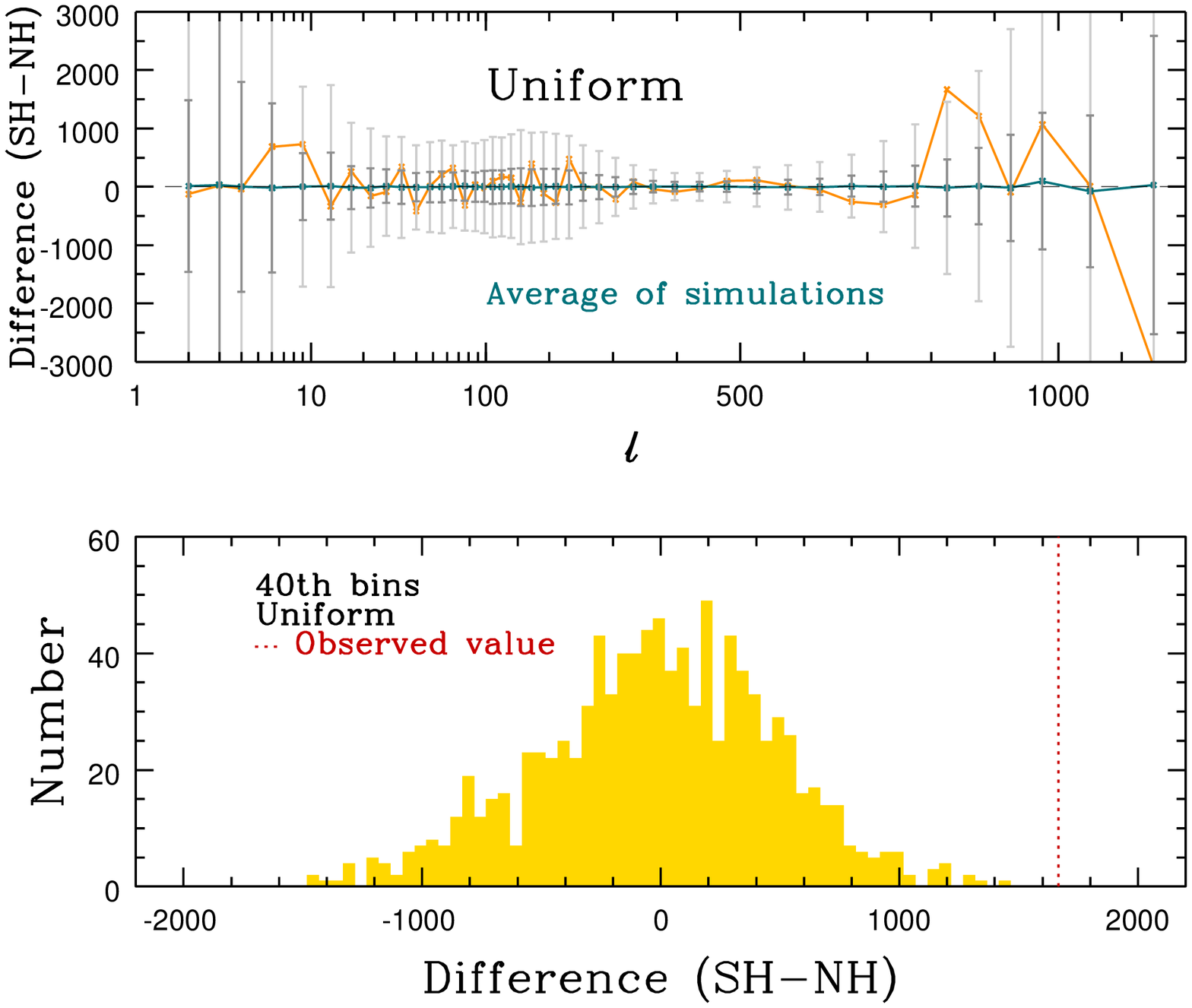}
\epsfxsize=4.25cm \epsfbox{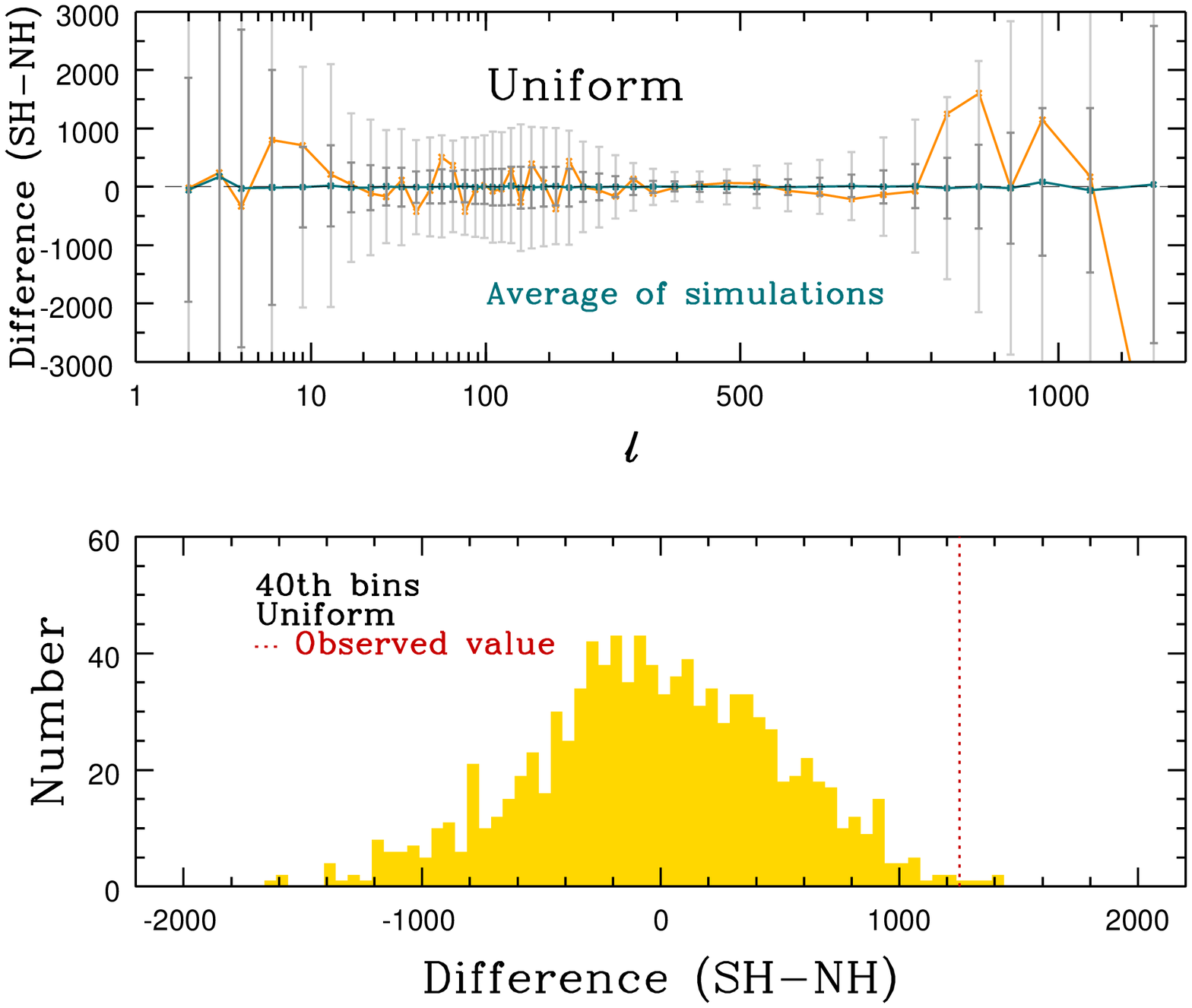}
\epsfxsize=4.25cm \epsfbox{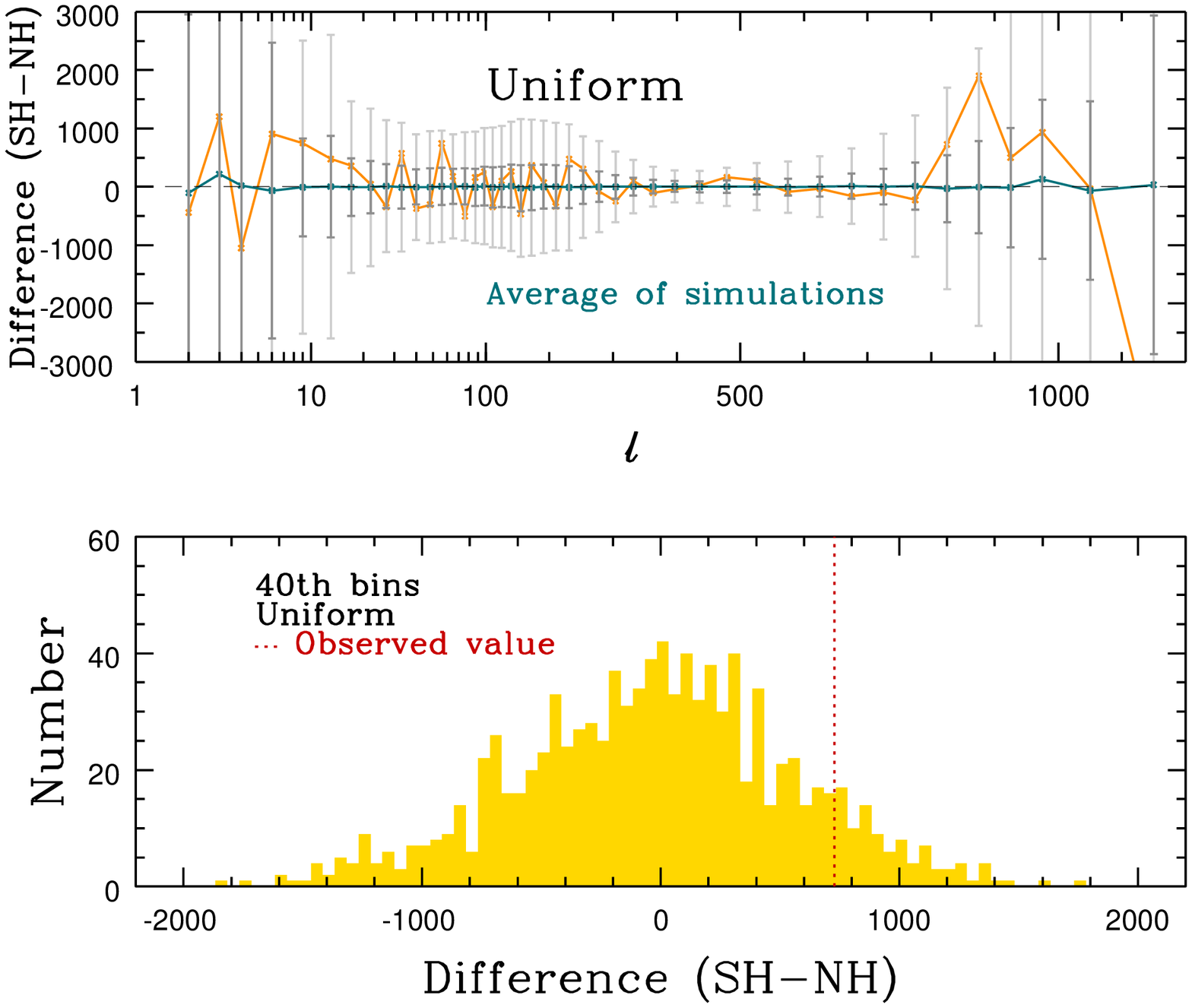} \\
\epsfxsize=4.25cm \epsfbox{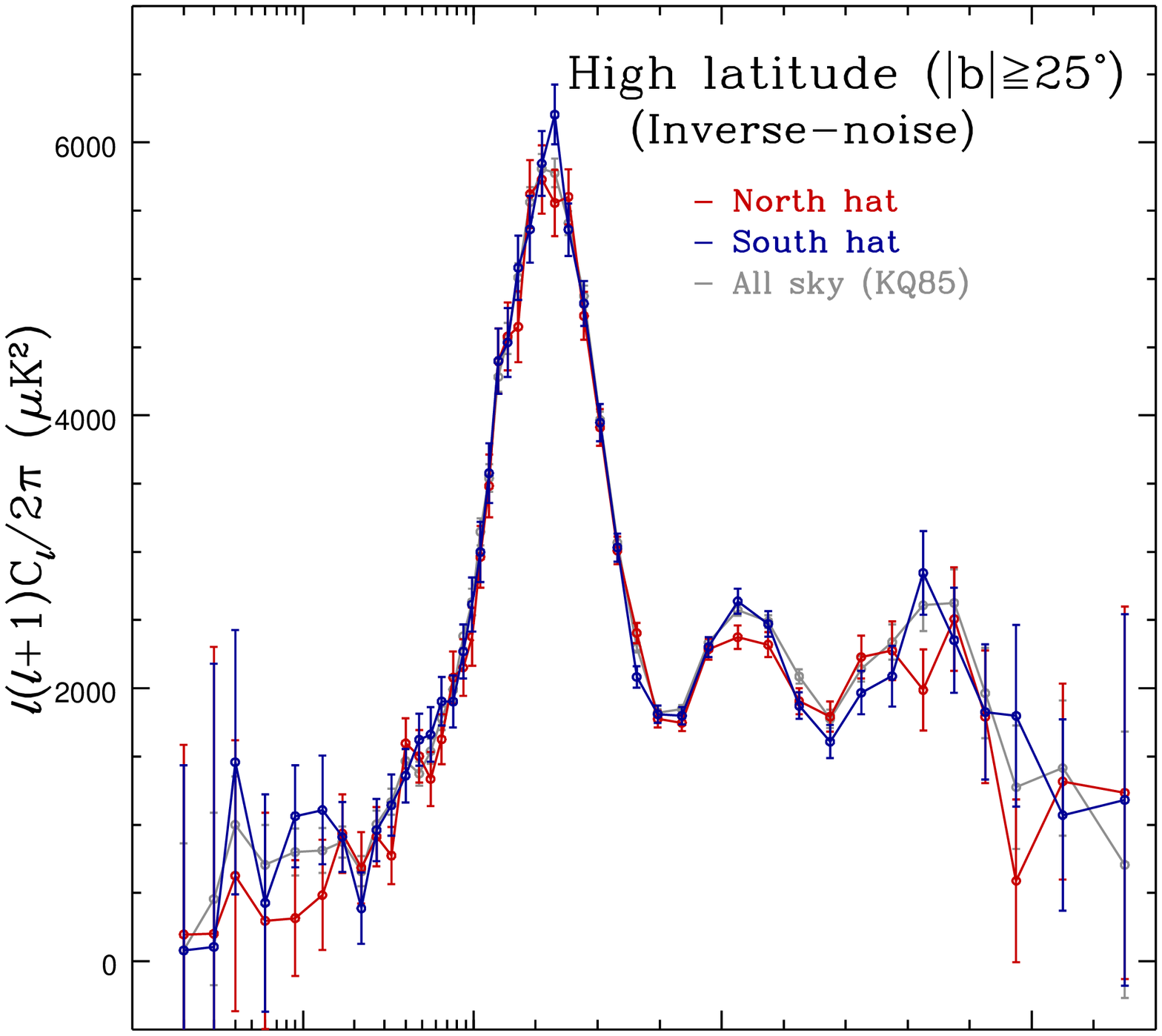} %
\epsfxsize=4.25cm \epsfbox{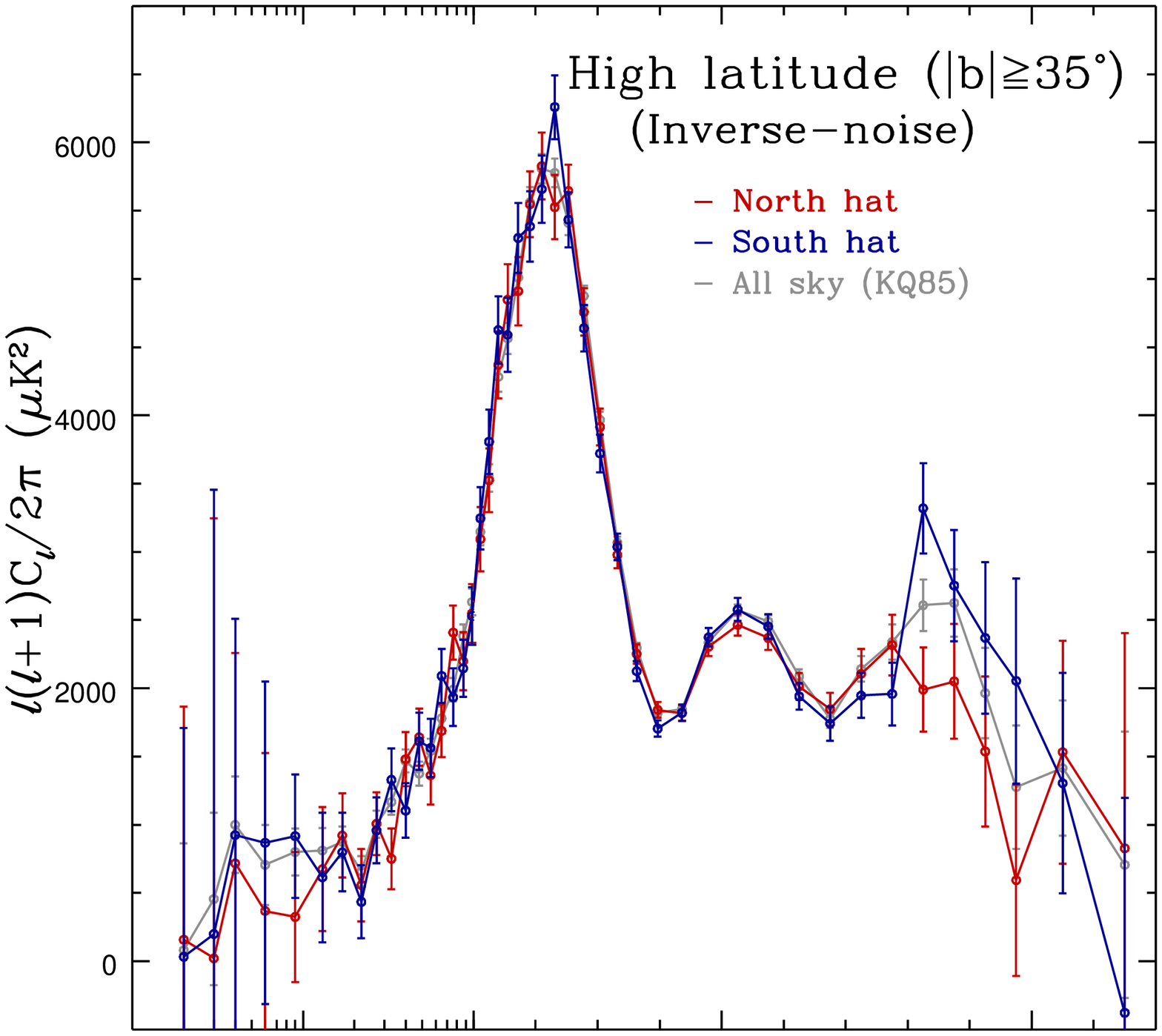}
\epsfxsize=4.25cm \epsfbox{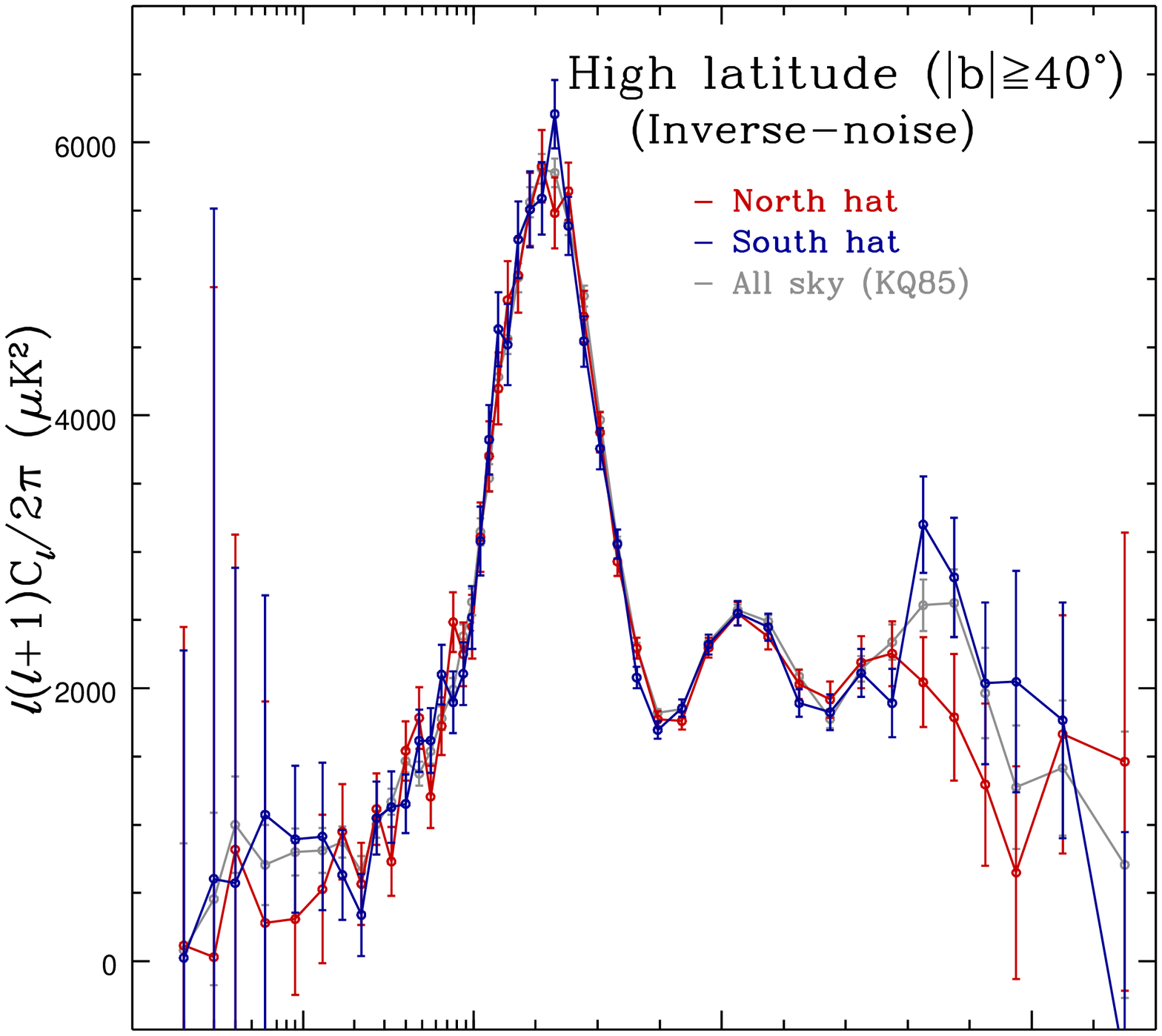}
\epsfxsize=4.25cm \epsfbox{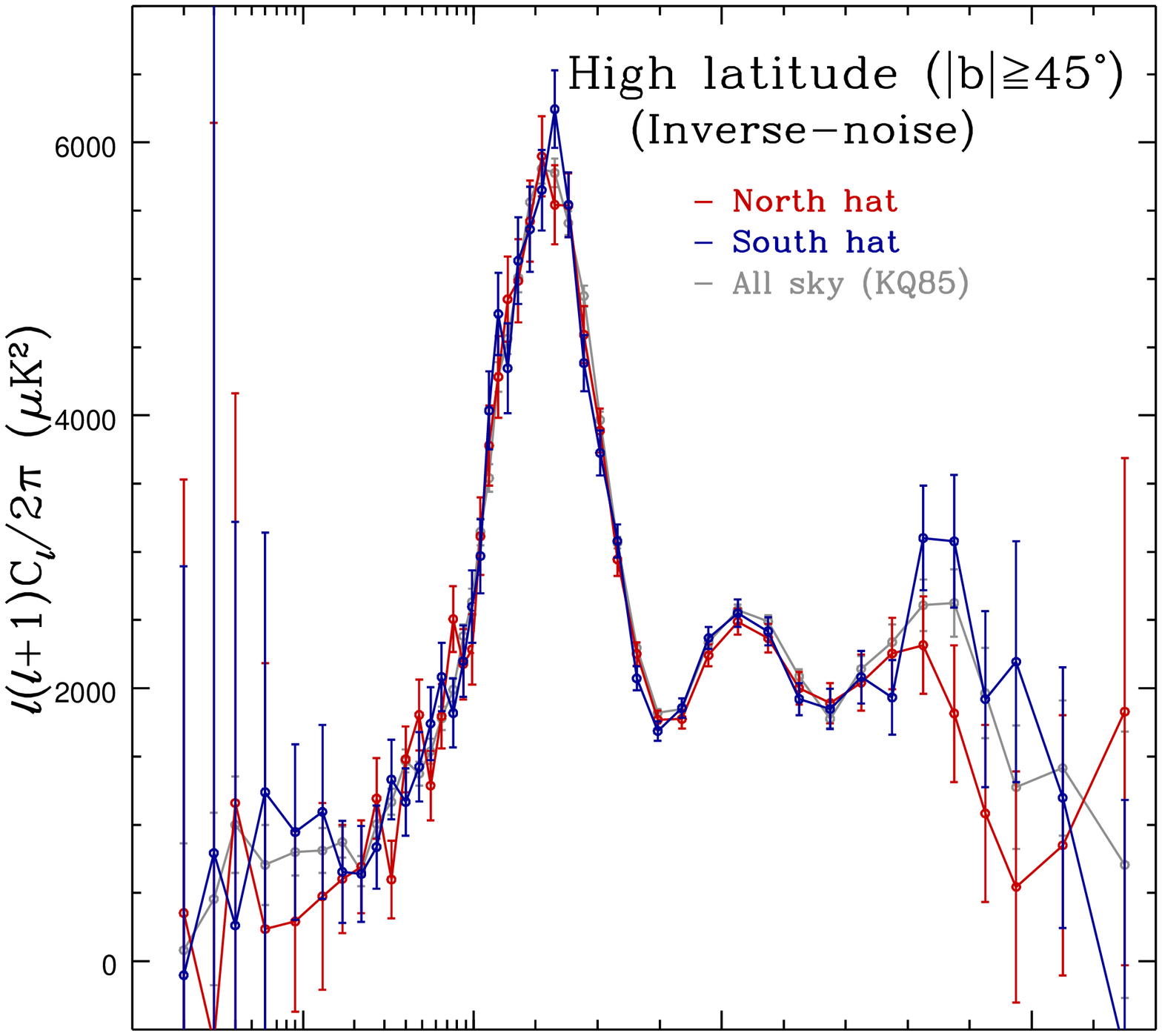} \\
\epsfxsize=4.25cm \epsfbox{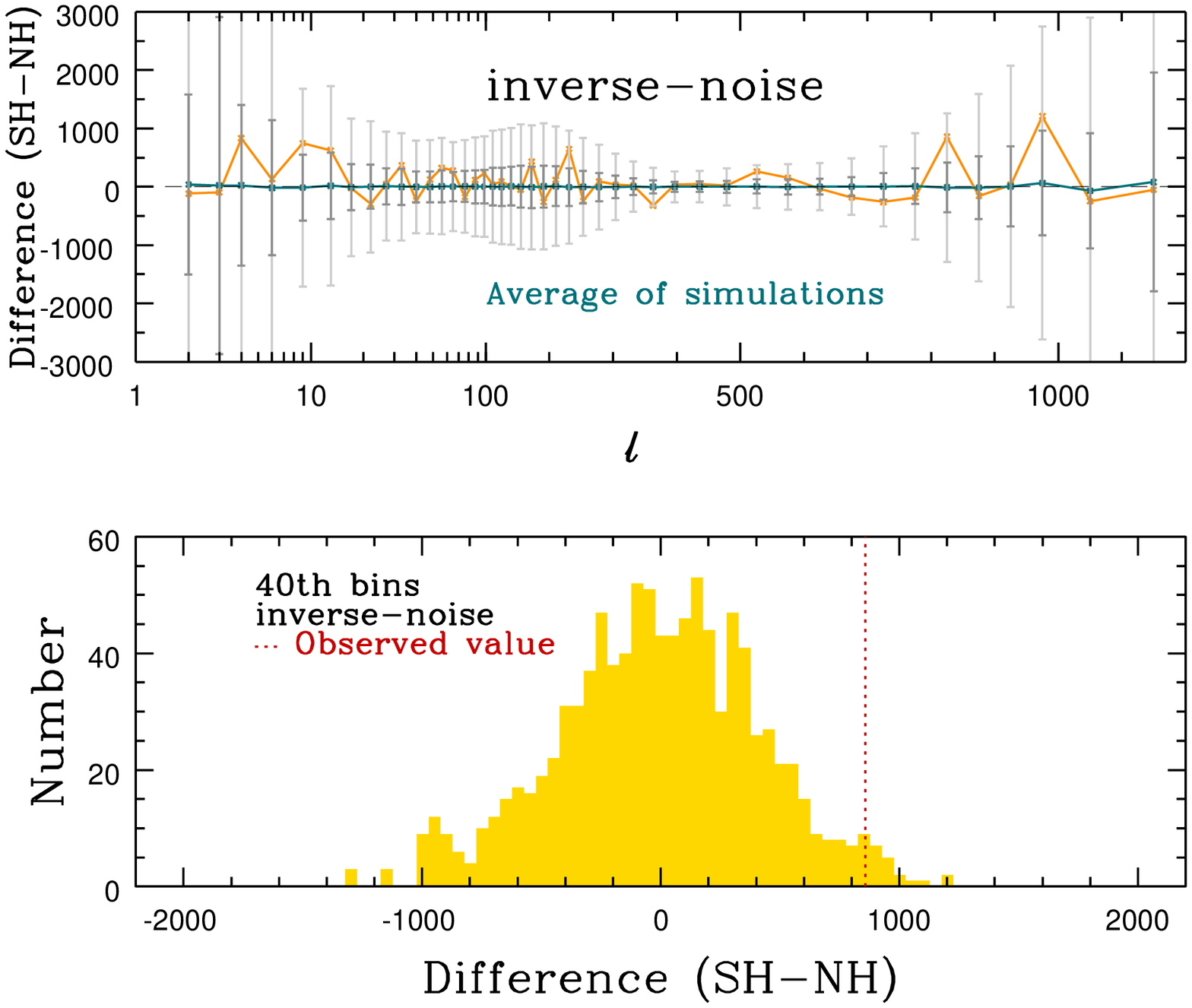} %
\epsfxsize=4.25cm \epsfbox{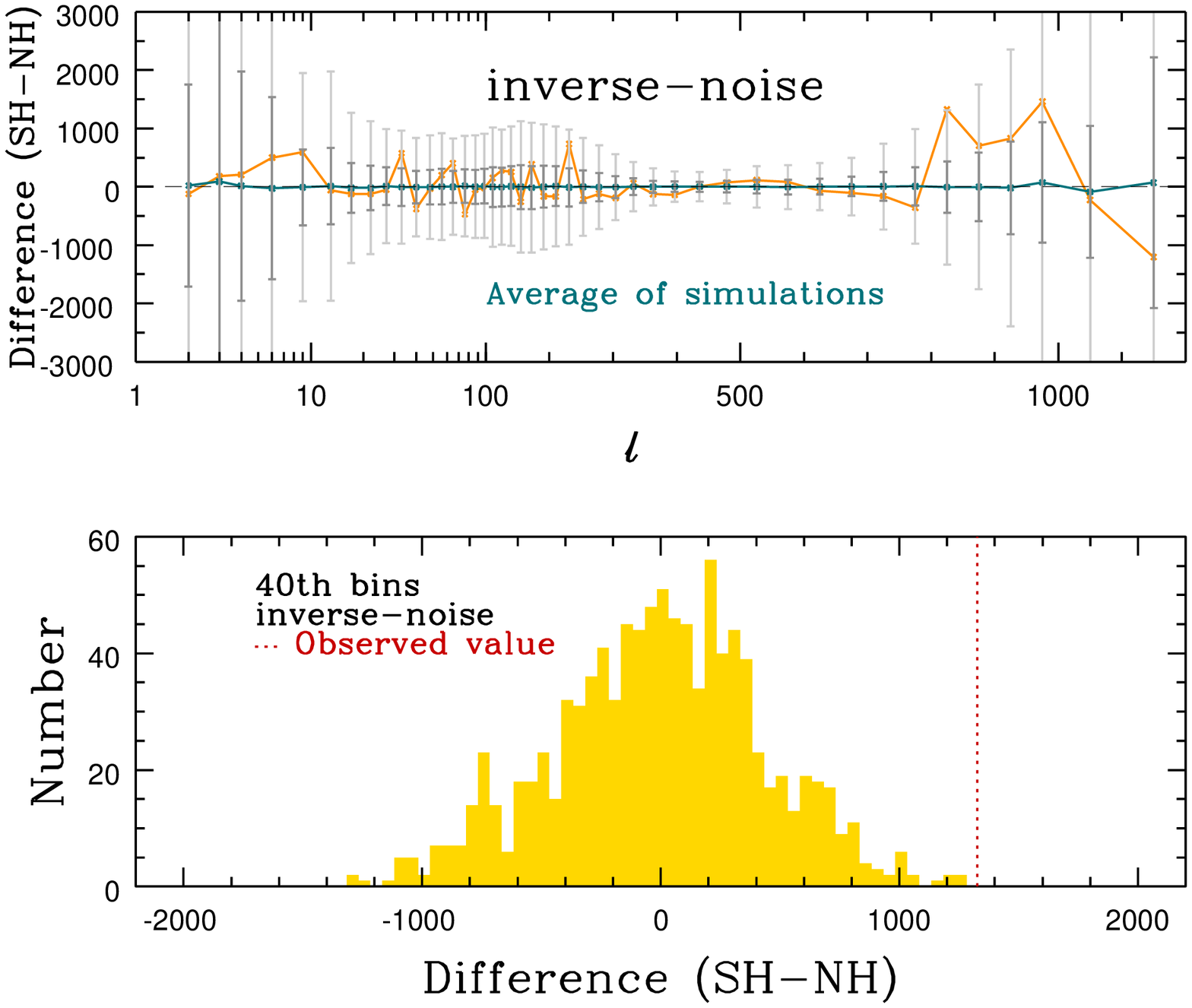}
\epsfxsize=4.25cm \epsfbox{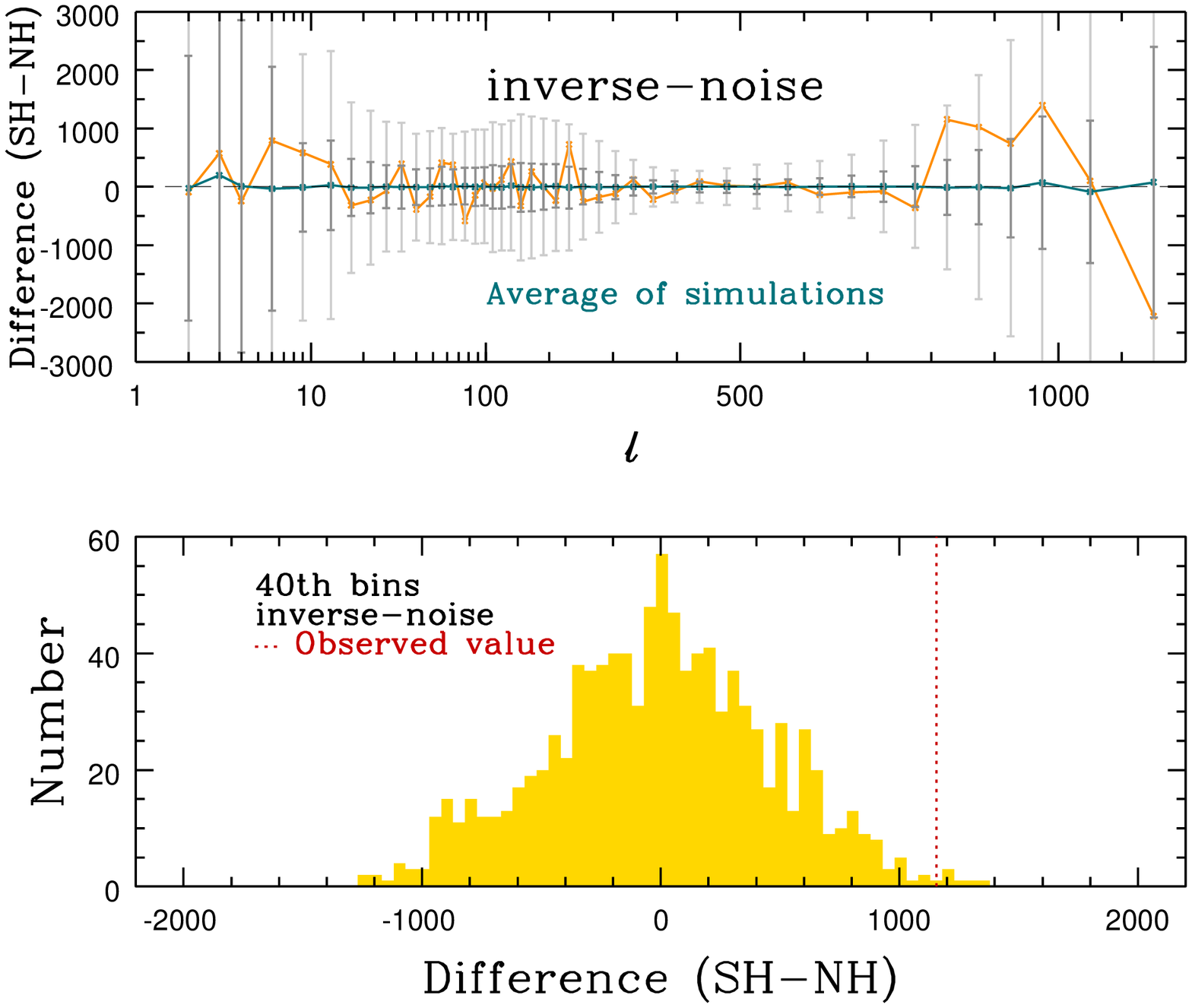}
\epsfxsize=4.25cm \epsfbox{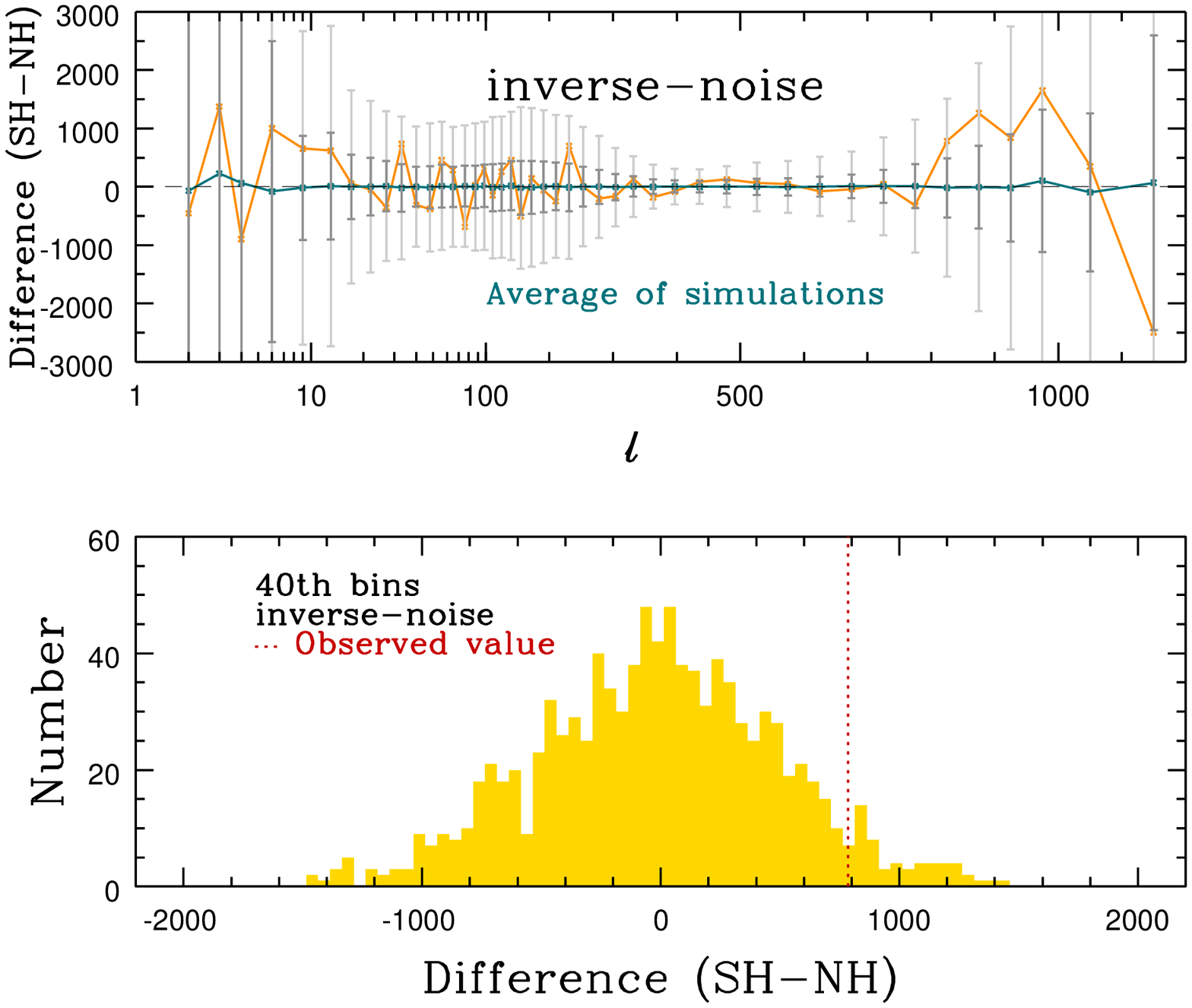}
\caption{Angular power spectra measured on the north and the south hat
         regions defined by different Galactic latitude cuts
         (from left to right, $|b|=25\deg$, $35\deg$, $40\deg$,
         $45\deg$), with a similar format given in Fig.\
         \ref{hats difference histogram} (the case of $|b|=30\deg$).
         Fractions of the sky area for north (south) regions 
         are 27.7\% (27.6\%), 20.6\% (20.5\%),
         17.2\% (17.3\%), 14.1\% (14.2\%), respectively,
         in increasing order of the latitude cut.
         Top panels are for the uniform weighting scheme
         while bottom panels for the inverse-noise weighting scheme.
         }
\label{fig:glat_cut_dep}
\end{figure*}

We also look into how our conclusion is sensitive to
the Galactic latitude cut. We compare the power spectrum measurements
for different latitude cuts, $|b|=25\deg$, $30\deg$, $35\deg$, $40\deg$,
and $45\deg$ (Fig.\ \ref{fig:glat_cut_dep}; the case of $|b|=30\deg$ is
presented in Fig.\ \ref{hats difference histogram}).
We notice that the north-south anomaly becomes the most significant in
the case of $|b|=35\deg$ cut, and no such an anomaly occurs in both
weighting schemes among the one thousand WMAP mock data sets
(see Table \ref{tab:counts}).
Furthermore, the statistical significance of this anomaly becomes weaker
as the latitude cut higher or lower than $|b|=35\deg$ is applied.
For $|b|=25\deg$ ($45\deg$), the $p$ value is $5.9$\% ($12$\%) 
in the inverse-noise weighting scheme.
Up to the Galactic latitude cut $|b|=40\deg$, the north-south anomaly is
maintained with a high statistical significance (see Table \ref{tab:counts}).
For higher latitude cuts like $|b| \ge 45\deg$, however,
the north-south anomaly in the power spectrum amplitude does not have
the statistical significance any more.

According to the two results shown above, the observed north-south anomaly
has weak dependences on the bin-width used in the power spectrum estimation
and the Galactic latitude cut, which implies that the north-south anomaly
is the realistic feature present in the WMAP data and supports our 
primary conclusion.

\section{Conclusions}
\label{sec:conclusions}

In this work we have compared the angular power spectra measured on
various local sky regions using the WMAP 7-year temperature anisotropy data.
The sky regions studied in this work are the low Galactic contamination
regions at high Galactic latitude (north hat and south hat), the strong
Galactic contamination regions at low Galactic latitude north and south
regions, the regions dominated by the WMAP instrument noise
(ecliptic plane regions), and the regions of low instrument noise
(ecliptic pole regions).

We found that the power spectra around the third peak measured
on the north hat and the south hat regions show an anomaly which is
statistically significant, deviating around $3\sigma$ from the
$\Lambda\textrm{CDM}$ model prediction.
We have tried to identify the cause of this anomaly by performing the
similar analysis on the low latitude regions and the regions with high
or low instrument noise.
Curiously, in the low Galactic latitude ($|b|<30\deg$) there appears
another but less statistically significant north-south anomaly around
the third peak, whose behavior is opposite to the one seen in the high
latitude regions and compensates the anomalies in the whole northern
and southern hemispheres.
At the present situation, we cannot draw any firm conclusion for
the origin of the observed anomaly.
However, we found that the observed north-south anomaly maintains with
the high statistical significance in the power spectra measured on
the regions with high instrument noise, and the anomaly becomes weaker
in the power spectra on the regions with low instrument noise.
Thus, in our present analysis the observed anomaly is significant on
the sky regions that are dominated by the WMAP instrument noise.
We have verified that the observed north-south anomaly around the third
peak has only weak dependences on the bin-width used in the power spectrum
estimation and the Galactic latitude cut, which strengthens our conclusion
(see Figs.\ \ref{fig:b30_1.2_1.5bin} and \ref{fig:glat_cut_dep}).

The location of the third peak in the angular power spectrum
corresponds to the angular scale that approaches the WMAP resolution
limit and that is dominated by the WMAP instrument noise.
Because the Planck satellite probes the CMB anisotropy with
higher angular resolution and sensitivity than the WMAP, we expect that
the origin of the observed anomaly will be identified in more detail
when the Planck data becomes publicly available \citep{Planck-etal-2011a}.

\begin{figure}[!t]
\centering
\epsfxsize=7.5cm \epsfbox{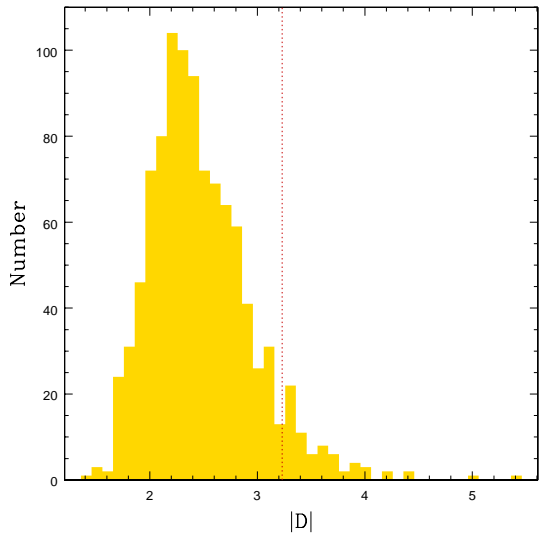}
\caption{A histogram of one thousand values of the maximum (among 45 bins)
         north-south power difference in unit of standard deviation,
         $|D|=\textrm{Max}(|C_b^{\textrm{\tiny NH}}-C_b^{\textrm{\tiny SH}}|/\sigma_b)$,
         where $C_b^\textrm{\tiny NH}$ ($C_b^\textrm{\tiny SH}$) is the $b$-th band power
         measured at the north (south) hat region ($|b| \ge 30\deg$),
         and $\sigma_b$ is the standard deviation of the power difference
         at the $b$-th multipole bin estimated from the WMAP simulation data
         sets, equivalent to the size of dark grey ($1\sigma$) error bars
         in the middle-left panel of Fig.\ \ref{hats difference histogram}.
         The vertical red dashed line indicates the measured value with
         $|D|=3.23$.
         }
\label{D_hat}
\end{figure}

It is possible that our detection of the north-south anomaly may be
strongly driven by {\it a posteriori} statistics based on the fact
that we have computed the power spectra on two different parts of the sky and
only then noticed a peculiar discrepancy between the two power amplitudes
in one bin around the third peak, neglecting the fact that there could have
been a similar discrepancy at any of the other multipole bins
(see \citealt{Zhang-Huterer-2010} for the related discussion).
One may argue that we may properly quantify the statistical
significance of the detected anomaly from the distribution of maximum
north-south differences (in unit of standard deviation) over all multipole
bins. The result for north and south hats ($|b|\ge 30\deg$) in the uniform
weighting scheme is shown in Fig.\ \ref{D_hat}.
In the histogram of one thousand values of the maximum (among 45 bins)
north-south power difference in unit of standard deviation, each estimated
from the WMAP mock observations, the detected north-south anomaly around
the third peak becomes statistically less significant; the probability
of finding cases with a deviation larger than the measured value is
$p=75/1000=7.5$\% which is slightly less than $2\sigma$ confidence limit.

However, it should be emphasized that the point described above is correct
{\it only} under the condition that the power spectrum at different multipoles
is equally affected by exactly the same CMB physics, Galaxy foreground,
and instrument noise properties, which is generally not true in our case
of harmonic space. The power spectrum at low multipoles is dominated by
the integrated Sachs-Wolfe effect and is more affected by the survey
geometry while the power spectrum at higher multipoles is more concerned
with the physics of acoustic oscillation and the WMAP instrument noise.
For example, it is not justified to consider all multipole bins in
quantifying the statistical significance of an anomaly detected in one
multipole bin {\it if} the anomaly originates from the phenomenon that is
influential at the local multipole range.
Therefore, our original interpretation of the detected north-south anomaly
around the third peak as statistically significant based on the analysis
focusing on the corresponding multipole bin is still valid
before the origin of the anomaly becomes unveiled.

As the Planck data will be soon available, we anticipate that our issue
of whether the anomaly is intrinsic one or due to the WMAP instrument noise
will be resolved by the Planck data.

\acknowledgments{
We acknowledge the use of the Legacy Archive for
Microwave Background Data Analysis
(LAMBDA). Support for LAMBDA is provided by the NASA Office
of Space Science. Some of the results in this paper have been derived
using the HEALPix and the CAMB softwares.
This work was supported by the Korea Research Foundation Grant funded
by the Korean Government (KRF-2008-341-C00022).}


\end{document}